\documentclass[5p,twocolumn]{elsarticle}

\usepackage{amssymb}

\usepackage{booktabs}
\usepackage{multirow}
\usepackage{bm}
\usepackage{amsmath}
\DeclareMathOperator*{\argmax}{argmax}
\DeclareMathOperator*{\argmin}{argmin}
\usepackage[ruled,vlined,linesnumbered]{algorithm2e}
\SetAlCapNameFnt{\footnotesize}
\SetAlCapFnt{\footnotesize}
\usepackage{subcaption}
\usepackage{lipsum}
\usepackage{xcolor}
\usepackage{xspace}
\usepackage{siunitx}
\usepackage{url}
\usepackage{placeins}
\usepackage{afterpage}
\usepackage{dblfloatfix}

\newtheorem{lemma}{Lemma}
\newtheorem{definition}{Definition}
\newtheorem{example}{Example}

\newcommand{\ttrkjoin}{\texttt{TTRKJoin}\xspace}

\journal{}

\begin{document}

\begin{frontmatter}



\title{ShallowBlocker: Improving Set Similarity Joins for Blocking}


\author[inst1,inst2]{Nils Barlaug}

\affiliation[inst1]{organization={Department of Computer Science, Norwegian University of Science and Technology},
            city={Trondheim},
            postcode={7034}, 
            country={Norway}}
\affiliation[inst2]{organization={Cognite},
            city={Lysaker},
            postcode={1366}, 
            country={Norway}}

\begin{abstract}
Blocking is a crucial step in large-scale entity matching but often requires significant manual engineering from an expert for each new dataset.
Recent work has show that deep learning is state-of-the-art and has great potential for achieving hands-off and accurate blocking compared to classical methods.
However, in practice, such deep learning methods are often unstable, offers little interpretability, and require hyperparameter tuning and significant computational resources.

In this paper,
we propose a hands-off blocking method based on classical string similarity measures: ShallowBlocker.
It uses a novel hybrid set similarity join combining absolute similarity, relative similarity, and local cardinality conditions with a new effective pre-candidate filter replacing size filter.
We show that the method achieves state-of-the-art pair effectiveness on both unsupervised and supervised blocking in a scalable way.
\end{abstract}




\end{frontmatter}



\section{Introduction}

Entity Matching (EM) is the task of identifying which records refer to the same real-world entity.
It is a core data integration task, and is done either within one data source (i.e., deduplication) or across data sources~\cite{christenDataMatchingConcepts2012, doanPrinciplesDataIntegration2012}.
One of the task's key characteristics is its quadratic nature.
The number of potential matches is quadratic in the number of records but the actual number of matches is typically linear in the number of records.
This has two important consequences:
1) It is often computationally infeasible to explicitly compare all potential pairs.
2) Most record pairs do not match because the ratio of non-matches to matches increases linearly with the number of records.
Therefore,
the problem is often solved in two steps --- blocking and then matching, where the blocking step generates a set of candidate record pairs and the matching step inspect the candidates to classify them as either match or non-match.
The idea is that the blocking procedure will return a sub-quadratic number of pairs in sub-quadratic time (ideally linear for both) while still recalling most matches,
paving the way for the matching procedure to achieve high precision using pairwise comparison in feasible time.

Blocking is a well-studied problem and there exists a multitude of techniques and methods~\cite{papadakisBlockingFilteringTechniques2021}.
Unfortunately,
it is often hard to solve the task in practice without expert knowledge.
One needs to be able to pick a suitable method and potentially hand-tune difficult to understand parameters for the data at hand.
Recently,
there has been substantial work on the use of deep learning for blocking \cite{ebraheemDistributedRepresentationsTuples2018, thirumuruganathanDeepLearningBlocking2021, zhangAutoBlockHandsoffBlocking2020}.
One of the potential benefits of such approaches is having a method that works satisfactorily across most datasets without expert handcrafting.
Of course,
deep learning is no silver bullet,
and often entails less interpretability, longer runtime, and tweaking training hyperparameters per dataset.
Additionally,
recent work highlights that deep learning methods does not necessarily outcompete traditional methods in terms of effectiveness \cite{papadakisHowReduceSearch2022}, despite earlier reports~\cite{thirumuruganathanDeepLearningBlocking2021}.

\paragraph{Proposed Method}

In this paper,
we propose ShallowBlocker, a novel blocking method based on set similarity joins.
The method relies on a new prefix filtering-based similarity join routine, \ttrkjoin, that we introduce.
The join routine combines constraints on absolute similarity, relative similarity, and local cardinality in order to exploit benefits of different similarity join types.
Furthermore, we leverage parallelization and propose a new pre-candidate filtering technique to get an efficient implementation.
To support hyperparameter selection, we perform analysis queries on randomly sampled records and known matches so that we can quickly estimate recall, number of retrieved pairs, and runtime for different hyperparameter configurations.
Finally, through carefully selected early cutoff on searches in \ttrkjoin we achieve approximate joins with quality guarantees.

ShallowBlocker consists of a strategy for choosing \ttrkjoin hyperparameters in both an unsupervised and supervised setting.
In the unsupervised setting the method tries to equally balance the pruning power of the three join constraints given a user-specified pair budget and degree of approximation.
While in the supervised setting the method optimizes the join hyperparameters according to an arbitrary user-specified objective function expressing the desired trade-off between recall, number of returned pairs, and runtime.

\paragraph{Contribution}
Our main contributions are:
\begin{itemize}
    \item We propose a new hands-off blocking method, ShallowBlocker, and show that it achieves state-of-the-art pair effectiveness for both unsupervised and supervised blocking in a scalable way.
    Importantly, it does not require elaborate dataset-specific tuning.
    For unsupervised use the user specifies pair budget and an optional approximation degree,
    while for supervised use the user can specify an arbitrary trade-off between recall, number of returned pairs, and runtime.
    \item We introduce a new expressive hybrid set similarity join primitive, $(\tau, \tau_r, k)$-join, suitable for blocking.
    \item We propose \ttrkjoin, an efficient $(\tau, \tau_r, k)$-join algorithm with a novel pre-candidate filtering technique and strong bounds.
    Furthermore, we describe a framework for quickly estimating the effect of different hyperparameters on \ttrkjoin and for performing approximate joins.
    ShallowBlocker introduces effective strategies for determining good hyperparameters for \ttrkjoin.
    \item We demonstrate that classical string similarity blocking methods still outperform deep learning-based blocking methods, such as DeepBlocker, on widely used benchmark datasets.
\end{itemize}

\paragraph{Outline}
We start by covering related work (Section~\ref{sec:related-work}) and problem statement (Section~\ref{sec:problem-statement}).
Then we introduce important set similarity join theory and techniques we will build upon (Section~\ref{sec:prefix-based-set-similarity-joins}).
In order to describe our proposed method we first discuss the new hybrid join primitive (Section~\ref{sec:hybrid-join}) followed by the proposed algorithm for it (Section~\ref{sec:ttrkjoin}),
a framework for estimating its behavior (Section~\ref{sec:estimation-framework}),
and how to do and interpret approximate joins (Section~\ref{sec:approximate-joins}) --- before we describe how everything goes together to form ShallowBlocker (Section~\ref{sec:shallowblocker}).
Finally,
we describe the experimental setup (Section~\ref{sec:experimental-setup}), go through the experiments with results (Section~\ref{sec:experiments}), and conclude (Section~\ref{sec:conclusion}).

\section{Related work}\label{sec:related-work}

Researchers have pursued many approaches for doing blocking,
and it is outside the scope of this paper to provide an extensive overview.
We cover only the most relevant and refer the reader to \citet{papadakisBlockingFilteringTechniques2021} for a detailed survey.

\subsection{Set Similarity Joins}

If we convert records to token sets (e.g., q-grams or words) we can cast blocking as a set similarity join problem,
where the goal is to find all pairs with similarity above some threshold.
SSJoin~\cite{chaudhuriPrimitiveOperatorSimilarity2006} and AllPairs~\cite{bayardoScalingAllPairs2007} proposed the highly effective prefix filtering technique for generating candidate pairs in addition to size filtering.
PPJoin~\cite{xiaoTopkSetSimilarity2009,xiaoEfficientSimilarityJoins2011} extends AllPairs with an additional candidate-time filter: the positional filter.
L2AP~\cite{anastasiuL2APFastCosine2014} specializes prefix filtering for Cosine similarity measure and introduce tighter bounds based on the Cauchy–Schwarz inequality.
There is a multitude of methods with more elaborate and aggressive filtering techniques,
both prefix-based and other~\cite{papadakisBlockingFilteringTechniques2021}.
However,
\citet{mannEmpiricalEvaluationSet2016} show that with efficient implementations the overhead of the filtering techniques often outweighs the benefits,
and that AllPairs and PPJoin is the best performing methods and state-of-the-art.

\subsection{Deep Learning}
The research community have increasingly focused on the use of deep learning for entity matching the last few years~\cite{barlaugNeuralNetworksEntity2021},
and some also target the blocking step.
DeepER~\cite{ebraheemDistributedRepresentationsTuples2018} uses pre-trained word embeddings and LSTMs to embed records.
It generates candidate pairs using random hyperplane Multi-Probe LSH,
and record embedding pairs are compared elementwise and then fed into a SVM to classify as match or no match.
The network and SVM is trained using labeled data.
AutoBlock~\cite{zhangAutoBlockHandsoffBlocking2020} improves over DeepER by using a novel attention mechanism and cross-polytope LSH.
DeepBlocker~\cite{thirumuruganathanDeepLearningBlocking2021} is a state-of-the-art deep learning-based blocker.
The authors explore a wide range of configurations, including different training strategies and different network types such as large pre-trained transformer models.
Despite being self-supervised, the method is more effective than AutoBlock.

\section{Problem Statement}\label{sec:problem-statement}

Let $A$ and $B$ be two sets of records,
and let $M \subseteq A \times B$ be all record pairs across $A$ and $B$ that refer to the same entity.
In other words, for all $(a,b) \in M$ the records $a$ and $b$ are references to the same entity.
The goal of entity matching/resolution is to determine $M$.
For blocking in particular, the goal is to find a superset $P$ of $M$ such that $|P| \ll |A \times B|$ in subquadratic time and memory.
Informally, the purpose is to remove large amounts of obvious non-matches so that a high-precision pair comparison downstream does not have to process a quadratic number of pairs.
If $A$ and $B$ is the same set we call the problem deduplication.

When we evaluate blocking methods we are interested in three main performance characteristics:
\begin{enumerate}
    \item \textbf{Recall}:
    To what degree the method is able to find true matches.
    We measure this with the classical recall measure $R = \frac{|P| \cap |M|}{|M|}$.
    \item \textbf{Pruning Power}:
    To what degree the method is able to discard potential pairs.
    The most direct measure of this is $|P|$, but that is hard to interpret and compare across datasets.
    Reduction rate ($1 - \frac{|P|}{|A \times B|}$) is a popular measure,
    but since $|M|$ is generally expected to be linear in $|A|$ and $|B|$ it will increase quickly towards 1 for larger datasets if the blocking method is effective,
    which makes comparison across dataset sizes difficult.
    Therefore, in this paper, we will report the empirical cardinality $\widetilde{k} = \frac{|P|}{\min(|A|, |B|)}$.
    \item \textbf{Efficiency}:
    How fast can it be done and with what computational resources.
    We measure this mainly through runtime, but memory consumption and special hardware requirements (e.g., GPU) is also of interest.
\end{enumerate}
In practice, the former and the two latter are in conflict, so blocking methods will typically let the user adjust the trade-off between them.

\section{Set Similarity Joins Using Prefix Filtering}\label{sec:prefix-based-set-similarity-joins}

Our method builds heavily on prefix filtering-based set similarity joins.
Therefore, we will now describe existing set similarity join building blocks we exploit in our method,
while the next sections will outline the join routine in our method.

We find it useful to introduce these concepts through the perspective of the popular similarity join method PPJoin~\cite{xiaoEfficientSimilarityJoins2008, xiaoEfficientSimilarityJoins2011},
as it offers a familiar frame of reference for the reader and let us introduce our join routine as an evolution of PPJoin.
Algorithm~\ref{alg:ppjoin} shows PPJoin similar to how it is presented by the authors.
It performs self-join with jaccard similarity on unweighted sets.
We will use this as a running example of the different building blocks as we go through them.
At the same time we will also describe the same ideas extended to nonself-joins, other set similarity measures, and weighted sets --- resulting in a more generic version of PPJoin at the end of the section that prepares us for the next sections.

\begin{algorithm}[tbhp]
    \caption{PPJoin$(\mathcal{A}, \tau_j)$}
    \label{alg:ppjoin}
    \footnotesize
    \SetKwFunction{KwVerify}{Verify}
    \KwIn{
        $\mathcal{A}$ is a collection of token sets. Each token set is already sorted by $\mathcal{O}$.
        $\tau_j$ is the Jaccard threshold.
    }
    \KwOut{Token set pairs $\{ (a, a') \mid a \cap a' \geq \tau_j \}$}
    Sort $\mathcal{A}$ by increasing size\;
    $I \gets [\emptyset]_t$ \tcp*[r]{Inverted token indices}
    $P \gets \emptyset$ \tcp*[r]{Pairs with Jaccard similarity at least $\tau_j$}
    \ForEach{$a \in \mathcal{A}$}{%
        $O \gets$ empty map, default value 0\;
        \For(\tcp*[f]{Prefix filter}){$i \gets 1$ \KwTo $|a| - \lceil \tau_o \cdot |a| \rceil + 1$}{\label{alg:ppjoin-prefix-filter}%
            $t \gets a[i]$\;
            $I'_t \gets \{ (a', j) \in I_t \mid \tau_j \cdot |a| \leq |a'| \}$ \tcp*{Size filter} \label{alg:ppjoin-size-filter}
            \ForEach{$(a', j) \in I'_t$}{%
                $o^* \gets O[a'] + 1 + \min(|a| - i, |a'| - j)$\;
                \eIf(\tcp*[f]{Pos.\ filter}){$o^* \geq \frac{\tau_j}{\tau_j+1}(|a| + |b|)$}{\label{alg:ppjoin-positional-filter}%
                    $O[a'] \gets O[a'] + 1$\;
                }{%
                    Remove $O[a']$\;
                }
            }
            $I_t \gets I_t \cup \{ (a',i) \}$\; \label{alg:ppjoin-self-join-index}
        }
        $P \gets P \cup \text{\KwVerify{$a,\mathcal{A},O,\tau_j$}}$ \;
    }
    \KwRet{$P$}\;
\end{algorithm}

\subsection{Set Similarity Join}
A set similarity join finds all set pairs with some set similarity measure above a user-provided threshold $\tau$\footnote{We will see in Section~\ref{sec:hybrid-join} that there are other types of similarity joins.}.
For the purpose of this paper a set contains tokens from some record.
More formally, given two collections of token sets $\mathcal{A}$ and $\mathcal{B}$ we want to find all pairs $(a, b) \in \mathcal{A} \times \mathcal{B}$ that have similarity above some threshold $\tau$, i.e., find $C = \{(a,b) \in A \times B \mid \text{sim}(a,b) \geq \tau\}$.
If $\mathcal{A} = \mathcal{B}$ we call it a self-join and usually ignore self-referring pairs $(a,a)$.

\begin{table*}[tbhp]
    \footnotesize
    \centering
    \begin{tabular}{llccccc}
        \toprule
        & & & Prefix size & \multicolumn{2}{c}{Size bounds} & Equivalent overlap \\
        \cmidrule{5-6}
        & Measure & $sim(a,b)$ & $\pi(x,\tau)$ & $\lambda_l(a,\tau)$ & $\lambda_u(a,\tau)$ & $\alpha(a,b,\tau)$ \\
        \midrule
        
        \addlinespace[0.5em]
        
        \multirow{4}{*}[-2em]{Unweighted} &
        Jaccard &
        $ \displaystyle \frac{|a \cap b|}{|a \cup b|} $ &
        $ \displaystyle |x| - \lceil \tau \cdot |x| \rceil + 1 $ &
        $ \displaystyle \tau \cdot |a| $ &
        $ \displaystyle \frac{|a|}{\tau} $ &
        $ \displaystyle \frac{\tau}{\tau + 1} (|a| + |b|) $ \\
        
        \addlinespace[0.8em]
        
        &
        Cosine &
        $ \displaystyle \frac{|a \cap b|}{\sqrt{|a| \cdot |b|}} $ &
        $ \displaystyle |x| - \lceil \tau^2 \cdot |x| \rceil + 1 $ &
        $ \displaystyle \tau^2 \cdot |a| $ &
        $ \displaystyle \frac{|a|}{\tau^2} $ &
        $ \displaystyle \tau \sqrt{|a| \cdot |b|} $ \\
        
        \addlinespace[0.8em]
        
        &
        Dice &
        $ \displaystyle \frac{2 \cdot |a \cap b|}{|a| + |b|} $ &
        $ \displaystyle |x| - \bigg\lceil \frac{\tau \cdot |x|}{2 - \tau} \bigg\rceil + 1 $ &
        $ \displaystyle \frac{\tau \cdot |a|}{2 - \tau} $ &
        $ \displaystyle \frac{(2 - \tau) \cdot |a|}{\tau} $ &
        $ \displaystyle \frac{\tau (|a| + |b|) }{2} $ \\
        
        \addlinespace[0.8em]
        
        &
        Overlap &
        $ \displaystyle |a \cap b| $ &
        $ \displaystyle |x| - \tau + 1 $ &
        $ \displaystyle \tau $ &
        $ \displaystyle \infty $ &
        $ \displaystyle \tau $ \\
        
        \addlinespace[0.5em]
        
        \midrule
        
        \addlinespace[0.5em]
        
        \multirow{4}{*}[-2em]{Weighted} &
        Jaccard &
        $ \displaystyle \frac{ \sum_{k} \min(a_k, b_k) }{ \sum_{k} \max(a_k, b_k) } $ &
        $ \displaystyle (1 - \tau) \cdot \omega(x) $ &
        $ \displaystyle \tau \cdot \omega(a) $ &
        $ \displaystyle \frac{\omega(a)}{\tau}$ &
        $ \displaystyle \frac{\tau}{\tau + 1} \big(\omega(a) + \omega(b)\big) $ \\
        
        \addlinespace[0.8em]
        
        &
        Cosine$^*$ &
        $ \displaystyle \sum_{k} a_k \cdot b_k $ &
        $ \displaystyle \omega(x) - \tau $ &
        $ \displaystyle \tau $ &
        $ \displaystyle \infty $ &
        $ \displaystyle \tau $ \\
        
        \addlinespace[0.8em]
        
        &
        Dice &
        $ \displaystyle \frac{ 2 \cdot \sum_{k} \min(a_k, b_k) }{ |a| + |b| } $ &
        $ \displaystyle ( 1 - \frac{\tau}{2 - \tau} ) \cdot \omega(x) $ &
        $ \displaystyle \frac{\tau}{2 - \tau} \cdot \omega(a) $ &
        $ \displaystyle \frac{(2 - \tau)}{\tau} \cdot \omega(a) $ &
        $ \displaystyle \frac{\tau}{2} \big(\omega(a) + \omega(b)\big)  $ \\
        
        \addlinespace[0.8em]
        
        &
        Overlap &
        $ \displaystyle \sum_{k} \min(a_k, b_k) $ &
        $ \displaystyle \omega(x) - \tau $ &
        $ \displaystyle \tau $ &
        $ \displaystyle \infty $ &
        $ \displaystyle \tau $ \\
        
        \addlinespace[0.5em]
        
        \bottomrule
    \end{tabular}
    \caption{
        Definitions of the set similarity measures together with prefix size, size bounds, and equivalent overlap for both unweighted and weighted sets.
        $^*$Requires every set $x$ to be normalized such that $\| x \|_2 = 1$.
    }
    \label{tab:similarity-measures}
\end{table*}

In this paper, we consider four of the most used set similarity measures: Jaccard, Cosine, Dice, and Overlap.
Table~\ref{tab:similarity-measures} list their definitions.
Note that Jaccard, Cosine, and Dice are normalized and produce a similarity score between 0 and 1,
while Overlap may be arbitrary large.

These similarity measures are for unweighted sets,
but our method will rely on weighted sets.
There are several ways to generalize these measures to weighted sets.
We use a straightforward approach where the overlap is generalized to $\sum_k \min(a_k, b_k)$,
where $a_k$ is the weight of the $k$th globally ranked token in $a$ according to some global ordering $\mathcal{O}$ if $a$ contains this token or zero otherwise\footnote{This is different from \cite{xiaoEfficientSimilarityJoins2011}, where the weight for a given token is assumed to be global. Here the weight can be different for the same token in different sets --- which is important for weighting schemes like TF-IDF.}.
The weighted counterpart to the four similarity measures are listed in Table~\ref{tab:similarity-measures}.
For convenience we define $x[i]$ to be the $i$th token in the weighted set $x$ according to $\mathcal{O}$,
$x[i..j]$ to be the subset containing the $i$th to the $j$th (inclusive) tokens of $x$,
$t.w$ the weight of a token $t$,
$t.k$ the global rank of a token $t$ according to the total ordering $\mathcal{O}$,
and the size $\omega(x) = \sum_{i=1}^{|x|} x[i].w$.

\subsection{Inverted Token Index}

Central to most set similarity joins (and PPJoin) are the use of inverted token indexes --- i.e., indexes of which token set contains a certain token.
A naive approach for finding all the token sets in $\mathcal{B}$ that are similar to some $a \in \mathcal{A}$ would be to look up all tokens of $a$ to find all token sets in $\mathcal{B}$ that have at least one common token and check if $sim(a,b) \geq \tau$ for all of them.
Finding all token set pairs above the threshold is then just a matter of repeating the procedure for all $a \in \mathcal{A}$\footnote{
In this case, sets in $\mathcal{A}$ are used to query against an index of $\mathcal{B}$.
Obviously, one could also do it the other way around --- and depending on the data it might be beneficial.
However, for simplicity, but without loss of generality, we assume $\mathcal{A}$ to be the query collection and $\mathcal{B}$ the index collection throughout the paper.
}.

This is the underlying idea of PPJoin, as seen in Algorithm~\ref{alg:ppjoin}.
Note that since it performs self-join it employs an optimization where the sets are indexed only after having been queried (see line \ref{alg:ppjoin-self-join-index}).
The problem with this approach is tokens that occur in a large number of token sets lead to an excessive number of retrieved token sets, and therefore token set comparisons, for pairs that are not very similar.
There are simply to many token set pairs that have at least one common token.
It is possible to improve on this by efficiently merging the inverted lists and avoid redundant work (e.g., ScanCount~\cite{liEfficientMergingFiltering2008}), but we still suffer from the same underlying problem.
Therefore, to achieve feasible runtimes we need to reduce the total number of lookups and token set comparisons through multiple pruning techniques --- also called filters.

\subsection{Prefix Filtering}

\begin{figure}[tbhp]
    \centering
    \includegraphics{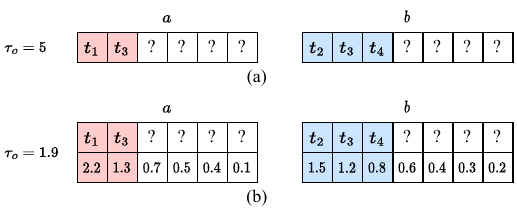}
    \caption{
        Example of the prefix filtering principle for (a) unweighted and (b) weighted token sets.
        The token sets are sorted by the total ordering $\mathcal{O}$.
        The highlighted tokens is the prefixes that must have at least one token common if the Overlap is at least $\tau_o$.
    }
    \label{fig:prefix-filtering}
\end{figure}

The key insight that enables prefix filtering is that once one have scanned through $\pi$ inverted lists without seeing some token set $b$ we know the intersection can be at most $|a| - \pi$ because we know $\pi$ of the tokens in $a$ does not exist in $b$.
In other words, we only need to probe $|a| - \tau_o + 1$ inverted lists to be certain we have found every $b \in \mathcal{B}$ with an overlap $|a \cap b|$ of at least $\tau_o$, and can safely prune the remaining inverted token lists.
We can provide an even stronger guarantee if we assume all token sets to be sorted in the same global total ordering $\mathcal{O}$.
This is known as the \textit{prefix filtering principle}~\cite{chaudhuriPrimitiveOperatorSimilarity2006} (adapted for this paper):
\begin{lemma}[Unweighted Prefix Filtering Principle]
    Let $a$ and $b$ be two token sets and assume some total token ordering $\mathcal{O}$.
    If ${|a \cap b| \geq \tau_o}$, then $\Big| a\big[1..(|a|-\tau_o + 1)\big] \cap b\big[1..(|b|-\tau_o + 1)\big] \Big| > 0$.
\end{lemma}
In other words, when denoting $x[1..\pi]$ as the $\pi$-prefix of $x$, if the overlap between $a$ and $b$ are at least $\tau_o$, then the $(|a| - \tau_o +1)$-prefix of $a$ and the $(|b| - \tau_o + 1)$-prefix of $b$ will have at least one token in common.
See Figure~\ref{fig:prefix-filtering}a for an example.
This not only means it is enough to probe the $(|a| - \tau_o + 1)$-prefix tokens of $a$ to find the token sets with overlap at least $\tau_o$ in $\mathcal{B}$,
but also that we only need to index the $(|b| - \tau_o + 1)$-prefix for all $b \in \mathcal{B}$.
This further reduces the indexing and query time as well as the memory footprint of the inverted index.
In the rest of the paper we will refer to such prefixes sufficient to find similar token sets of $x$ above some threshold as simply the prefix of $x$.

\paragraph{Token Ordering}
The real power of the prefix filtering principle lies in our ability to freely choose the token ordering $\mathcal{O}$.
We order by increasing frequency of occurrence among token sets --- essentially ignoring the most frequent tokens.
If $\tau_o$ is large enough and the distribution of tokens skewed enough we effectively avoid the problem of frequently occurring tokens.

\paragraph{Other Similarity Measures}
Utilizing this is straightforward when we have an Overlap similarity threshold.
The trick to extending to other similarity measures is that we can bound the overlap $|a \cap b|$ from below using $|a|$ and the similarity measure at hand.
Assume, for example, a Jaccard threshold $\tau_j$.
It can be shown easily that
\begin{gather*}
    \text{sim}_j(a, b) = \frac{|a \cap b|}{|a \cup b|} \geq \tau_j \\
    \Downarrow \\
    |a \cap b| \geq \tau_j \cdot |a|
\end{gather*}
Which means we can simply replace $\tau_o$ with $\big\lceil \tau_j \cdot |a| \big\rceil$ and only query the $(|a| + \big\lceil \tau \cdot |a| \big\rceil + 1)$-prefix for $a \in \mathcal{A}$ and only index the $(|b| + \big\lceil \tau \cdot |b| \big\rceil + 1)$-prefix of $b \in \mathcal{B}$.
See line~\ref{alg:ppjoin-prefix-filter} in Algorithm~\ref{alg:ppjoin}.
One can bound the overlap for the other similarity measures in a similar fashion.
Table~\ref{tab:similarity-measures} lists $\pi(x, \tau)$, the prefix size for token set $x$ with respect to some similarity threshold $\tau$, for all four similarity measures.

\paragraph{Weighted Sets}
Prefix filtering can easily be extended to weighted sets where the overlap between $a$ and $b$ is defined as $\sum_k \min(a_k, b_k)$.
We generalize the prefix filtering principle into a weighted version.
\begin{lemma}[Weighted Prefix Filtering Principle]
    Let $a$ and $b$ be two weighted token sets and assume some total token ordering $\mathcal{O}$.
    Furthermore, let
        $\phi(x, p) = \argmin_i \Big( \omega \big( x[1..i] \big) > p \Big)$.
    If
        ${\sum_k \min(a_k, b_k) \geq \tau_o}$
    then
        $\sum_k \min \Big( a\big[1..\phi(a, \omega(a) - \tau_o)\big]_k , b\big[1..\phi(b, \omega(b) - \tau_o)\big]_k \Big) > 0$.
\end{lemma}

So instead of having to only consider the ($|a| - \tau_o + 1$)-prefix to ensure we find every $b$ with overlap $|a \cap b|$ of at least $\tau_o$,
we have to consider the smallest $\pi$-prefix where the total weight of the $\pi$ tokens in the prefix are larger than $\omega(a) - \tau_o$ to ensure we find every $b$ with overlap $\sum_k \min(a_k, b_k)$ of at least $\tau_o$.
The equivalent applies for which tokens of $b \in \mathcal{B}$ we need to index.
See Figure~\ref{fig:prefix-filtering}b for an example.

We extend to other similarity measures the same way as for the unweighted case.
Table~\ref{tab:similarity-measures} lists the prefix size $\pi(x, \tau)$ for all four similarity measures.
The prefix size of a weighted set with respect to some similarity threshold is, instead of being the sufficient number of tokens, the total weight the prefix need to exceed.

\subsection{Size Filter}
A simple, yet effective, technique is to crop inverted lists using the size of the token sets.
We note that if $|a \cap b| \geq \tau_o$, then $|b| \geq \tau_o$.
In other words,
$\tau_o$ is a lower bound on the size of $b$.
Therefore, if we sort the inverted lists for the tokens of $\mathcal{B}$ by increasing size of $b$ we can efficiently skip the beginning of the list until $b \geq \tau_o$.

Generalizing to other set similarity measures is a matter of bounding $|b|$ using $\tau$ and $|a|$.
For example, for a Jaccard threshold $\tau_j$ we have that
\begin{gather*}
    \text{sim}_j(a,b) = \frac{|a \cap b|}{|a \cup b|} \geq \tau_j \\
    \Downarrow \\
    |b| \geq \tau_j \cdot |a|
\end{gather*}
We can see this in use at line~\ref{alg:ppjoin-size-filter} in Algorithm~\ref{alg:ppjoin}.
For the normalized set similarity measures we can also identify an upper bound and stop traversal of the ordered inverted lists early,
which is useless in the presence of the indexing trick for self-join in Algorithm~\ref{alg:ppjoin},
but will be useful for nonself-joins.
Table~\ref{tab:similarity-measures} lists the upper and lower size bounds, $\lambda_u(a, \tau)$ and $\lambda_l(a, \tau)$ for the different similarity measures.

The extension to weighted sets when the overlap is defined as $\sum_k \min(a_k, b_k)$ is straightforward,
follows the same derivation, and ends up with the same bounds (see Table~\ref{tab:similarity-measures}).
The only difference being the size bounds being on total weight instead of cardinality.

\subsection{Positional Filter}

\begin{figure}[tbhp]
    \centering
    \includegraphics{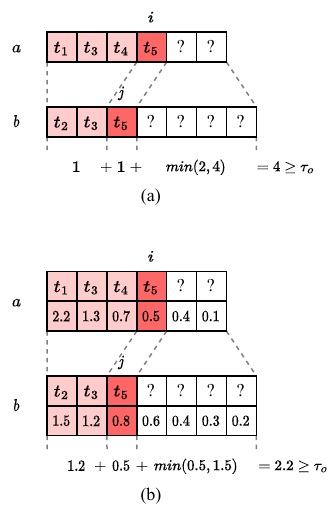}
    \caption{
        Example of the positional filtering principle for (a) unweighted and (b) weighted tokens sets.
        The token sets are sorted by the total ordering $\mathcal{O}$.
        The example show how we can use the fact that $a[i] = b[j]$ and that we know the overlap between $a[1..(i-1)]$ and $b[1..(j-1)]$ to calculate an upper bound on the overlap between $a$ and $b$.
    }
    \label{fig:positional-filtering}
\end{figure}

While prefix filtering enables us to reduce the number of tokens we need to probe,
and size filtering let us ignore parts of the inverted lists,
positional filtering let us dismiss many explicit pairs without calculating their similarity.
The key idea is that if we know $a$ and $b$ have a common token on position $i$ and $j$ (such that $a[i] = b[j]$),
and we know the overlap between the prefixes $a\big[1..(i-1)\big]$ and $b\big[1..(j-1)\big]$,
we can bound their total overlap because $a\big[(i+1)..|a|\big] \cap b\big[(j+1)..|b|\big] \leq \min(|a|-i, |b|-j)$.
This is known as the \textit{positional filtering principle}~\cite{xiaoEfficientSimilarityJoins2011} (adapted for this paper):
\begin{lemma}[Unweighted Positional Filtering Principle]
    Let $a$ and $b$ be two token sets and $x[i..j]$ be the $i$th to the $j$th (inclusive) ranked tokens of a token set $x$ according to some total ordering $\mathcal{O}$.
    If $|a \cap b| \geq \tau_o$ and $a[i] = b[j]$, then $\Big| a\big[1..(i-1)\big] \cap b\big[1..(j-1)\big] \Big| + 1 + \min\big(|a|-i, |b|-j\big) \geq \tau_o$.
\end{lemma}

See Figure~\ref{fig:positional-filtering}a for an example.
We utilize this principle by keeping track of how many common tokens we have seen between $a$ and all token sets in $\mathcal{B}$ as we traverse the inverted lists for $a$'s prefix in order of $\mathcal{O}$.
Additionally, we store the position $j$ of tokens within the token sets in the inverted lists for $\mathcal{B}$.
Thereby,
when we encounter token set $b$ as we probe the inverted list of token $a[i]$ and know the position of the token in $b$ to be $j$ we can simply check whether $\Big| a\big[1..(i-1)\big] \cap b\big[1..(j-1)\big] \Big| + 1 + \min(|a|-i, |b|-j) \geq \tau_o$ holds and discard the token set pair immediately if it does not.

\paragraph{Other Similarity Measures}
If we bound $|a \cap b|$ from below by some expression $\alpha(a,b,\tau)$ using $|a|$, $|b|$, and some similarity threshold $\tau$ we can use $\alpha(a,b,\tau)$ instead of $\tau_o$ in the inequality since $|a \cap b| \geq \alpha(a,b,\tau) \geq \tau_o$.
We call $\alpha(a,b,\tau)$ an equivalent overlap.
For Jaccard, one can derive $\alpha(a,b,\tau) = \frac{\tau}{\tau + 1}(|a| + |b|)$,
and we see this being used on line~\ref{alg:ppjoin-positional-filter} in Algorithm~\ref{alg:ppjoin}.
Table~\ref{tab:similarity-measures} lists equivalent overlaps for all similarity measures.

\paragraph{Weighted Sets}
The positional filtering principle can be extended to weighted sets:
\begin{lemma}[Weighted Positional Filtering Principle]
    Let $a$ and $b$ be two weighted token sets and assume some total token ordering $\mathcal{O}$.
    If ${\sum_k \min(a_k, b_k) \geq \tau_o}$ and ${a[i] = b[j]}$,
    then
    \begin{equation*}
        \begin{split}
            &\sum_k \min \Big( a\big[1..(i-1)\big]_k , b\big[1..(j-1)\big]_k \Big) + \min(a[i],b[j]) \\
            &+ \min\bigg(\omega\Big(a\big[(i+1)..|a|\big]\Big), \omega\Big(b\big[(j+1)..|b|\big]\Big)\bigg) \geq \tau_o
        \end{split}
    \end{equation*}
\end{lemma}
Figure~\ref{fig:positional-filtering}b shows an example.
This means we can utilize positional filtering the same way as before by using the accumulated overlap with different token sets $b \in \mathcal{B}$ as we traverse the prefix of $a$,
but instead of storing the position $j$ we store $\omega\Big(b\big[(j+1)..|b|\big]\Big)$.
One problem with this naive extension of positional filtering is that in order to calculate $\min(a[i], b[j])$ we must either look up $b[j]$ or grow the index to include it.
The PPJoin authors simply suggest to store $\omega\Big(b\big[j..|b|\big]\Big)$ and ignores the $\min(a[i], b[j])$ term \cite{xiaoEfficientSimilarityJoins2011}.
This effectively exploits the fact that
\begin{equation*}
        \begin{split}
            &\min(a[i],b[j]) + \min\bigg(\omega\Big(a\big[(i+1)..|a|\big]\Big), \omega\Big(b\big[(j+1)..|b|\big]\Big)\bigg) \\
            & \leq \min\bigg(\omega\Big(a\big[i..|a|\big]\Big), \omega\Big(b\big[j..|b|\big]\Big)\bigg)
        \end{split}
\end{equation*}
In our experience, this trade-off between the overhead of accessing/storing $b[j]$ and the tighter bound with $\min(a[i], b[j])$ favors their approach as long as the verification routine is fast.
Therefore, we assume this form of positional filtering from here on.
We can derive equivalent overlaps as in the unweighted case to support other similarity measures --- see Table~\ref{tab:similarity-measures}.

\subsection{Generalized PPJoin}\label{sec:gppjoin}

\begin{figure}[tbhp]
    \centering
    \includegraphics{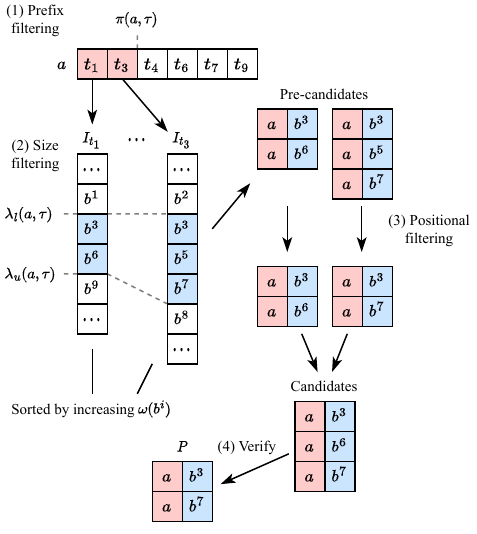}
    \caption{
        Illustration of the filtering steps in PPJoin.
        Inspired by \citet{mannEmpiricalEvaluationSet2016}.
    }
    \label{fig:ppjoin}
\end{figure}

\begin{algorithm}[tbh]
    \caption{BuildIndex$(\mathcal{B}, \tau)$}
    \label{alg:build-index}
    \footnotesize
    \SetKw{Continue}{\textbf{continue}}
    \SetKw{Break}{\textbf{break}}
    \KwIn{
        $\mathcal{B}$ is a collection of weighted token sets already sorted by $\mathcal{O}$.
        $\tau$ is the similarity threshold.
    }
    \KwOut{Inverted token indices $I$}
    
    $I \gets [\emptyset]_t$\;
    Sort $\mathcal{B}$ by increasing $\omega(b)$\label{alg:build-index-sort}\;
    \ForEach{$b \in \mathcal{B}$}{\label{alg:build-index-loop}%
        $j \gets 1$\;
        $s_b \gets \omega(b)$\;
        \While{$s_b \geq \sigma(b, \tau)$}{%
            $I_t \gets I_t \cup \{ (b, s_b) \}$\;
            $s_b \gets s_b - b[j].w$\;
            $j \gets j + 1$\;
        }
    }
    \KwRet{I}\;
\end{algorithm}

\begin{algorithm}[tbh]
    \caption{GPPJoin$(\mathcal{A}, \mathcal{B}, \tau)$}
    \label{alg:generalized-ppjoin}
    \footnotesize
    \KwIn{
        $\mathcal{A}$ and $\mathcal{B}$ are collections of weighted token sets already sorted by $\mathcal{O}$.
        $\tau$ is the similarity threshold.
    }
    \KwOut{Token set pairs $\{ (a, b) \mid sim(a, b) \geq \tau \}$}
    \SetKwFunction{BuildIndex}{BuildIndex}
    $I \gets$ \BuildIndex{$\mathcal{B}, \tau$}\tcp*[r]{Inverted token indices}
    $P \gets \emptyset$ \tcp*[r]{Pairs with similarity at least $\tau$}
    \ForEach{$a \in \mathcal{A}$}{%
        $O \gets$ empty map, default value 0\;
        $i \gets 1$\;
        $p_a \gets 0$\;
        \While(\tcp*[f]{Prefix filter}){$p_a \leq \pi(a, \tau)$}{%
            $t,w \gets a[i]$\;
            \DontPrintSemicolon\tcp*{Size filter}
            $I'_t \gets \{ (b, s_b) \in I_t \mid \lambda_l(a,\tau) \leq \omega(b) \leq \lambda_u(a,\tau) \}$\;
            \ForEach{$(b, p_b) \in I'_t$}{%
                $o^* \gets O[b] + \min(\omega(a) - p_a, s_b)$\;
                \eIf(\tcp*[f]{Positional filter}){$o^* \geq \alpha(a,b,\tau)$}{%
                    $P[b] \gets O[b] + w$\;
                }{%
                    Remove $O[b]$\;
                }
            }
            $i \gets i + 1$\;
            $p_a \gets p_a + w$\;
        }
        $P \gets P \cup \text{\KwVerify{$a,\mathcal{B},P,\tau$}}$ \;
    }
    \KwRet{$P$}\;
\end{algorithm}

Figure~\ref{fig:ppjoin} illustrates how the different filtering techniques come together.
We call the prefix and size filter for pre-candidate filters.
They prune the implicit pair space and limit which explicit pairs are materialized.
The positional filter is a candidate filter,
which means it prunes explicit pairs before they are undergo the \texttt{Verify} routine to check if they are above the similarity threshold.
While Algorithm~\ref{alg:ppjoin} depicts PPJoin similar to how the authors do \cite{xiaoEfficientSimilarityJoins2008, xiaoEfficientSimilarityJoins2011},
we also provide a fully generalized version for the reader in Algorithm~\ref{alg:generalized-ppjoin} (and \ref{alg:build-index}) that performs nonself-join for weighted sets with any supported set similarity measure using all the extensions described above.
This will serve as a starting point for describing our method and contrasting it to existing methods.

\subsection{Improved Prefix Filtering for Weighted Cosine}\label{sec:l2ap-bound}
In Table~\ref{tab:similarity-measures} we presented the prefix for weighted cosine similarity one gets by following the PPJoin authors~\cite{xiaoEfficientSimilarityJoins2011} instructions for extending to weighted similarity --- which is close to how prefix filtering is done in AllPairs~\cite{bayardoScalingAllPairs2007}.
The bound is a natural extension of the unweighted case\footnote{However, note that it is necessary to assume normalized (unit length) vectors in order to bound prefix/suffix and size when performing prefix filtering and size filtering. This simply means one must take care to normalize all token sets before performing the similarity join.}.
Unfortunately,
prefix and size filtering with these bounds are not very effective because the bounds loosen as the token sets get bigger.
To see why, we will look at an example.
\begin{example}
    Assume a normalized weighted set $x$ of $n$ unique tokens with weight $\sqrt{\frac{1}{n}}$,
    which then has size $\omega(x) = \sum_n \sqrt{1/n} = \sqrt{n}$.
    The (relative) prefix size of $x$ for weighted cosine similarity expressed in fraction of the tokens is
    \begin{equation*}
        \frac{\omega(x) - \tau}{\omega(x)} = 1 - \frac{\tau}{\omega(x)} = 1 - \frac{\tau}{\sqrt{n}}
    \end{equation*}
    We observe that the relative prefix size goes towards $1$ as $n$ increases.
    In other words, prefix filtering is less and less effective as the token set grow and will eventually have no effect.
    Contrast this to jaccard similarity,
    which has the relative prefix size
    \begin{equation*}
        \frac{(1 - \tau) \cdot \omega(x)}{\omega(x)} = 1 - \tau
    \end{equation*}
    The relative prefix size is independent of the number of tokens.
    If $\tau = 0.5$ then prefix filtering will will avoid lookup on half of the tokens no matter how many tokens the set contains.
    An equivalent example could be made for size filtering.
\end{example}

L2AP~\cite{anastasiuL2APFastCosine2014} introduces tighter bounds for cosine similarity in AllPairs.
The key is to utilize the Cauchy-Schwarz inequality:
\begin{equation}
    \begin{split}
        \text{dot}(a,b) &\leq \lVert a \rVert_2 \times \lVert b \rVert_2 \\
        \sum_k a_k \cdot b_k &\leq \sqrt{\sum_k a_k^2} \times \sqrt{\sum_k b_k^2}
    \end{split}
\end{equation}
We can then infer a suffix size bound:
\begin{equation}
    \begin{split}
        \tau &\leq \text{dot}(a,b) \\
        &= \text{dot}\Big(a\big[1..(i-1)\big],b\Big) + \text{dot}\Big(a\big[i..|a|\big],b\Big) \\
        &= \text{dot}\Big(a\big[i..|a|\big],b\Big) \\
        &\leq \left\lVert a\big[i..|a|\big] \right\rVert_2 \times \lVert b \rVert_2 \\
        &\leq \left\lVert a\big[i..|a|\big] \right\rVert_2 
    \end{split}
\end{equation}
Which is equivalent to a prefix size bound of
\begin{equation}
    \left\lVert a\big[1..(i-1)\big] \right\rVert_2 \geq \sqrt{1 - \tau^2}
\end{equation}
Note that the bound is on the $L^2$ norm of the prefix instead of the $L^1$ norm as in PPJoin.
Importantly, the bound is much more effective because it does not grow with the number of tokens.

\section{Expressive Hybrid Join Primitive}\label{sec:hybrid-join}
\newcommand{\dynthresh}{\widetilde{\tau}}

This section will describe a new flexible join primitive that will be the core building block of our proposed methods.
We first discuss strength and weaknesses of three types of similarity join conditions,
and argue that they complement each others weaknesses and there is merit to combining them.
Therefore, we propose a new hybrid join type and will follow up in the next section with an efficient algorithmic realization of such a join.

\subsection{Similarity Join Conditions}

All similarity joins have an inclusion condition specifying which pairs should be returned or not.
We will now briefly discuss three types of the similarity join conditions used for entity matching.

\begin{figure*}[tbhp]
    \centering
    \includegraphics{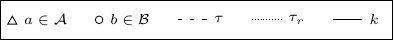} \\ \vspace{1em}
    \renewcommand{\thesubfigure}{a-1}
    \begin{subfigure}{0.3\textwidth}
        \includegraphics{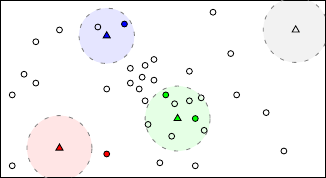}
        \caption{$\tau$-join with lower $\tau$}
        \label{fig:join-constraints-lower-tau}
    \end{subfigure}
    \hspace{0.01\textwidth}
    \renewcommand{\thesubfigure}{b-1}
    \begin{subfigure}{0.3\textwidth}
        \includegraphics{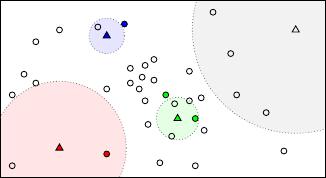}
        \caption{$\tau_r$-join with lower $\tau_r$}
        \label{fig:join-constraints-lower-tau-r}
    \end{subfigure}
    \hspace{0.01\textwidth}
    \renewcommand{\thesubfigure}{c-1}
    \begin{subfigure}{0.3\textwidth}
        \includegraphics{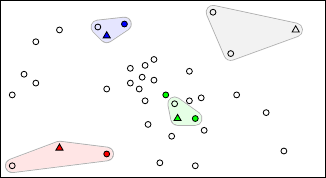}
        \caption{$k$-join with lower $k$}
        \label{fig:join-constraints-lower-k}
    \end{subfigure}
    
    \vspace{1em}
    
    \renewcommand{\thesubfigure}{a-2}
    \begin{subfigure}{0.3\textwidth}
        \includegraphics{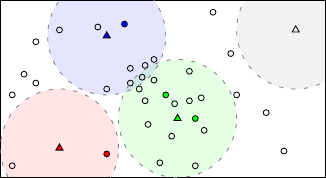}
        \caption{$\tau$-join with higher $\tau$}
        \label{fig:join-constraints-higher-tau}
    \end{subfigure}
    \hspace{0.01\textwidth}
    \renewcommand{\thesubfigure}{b-2}
    \begin{subfigure}{0.3\textwidth}
        \includegraphics{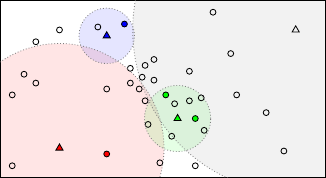}
        \caption{$\tau_r$-join with higher $\tau_r$}
        \label{fig:join-constraints-higher-tau-r}
    \end{subfigure}
    \hspace{0.01\textwidth}
    \renewcommand{\thesubfigure}{c-2}
    \begin{subfigure}{0.3\textwidth}
        \includegraphics{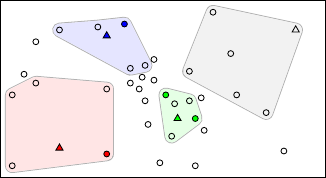}
        \caption{$k$-join with higher $k$}
        \label{fig:join-constraints-higher-k}
    \end{subfigure}
    \caption{
    Conceptual illustration of weaknesses of join conditions for $\tau$-join, $\tau_r$-join, and $k$-join.
    Triangle points are sets from $\mathcal{A}$, circle points are sets from $\mathcal{B}$, point color indicate matches, area color indicate inclusion according to the join condition, and euclidean proximity is similarity.
    }
    \label{fig:join-constraints-weaknesses}
\end{figure*}

\paragraph{1. Absolute Similarity Threshold ($\tau$-join)}
A $\tau$-join returns all set pairs $(a,b)$ across the set collections $\mathcal{A}$ and $\mathcal{B}$ such that $sim(a,b) \geq \tau$.
This type of join for have received a substantial amount of work~\cite[e.g.,][]{mannEmpiricalEvaluationSet2016},
and PPJoin~\cite{xiaoEfficientSimilarityJoins2008,xiaoEfficientSimilarityJoins2011} is an example of a $\tau$-join.
The main strength of absolute similarity thresholds is their ability to consistently filter out low similarity pairs.
However, $\tau$-joins do not handle varying similarity density well.
Some sets in $\mathcal{A}$ may be quite similar to many sets in $\mathcal{B}$,
while other sets in $\mathcal{A}$ may only be modestly similar to a few sets in $\mathcal{B}$.
The problem is that even though a lower $\tau$ will avoid a substantial number of false positives from sets with high similarity density it will struggle to recall neighbors of sets with low similarity density.
While a higher $\tau$ will have the exact opposite problem.
See Figure~\ref{fig:join-constraints-lower-tau} and \ref{fig:join-constraints-higher-tau} for an illustration of how there might not be any $\tau$ that provide an acceptable solution.

\paragraph{2. Relative Similarity Threshold ($\tau_r$-join)}
A $\tau_r$-join returns the pairs $(a,b)$ for which $\text{sim}(a,b) \geq \tau_r S^*$,
where $S^* = \max_{b' \in \mathcal{B}} sim(a, b')$.
This type of join has not received much attention,
but a very similar variant\footnote{They formulated it with similarity distance instead of similarity.} was recently studied and motivated by \citet{liAutoFuzzyJoinAutoProgramFuzzy2021}.
The strength of relative similarity thresholds is that they can exploit relative differences in similarity to filter more aggressively when there exists high-similarity matches.
While normally one would be interested in $b$ with a similarity to $a$ of 0.6,
one might want to dismiss it if there exists another $b'$ with similarity 0.99.
The main challenge is sets that are not highly similar to any other set.
As $S^*$ decreases the relative similarity of all other sets will converge towards each other.
Therefore, setting $\tau_r$ to a value high enough to get acceptable recall might also lead to certain sets in $\mathcal{A}$ without any similar sets in $\mathcal{B}$ generate an unacceptable number of pairs.
See Figure~\ref{fig:join-constraints-lower-tau-r} and \ref{fig:join-constraints-higher-tau-r} for an illustration.

\paragraph{3. Local Cardinality Threshold ($k$-join)}
A $k$-join returns pairs $(a,b)$ such that $b$ is within the $k$ most similar records to $a$.
These join types have received increased attention recently for use with deep learning embeddings~\cite[e.g.,][]{thirumuruganathanDeepLearningBlocking2021}.
Note that this is different from joins with global cardinality constraints such as top-$k$ similarity join~\cite{xiaoTopkSetSimilarity2009, yangAdaptiveTopkOverlap2020}, which return the $k$ most similar pairs across all sets, and is outside the scope of this paper.
The obvious benefit of a local cardinality threshold is that it bounds the number of returned pairs and is not prone to the same scenarios that would blow up the number of returned pairs like for $\tau$-join and $\tau_r$-join.
It is also adaptive in the sense that it will be able to pick up both low and high similarity matches depending on the similarity density.
The downside is that the number of returned pairs is static and it is not able to leverage the similarity values to reduce the number of pairs when the data suggests so.
When all sets in $\mathcal{B}$ is very dissimilar to some set $a$ it might be unnecessary to still insist on returning $k$ pairs for $a$ --- maybe $a$ have no matches.
See Figure~\ref{fig:join-constraints-lower-k} and \ref{fig:join-constraints-higher-k} for an illustration of how a $k$-join might get unsatisfactory recall when $k$ is set low but might return too many pairs that are unlikely to match when $k$ is set higher.

\subsection{Hybrid Join Type}

\begin{table}[tbhp]
    \footnotesize
    \centering
    \begin{tabular}{lp{33mm}p{33mm}}
        \toprule
        Join\\Type & Strength & Weakness \\
        \midrule
        $\tau$-join
        & Consistently filter out low similarity pairs
        & Can not adapt to varying similarity density \\ \addlinespace[0.5em]
        $\tau_r$-join
        & Filters more aggressively when a set sticks out as more similar
        & Too forgiving when sets do not stick out with high similarity \\ \addlinespace[0.5em]
        $k$-join
        & Bounds the number of pairs and pick up both low and high similarity matches
        & Returns unnecessarily many pairs when all have low similarity \\
        \bottomrule
    \end{tabular}
    \caption{Strength and weaknesses of the join condition for $\tau$-join, $\tau_r$-join, and $k$-join.}
    \label{tab:join-strength-weakness}
\end{table}

Table~\ref{tab:join-strength-weakness} summarizes the strengths and weaknesses of the three join conditions discussed in the previous subsection.
We argue that the strengths of the three individual conditions complement the weaknesses of the others.
Therefore,
we propose a hybrid join type incorporating all three: $(\tau, \tau_r, k)$-join.
\begin{definition}[$(\tau,\tau_r,k)$-join]
Let $\mathcal{A}$ and $\mathcal{B}$ be two collections of sets, and $\text{sim}(a,b)$ be some set similarity measure.
Furthermore,
let $\psi_a(b) : \mathcal{B} \to \{1,2,\dots,|\mathcal{B}|\}$ be a bijective function that order all $b \in \mathcal{B}$ by decreasing similarity to $a$ so that $\text{sim}(a,b) > \text{sim}(a,b')$ if $\psi_a(b) < \psi(b')$.
Finally let $S_a^* = \max_{b \in \mathcal{B}} sim(a, b)$.
A $(\tau, \tau_r, k)$-join over $\mathcal{A}$ and $\mathcal{B}$ with respect to some similarity measure $sim$ is a join that return all pairs $(a,b)$ such that
\begin{equation*}
    \text{sim}(a,b) \geq \tau \land \text{sim}(a,b) \geq \tau_r S_a^* \land \psi_a(b) \leq k
\end{equation*}
\end{definition}
\begin{figure}[tbhp]
    \centering
    \includegraphics{illustrations/join_constraints_legend.pdf}
    
    \vspace{1em}
    
    \includegraphics{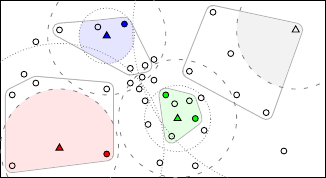}
    \caption{Conceptual illustration of $(\tau,\tau_r,k)$-join.}
    \label{fig:join-constraints-all}
\end{figure}
In other words,
a $(\tau,\tau_r,k)$-join requires all three join condition to be met at once.
Figure~\ref{fig:join-constraints-all} illustrates how the different conditions can complement each other the same way Figure~\ref{fig:join-constraints-weaknesses} illustrated weaknesses of the individual conditions.
This join type is significantly more expressive,
but the number of hyperparameters also increases,
which makes it more challenging to choose hyperparameters.
This will be a central part of our proposed method and will be addressed in later sections.
For now, assume that they are provided.

\section{Efficient $(\tau, \tau_r, k)$-join Algorithm}\label{sec:ttrkjoin}

\begin{algorithm}[tbhp]
    \caption{BuildTTRKIndex$(\mathcal{B}, \tau; sim)$}
    \label{alg:build-ttrk-index}
    \footnotesize
    \SetKw{Continue}{\textbf{continue}}
    \SetKw{Break}{\textbf{break}}
    \KwIn{
        $\mathcal{B}$ is a collection of weighted token sets already sorted by $\mathcal{O}$.
        $\tau$ is the similarity threshold.
    }
    \KwOut{
    Inverted token indices $I^+, I^-$ for token sets with above and below average prefix weight.
    }
    
    $I \gets [\emptyset]_t$\;
    $P \gets [0]_t$\;
    \ForEach{$b \in \mathcal{B}$}{%
        $j \gets 1$\;
        $p_b \gets 0$\;
        \While{$s_b \geq \sigma(b, \tau)$}{%
            $I_t \gets I_t \cup \{ (b, j, p_b) \}$\;
            $P_t \gets P_t + p_b$\;
            $p_b \gets p_b + b[j].w$\;
            $j \gets j + 1$\;
        }
    }
    $I^+ \gets [\emptyset]_t$\;
    $I^- \gets [\emptyset]_t$\;
    \ForEach{$I_t \in I$}{%
        $\overline{p}_b \gets P_t / |I_t|$\;
        $I^+_t\!.\text{minPrefix} \gets \infty$\;
        $I^-_t\!.\text{minPrefix} \gets \infty$\;
        \ForEach{$(b, j, p_b) \in I_t$}{%
            $s_b \gets \omega(b) - p_b$\;
            \eIf{$p_b \geq \overline{p}_b$}{%
                $I^+_t \gets I^+_t \cup \{ (b, j, s_b) \}$\;
                $I^+_t\!.\text{minPrefix} \gets \min(I^+_t\!.\text{minPrefix}, p_b$)\;
            }{%
                $I^-_t \gets I^-_t \cup \{ (b, j, s_b) \}$\;
                $I^-_t\!.\text{minPrefix} \gets \min(I^-_t\!.\text{minPrefix}, p_b$)\;
            }
        }
        Sort $I^+_t$ and $I^-_t$ by increasing $s_b$\;
    }

    \KwRet{$I^+, I^-$}\;
\end{algorithm}

\begin{algorithm}[tbh]
    \caption{TTRKJoin$(\mathcal{A}, \mathcal{B}, \tau, \tau_r, k, \rho^*;sim)$}
    \label{alg:ttrk-join}
    \footnotesize
    \SetKw{Continue}{\textbf{continue}}
    \SetKw{Break}{\textbf{break}}
    \KwIn{
        $\mathcal{A}$ and $\mathcal{B}$ are collections of weighted token sets already sorted by $\mathcal{O}$.
        $\tau$ and $\tau_r$ are the similarity threshold and the relative similarity threshold.
        $k$ is the maximum number of neighbors per token set.
        $\rho^*$ is the maximum traversal rank.
    }
    \KwOut{Results of a $(\tau,\tau_r,k)$-join with a maximum traversal rank of $\rho^*$}
    
    \SetKwFunction{BuildTTRKIndex}{BuildTTRKIndex}
    \SetKwFunction{PartialSim}{PartialSim}
    
    $I^+,I^- \gets$ \BuildTTRKIndex{$\mathcal{B}, \tau$}\;
    $P \gets \emptyset$\;
    
    \ForEach{$a \in \mathcal{A}$}{\label{alg:ttrk-join-loop-a}%
        $Q \gets$ empty min heap\;
        $V \gets \emptyset$\;
        $\dynthresh \gets \tau$\;\label{alg:init-dynthresh}%
        $i \gets 1$\;
        $\rho \gets 0$\;
        $s_a \gets \omega(a)$\;
        \While(\tcp*[f]{Prefix filter}){$s_a \geq \sigma(a, \dynthresh)$}{\label{alg:ttkr-join-prefix-filter}%
            $t,w \gets a[i]$\;
            \ForEach{$I_t \in [I^-_t,I^+_t]$}{\label{alg:loop-partitioned-index}%
                start $\gets$ binary search first $I_t[start] \geq \lambda_l(a,\dynthresh)$\;
                end $\gets$ binary search last $I_t[end] \leq \lambda_u(a,\dynthresh)$\;
                $\rho \gets \rho + (start - 1)$\;\label{alg:ttkr-join-rho-start}
                \DontPrintSemicolon\tcp*{PPS filter}
                \ForEach{$(b, j, s_b) \in I_t[start..end]$}{%
                    $\rho \gets \rho + 1$\;
                    \lIf{$\rho > \rho^*$}{\Break loop at line~\ref{alg:ttkr-join-prefix-filter}}
                    \DontPrintSemicolon\tcp*{Positional filter}
                    \If{$\min(s_a, s_b) \geq \alpha(a,b,\dynthresh)$}{\Continue}
                    \lIf{$b \in V$}{\Continue\label{alg:ttrk-join-check-v}}
                        $S \gets \PartialSim(a, b, i, j, s_a, s_b, \dynthresh)$\label{alg:ttrk-join-sim}\;
                        $V \gets V \cup \{b\}$\label{alg:ttrk-join-update-v}\;
                        \If{$S \geq \dynthresh$}{%
                            $Q.push(b, S)$\;
                            \lIf{$|Q| > k$}{%
                                $Q.pop()$%
                            }
                            \lIf{$|Q| = k$}{%
                                $\dynthresh \gets \max(\dynthresh, Q.top.sim)$\label{alg:ttrk-join-q-tighten}\hspace{-2pt}
                            }
                            \If{$\tau_r \cdot S > \dynthresh$}{\label{alg:ttrk-join-taur-tighten-start}%
                                $\dynthresh \gets \tau_r \cdot S$\label{alg:ttrk-join-taur-tighten-end}\;
                                \lWhile{$Q.top.sim < \dynthresh$}{%
                                    $Q.pop()$\label{alg:ttrk-join-taur-prune}%
                                }%
                            }
                        }
                }
                $\rho \gets \rho + (|I_t| - end)$\;\label{alg:ttkr-join-rho-end}
            }
            $s_a \gets s_a - w$\;
            $i \gets i + 1$\;
        }
        $P \gets P \cup Q$\;
    }
    
    \KwRet{$P$}\;
\end{algorithm}

\begin{algorithm}[tbhp]
    \caption{PartialSim$(a,b,i,j,s_a,s_b,\tau; sim)$}
    \label{alg:partial-sim}
    \footnotesize
    \SetKw{Continue}{\textbf{continue}}
    \SetKw{Break}{\textbf{break}}
    \KwIn{
    }
    \KwOut{}
    
    \If{$a[|a|] > b[|b|]$}{
        $\text{swap}(a,b); \text{swap}(i,j); \text{swap}(s_a,s_b)$\;
    }
    
    $o \gets \alpha(a,b,\dynthresh)$\;
    
    intersection $\gets 0$ \tcp*{For $\sum_k \min(a_k, b_k)$}
    union $\gets (\omega(a) - s_a) + (\omega(b) - s_b)$ \tcp*{For $\sum_k \max(a_k, b_k)$}
    dotProduct $\gets 0$\;
    
    \While{$i \leq |a|$}{
        \While{$a[i] > b[j]$}{
            $s_b \gets s_b - b[j].w$\;
            \lIf{$s_b < o$}{
                \KwRet{$0$}%
            }
            $\text{union} \gets \text{union} + b[j].w$\;
            $j \gets j + 1$\;
        }
        \If{$a[i] = b[j]$}{
            $o \gets o - \min(a[i].w, b[i].w)$\;
            $s_a \gets s_a - a[i].w$\;
            $s_b \gets s_b - b[i].w$\;
            \lIf{$\min(s_a, s_b) < o$}{
                \KwRet{$0$}%
            }
            $\text{intersection} \gets \text{intersection} + \min(a[i].w, b[i].w)$\;
            $\text{union} \gets \text{union} + \max(a[i].w, b[i].w)$\;
            $\text{dotProduct} \gets \text{dotProduct} + a[i].w \cdot b[i].w$\;
            $j \gets j + 1$\;
        }
        $i \gets i + 1$\;
    }
    \While{$j \leq |b|$}{
            $s_b \gets s_b - b[j].w$\;
            \lIf{$s_b < o$}{
                \KwRet{$0$}%
            }
            $\text{union} \gets \text{union} + b[i].w$\;
            $j \gets j + 1$\;
    }
    
    \KwRet{$\text{sim}(\text{intersection} / \text{union} / \text{dotproduct} / \omega(a) / \omega(b))$}\;
\end{algorithm}

We will now propose an efficient algorithm, \ttrkjoin, for performing $(\tau,\tau_r,k)$-joins with weighted Jaccard, Cosine, Dice, or Overlap similarity.
The join routine is prefix filtering-based and is outlined in Algorithm~\ref{alg:ttrk-join}.
Even though \ttrkjoin is significantly different than PPJoin,
we will for the benefit of the reader motivate and describe \ttrkjoin as a series of changes to PPJoin\footnote{Specifically, the generalized version, \texttt{GPPJoin}, presented in Section \ref{sec:gppjoin}.} --- as this is a well-known and state-of-the-art method~\cite{mannEmpiricalEvaluationSet2016}.
First, we describe how we handle the additional $\tau_r$ and $k$ join conditions.
Then, we describe how we incorporate a better cosine prefix bound from L2AP~\cite{anastasiuL2APFastCosine2014},
before we propose a new effective filtering technique to replace size filtering.
Following,
we introduce an additional parameter to the algorithm that will be used to achieve approximate joins in Section \ref{sec:approximate-joins}.
Lastly, we cover some implementation details.

\subsection{Join Conditions}

While it is possible to naively extend \texttt{GPPJoin} to handle the $\tau_r$ and $k$ conditions by simply applying them to candidates from the \texttt{Verify} routine,
this would not exploit the constraints put on the similarity to prune the search space.
Therefore, we evolve \texttt{GPPJoin} in two main ways --- one for $k$ and one $\tau_r$.

\subsubsection{Local Cardinality Threshold $k$}
We keep track of the $k$ most similar token sets in a minimum priority queue $Q$ as we probe the inverted lists for $a$.
Instead of accumulating the overlap in $O$ for all the candidates in the prefix of $a$,
we eagerly compute the similarity (line~\ref{alg:ttrk-join-sim}) to each $b$ on first encounter in order to maintain $Q$.
We avoid redundant work by keeping track of already encountered token sets in $V$ (line~\ref{alg:ttrk-join-check-v} and \ref{alg:ttrk-join-update-v}).
Furthermore, let $\dynthresh$ be the similarity threshold used by prefix, size, and positional filtering.
It is initially set to $\tau$,
but $Q$ let us tighten $\dynthresh$ to $\max(\dynthresh, Q.top.sim)$ when $Q$ have been updated and $|Q| = k$ because we know that we are not interested in token sets $b$ with a similarity to $a$ lower than those among the top $k$ we have already found (line~\ref{alg:ttrk-join-q-tighten}).
Effectively, the three filters (prefix, size, positional) are now dynamic and tighten as we discover token sets with higher similarity.

\subsubsection{Relative Similarity Threshold $\tau_r$}
Every time we compute a new similarity $S$ to a set $b \in \mathcal{B}$ we try to tighten $\dynthresh$ with $\tau_r S$ (line~\ref{alg:ttrk-join-taur-tighten-start}-\ref{alg:ttrk-join-taur-tighten-end}).
If it is tightened we also make sure to prune $Q$ with the new threshold (line~\ref{alg:ttrk-join-taur-prune}).

\begin{table*}[tbhp]
    \footnotesize
    \centering
    \begin{tabular}{lcccccc}
        \toprule
        & & Norm & Lower query suffix bound & \multicolumn{2}{c}{Index suffix bounds} & Equivalent overlap \\
        \cmidrule(rl){5-6}
        Measure & $sim(a,b)$ & $l$ & $\sigma(x,\tau)$ & $\lambda_l(a,\tau)$ & $\lambda_u(a,\tau)$ & $\alpha(a,b,\tau)$ \\
        \midrule
        
        \addlinespace[0.5em]
        
        Jaccard &
        $ \displaystyle \frac{ \sum_{k} \min(a_k, b_k) }{ \sum_{k} \max(a_k, b_k) } $ &
        $ \displaystyle 1 $ &
        $ \displaystyle \tau \cdot \omega(x) $ &
        $ \displaystyle \tau \cdot (\omega(a) + p_b) $ &
        $ \displaystyle \frac{s_a}{\tau} - p_a - p_b$ &
        $ \displaystyle \frac{\tau}{\tau + 1} \big(\omega(a) + \omega(b)\big) $ \\
        
        \addlinespace[0.8em]
        
        Cosine$^*$ &
        $ \displaystyle \sum_{k} a_k \cdot b_k $ &
        $ \displaystyle 2 $ &
        $ \displaystyle \tau^2 $ &
        $ \displaystyle \frac{\tau^2}{s_a} $ &
        $ \displaystyle \infty $ &
        $ \displaystyle \tau $ \\
        
        \addlinespace[0.8em]
        
        Dice &
        $ \displaystyle \frac{ 2 \cdot \sum_{k} \min(a_k, b_k) }{ |a| + |b| } $ &
        $ \displaystyle 1 $ &
        $ \displaystyle \frac{\tau}{2 - \tau} \cdot \omega(x) $ &
        $ \displaystyle \frac{\tau}{2 - \tau} \cdot (\omega(a) + p_b) $ &
        $ \displaystyle \frac{(2 - \tau)}{\tau} \cdot s_a - p_a - p_b $ &
        $ \displaystyle \frac{\tau}{2} \big(\omega(a) + \omega(b)\big)  $ \\
        
        \addlinespace[0.8em]
        
        Overlap &
        $ \displaystyle \sum_{k} \min(a_k, b_k) $ &
        $ \displaystyle 1 $ &
        $ \displaystyle \tau $ &
        $ \displaystyle \tau $ &
        $ \displaystyle \infty $ &
        $ \displaystyle \tau $ \\
        
        \addlinespace[0.5em]
        
        \bottomrule
    \end{tabular}
    \caption{
        Definitions of the weighted set similarity measures together with suffix bounds and equivalent overlap we use in \ttrkjoin.
        $^*$Requires every set $x$ to be normalized such that $\| x \|_2 = 1$.
    }
    \label{tab:ttrk-bounds}
\end{table*}

\subsection{Incorporating $L^2$ Based Cosine Bounds}
In Section~\ref{sec:l2ap-bound} we saw that the bounds for weighted cosine from PPJoin~\cite{xiaoEfficientSimilarityJoins2011} get more loose as the number of tokens increases.
L2AP~\cite{anastasiuL2APFastCosine2014} offers a tighter bound on the prefix,
but on the $L^2$ norm of the prefix instead of $L^1$.
We want a unified algorithm that can handle all four similarity measures (Jaccard, Cosine, Dice, Overlap) while also exploiting this bound.

We generalize the size function $\omega$ to be the $l$-norm sum:
\begin{equation}
    \omega(x) = \lVert x \rVert^l_l = \sum_k x_k^l
\end{equation}
Furthermore, let all prefixes and suffixes be $l$-norm sums.
In this setup, Cosine has ${l = 2}$ while Jaccard, Dice, and Overlap have ${l = 1}$.
The prefix size for cosine is then $1 - \tau^2$, while they remain the same for the other similarity measures.
To simplify the bounds and the algorithm we operate with suffix bounds instead of prefix bounds.
Table~\ref{tab:ttrk-bounds} lists the suffix bounds for the four similarity measures, as well as the norm to use.
Note that we can not do size filtering for cosine with this definition of size because it will always be $1$ for normalized weighted token sets.
Luckily, we will not be using size filtering.

\subsection{Prefix-Partitioned Suffix Filtering}
PPJoin (and many other set similarity joins) accumulates overlap for candidates over all tokens in the prefix before computing all similarities.
We compute similarities eagerly on first occurrence of a candidate in an inverted list and simply ignore the same candidate if it shows up in subsequent inverted lists.
This allows us to make use of the local cardinality threshold $k$ and relative threshold $\tau_r$ to prune more aggressively by increasing the dynamic threshold $\dynthresh$ during the search.
Moreover, it also opens up alternative ways than size filtering to trim the inverted lists.
We will now describe our proposed pre-candidate filter, the Prefix-Partitioned Suffix (PPS) Filter, in three steps.

\subsubsection{Suffix Instead of Size}

The key observation is that when we encounter token set $b$ for the first time when looking up $a$'s $i$th token and that token has position $j$ in $b$ we know that there is no overlap between $a[1..(i-1)]$ and $b[1..(j-1)]$.
To simplify notation,
let the prefix and suffix of $a$ and $b$ be
$p_a = \omega(a[1..(i-1)])$,
$s_a = \omega(a[i..|a|])$,
$p_b = \omega(b[1..(i-1)])$,
and $p_b = \omega(b[j..|b|])$.

The overlap between $a$ and $b$ is upper bounded by the suffix $s_b$ (and $s_a$).
Therefore, we can sort by and threshold $s_b$ instead of $w(b)$ in size filtering with the same bounds.
We call this a suffix filter\footnote{Not to be confused with the suffix filter in PPJoin+~\cite{xiaoEfficientSimilarityJoins2011}, which performs candidate (not pre-candidate) filtering.} --- in contrast to size filter.
It makes intuitively sense to filter on the suffix of sets in $b \in \mathcal{B}$ because the suffix of $b$ reflects the remaining tokens $a$ can still overlap with.
In effect,
the suffix filter moves part of the pruning power of the positional filter from candidate filtering to pre-candidate filtering.
It has significantly tighter lower bounds than the size filter because $s_b \leq \omega(b)$,
but also has looser upper bounds.
However,
reusing the bounds from size filtering is naive,
and we can do better with the extra positional information we have available.

\subsubsection{Position-Enhanced Index Suffix Bounds}
Tighter bounds can be inferred by using the fact that $a[1..(i-1)]$ and $b[1..(j-1)]$ will not have any overlap.
Take Jaccard as an example (with threshold $\tau_j$):
\begin{equation}
    \begin{split}
        \frac{\sum_k \min(a_k, b_k)}{\sum_k \max(a_k, b_k)} &\geq \tau_j \\
        \frac{\sum_k \min(a_k, b_k)}{\omega(a) + \omega(b) - \sum_k \min(a_k, b_k)} &\geq \tau_j \\
        (\tau_j + 1) \sum_k \min(a_k, b_k) &\geq \tau_j (\omega(a) + \omega(b)) \\
        (\tau_j + 1) \sum_{k \geq k_s} \min(a_k, b_k) &\geq \tau_j (\omega(a) + \omega(b)) \\
    \end{split}
\end{equation}
where $k_s = \max(a[i].k, b[j].k)$.
We can use the fact that $\sum_{k \geq k_s} \min(a_k, b_k)$ is upper bounded by $s_b$ and $s_a$ to get a lower and upper bound of $s_b$:
\begin{equation}
    \begin{split}
        (\tau_j + 1) \sum_{k \geq k_s} \min(a_k, b_k) &\geq \tau_j (\omega(a) + \omega(b)) \\
        (\tau_j + 1) s_b &\geq \tau_j (\omega(a) + \omega(b)) \\
        (\tau_j + 1) s_b &\geq \tau_j (\omega(a) + p_b + s_b) \\
        s_b &\geq \tau_j (\omega(a) + p_b) \\
    \end{split}
\end{equation}
\begin{equation}
    \begin{split}
        (\tau_j + 1) \sum_{k \geq k_s} \min(a_k, b_k) &\geq \tau_j (\omega(a) + \omega(b)) \\
        (\tau_j + 1) s_a &\geq \tau_j (\omega(a) + \omega(b)) \\
        (\tau_j + 1) s_a &\geq \tau_j (p_a + s_a + p_b + s_b) \\
        s_b &\leq \frac{s_a}{\tau_j} - p_a - p_b \\
    \end{split}
\end{equation}
Bounds for the other similarity measures can be inferred in similar fashion\footnote{We use the Cauchy-Schwarz inequality for Cosine.}.
They are listed in Table~\ref{tab:ttrk-bounds}.

The bounds rely on $p_a$, $s_a$, and $p_b$.
The attentive reader might have realized that,
while $p_a$ and $s_a$ is known when applying the suffix filter,
we do not actually know $p_b$ because it might (and most likely will) differ between the different entries in the inverted list $I_t$.
We must replace it with a lower bound.
Instead of simply using zero we can use the minimum $p_b$ in $I_t$ --- denoted $I_t.\text{minPrefix}$.

\subsubsection{Prefix-Partitioned Index}
For Jaccard and Dice, where the bounds rely on $p_b$,
higher $I_t.\text{minPrefix}$ results in tighter suffix bounds.
This makes intuitive sense since a larger prefix of index sets $b \in I_t$ that does not overlap with the query $a$ means a smaller fraction of each $b$'s size can overlap with $a$.
Obviously, with large inverted lists it is likely that $I_t.\text{minPrefix}$ will be low.

In order to prune sets with large prefixes more aggressively we partition each $I_t$ into two inverted lists $I_t^+$ and $I_t^-$,
containing sets with prefixes above and below the average prefix size $\overline{p}_b$.
Both $I_t^-$ and $I_t^+$ are still sorted by $s_b$ and we can crop them the same way we would for a suffix sorted $I_t$,
but now $I_t^+.\text{minPrefix}$ will be at least $\overline{p}_b$.
The additional work at indexing time is negligible but it introduces some overhead at query time because twice as many lists need to be cropped.
Assuming an appropriate spread in prefix sizes the increased pruning of $I_t^+$ can make up for it and more.

While other more intricate partitioning schemes are possible, we find this to be a reasonable approach that exploits additional positional information without severely increasing the complexity or overhead compared to a standard sorted inverted list.
We deem further exploration of alternative approaches outside the scope of this paper and hope to address it in future work.

\subsubsection{Putting it Together}
\begin{figure}[htbp]
    \centering
    \includegraphics{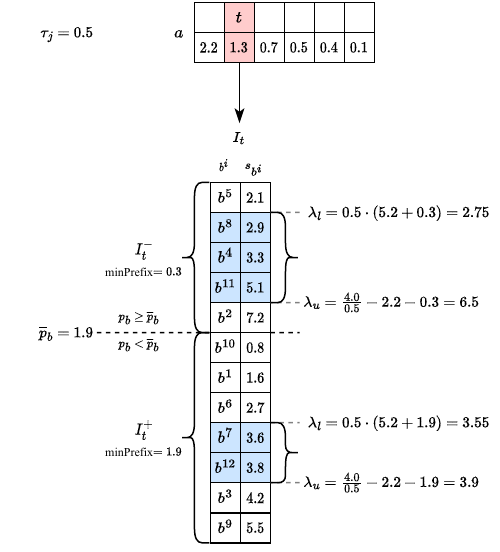}
    \caption{An example of performing Prefix-Partitioned Suffix Filtering when looking up token $t$ of token set $a$ and looking for matches with a Jaccard similarity of at least 0.5.}
    \label{fig:pps-filtering-example}
\end{figure}
Algorithm~\ref{alg:build-ttrk-index} outlines the construction of the index with prefix-partitioned inverted lists sorted by suffix size.
We first construct inverted lists for all tokens and then partition them on prefix and sort by suffix afterwards.
The actual implementation partitions $I_t$ in-place and reuses its memory for $I_t^-$ and $I_t^+$.
The runtime complexity is the same as for building \texttt{GPPJoin} index (Algorithm~\ref{alg:build-index}):
$O(T + |\mathcal{T}| \text{df}^* \log \text{df}^*)$, where $\text{df}^*$ is the maximum frequency of any token among the token sets in $\mathcal{B}$.
We use the partitioned lists at line~\ref{alg:loop-partitioned-index} in Algorithm~\ref{alg:ttrk-join}.
Note that we crop and traverse $I_t^-$ first because it is (heuristically) more likely to contain high-similarity sets.
Figure~\ref{fig:pps-filtering-example} show an example of performing Prefix-Partitioned Suffix Filtering.

\subsection{Early Traversal Rank Cutoff ($\rho^*$)}
The filtering techniques employed allow us to prune the search space aggressively by skipping and cropping the inverted lists and avoid full similarity comparison, but sometimes this is not enough.
This is especially true when the thresholds are set conservatively (low $\tau$ and $\tau_r$, high $k$).
Therefore,
we introduce an additional parameter to stop the search for each $a \in \mathcal{A}$ early.
\begin{definition}
Assume we are querying $a$ and $I_{a[1]}^-, I_{a[1]}^+, \dots, I_{a[|a|]}^-, I_{a[|a|]}^+$ are all the inverted lists of the tokens in $a$.
To simplify notation, let ${I_{a[n]}^- = I_{2n-1}}$ and ${I_{a[n]}^+ = I_{2n}}$.
The traversal rank $\rho$ for the $q$th entry in $I_n$ is $q + \sum_{i=1}^{n-1} |I_{a[i]}|$.
\end{definition}
The parameter $\rho^*$ is the \textit{maximum traversal rank} and Algorithm~\ref{alg:ttrk-join} will only consider entries of the inverted indices with $\rho \leq \rho^*$.
Note that the traversal rank of an entry is stable across different values of $\tau$, $\tau_r$, and $k$.
Thus, $\rho^*$ will always prune the same entries.
This requires some extra bookkeeping (line~\ref{alg:ttkr-join-rho-start} and \ref{alg:ttkr-join-rho-end}) to handle the PPS filter.

Setting $\rho^* = \infty$ yields a $(\tau,\tau_r,k)$-join, but lower values might yield different results.
This parameter is the key to perform approximate $(\tau,\tau_r,k)$-joins.
We postpone the discussion about how to interpret and automatically set this parameter to Section~\ref{sec:approximate-joins}.

\subsection{Exploiting Parallelization}
State-of-the-art deep learning methods exploit modern GPUs to train and run their neural networks in a massively parallel fashion.
While there is no trivial way to leverage GPUs for our join algorithm,
it is relatively easy to make use of multiple CPU cores since the main loop on line~\ref{alg:ttrk-join-loop-a} is embarrassingly parallelizable.
Additionally,
we can parallelize partitioning and sorting of the inverted lists in the \texttt{BuildTTRKIndex} routine (Algorithm~\ref{alg:build-ttrk-index}).

\subsection{Fast Verification}
We adapt the verification routine from \citet{mannEmpiricalEvaluationSet2016} to weighted sets and other similarity measures in order to form a fast similarity routine.
See \texttt{PartialSim} in Algorithm~\ref{alg:partial-sim}.
Since we compute the similarity when encountering first common token $t$ and we know the position of $t$ in $a$ and $b$\footnote{
Note that \texttt{BuildTTRKIndex} stores both the position $j$ and the suffix size $s_b$ of the token $t$ in $b$ in the inverted list.
} as well as their suffixes $s_a$ and $s_b$,
we can skip the tokens $a[1..(i-1)]$ and $b[1..(j-1)]$ (remembering to take care of the Jaccard denominator).
Furthermore,
we can keep track over the remaining suffix weight of the token sets, as well as the remaining necessary overlap, to detect early that we can not meet the similarity threshold.
Note that Algorithm~\ref{alg:partial-sim} computes the terms for all similarity measures to give a complete picture,
but in practice we can compute only the ones necessary for the measure at hand.

\section{Join Behaviour Estimation Framework}\label{sec:estimation-framework}

\ttrkjoin have four hyperparamters that are hard to set.
No matter the techniques we employ to make the algorithm efficient,
it will still be very resource intensive to explore different hyperparameter configurations by simply running the algorithm with a large number of different configurations.
So central to the method we will propose is the ability to estimate its behaviour with different hyperparameters without running the full algorithm.
We are interested in estimating the three key characteristics of blocking through their main measure: 1) Recall, 2) Number of retrieved pairs, and 3) Runtime.
In this section,
we will present a framework for estimating these properties through analysis of select or sampled queries.

\subsection{Recall Conditions}
The only reliable way to estimate recall (at least in the general case) is to evaluate against known matches.
Assume we have a representative subset of known matches $\widetilde{M} \subseteq M$.
Instead of running \texttt{TTRKJoin} with every hyperparamter configuration of interest to find the recall of $\widetilde{M}$,
assume for each known match $m \in \widetilde{M}$ we know the maximum $\tau$ and $\tau_r$ and the minimum $k$ and $\rho*$ that is necessary to recall it.
We call those four values the recall conditions of $m$ --- denoted $\theta_m$.
Given a hyperparameter configuration $(\tau, \tau_r, k, \rho^*)$ we can answer whether a match $m$ would be recalled by \texttt{TTRKJoin} in constant time by checking whether the hyperparameters respect the recall conditions $\theta_m$.
Furthermore, if we have the recall conditions $\Theta = \{\theta_m\}$ for all $m \in \widetilde{M}$ we can trivially estimate the recall in $O(|\widetilde{M}|)$ time.
For this strategy to be effective we need to able to find the recall conditions efficiently.

\subsubsection{Finding Recall Conditions}

\let\oldnl\nl
\newcommand{\nonl}{\renewcommand{\nl}{\let\nl\oldnl}}
\begin{algorithm}[tbhp]
    \caption{FindRecallCond$(\mathcal{A}, I^\pm, \widetilde{M}, D, [\tau, \tau_r, k, \rho^*]; sim)$}
    \label{alg:find-recall-cond}
    \footnotesize
    \SetKw{Continue}{\textbf{continue}}
    \SetKw{Break}{\textbf{break}}
    \SetKwFunction{TTRKJoin}{TTRKJoin}
    \SetKwBlock{DummyBlock}{\vspace{-1em}}{}
    \KwIn{
        $\mathcal{A}$ is a collection of weighted token sets already sorted by $\mathcal{O}$.
        $I^\pm$ is a PPS index of $\mathcal{B}$.
        $\widetilde{M} \subseteq \mathcal{A} \times \mathcal{B}$ are known matches.
        $D$ are regularization parameter values of interest.
        $\tau, \tau_r, k, \rho^*$ are the most permissive values of the \TTRKJoin parameters that the conditions will cover (defaults to $[0,0,\infty,\infty]$).
    }
    \KwOut{$\Theta^d$, the strictest join conditions ($\tau, \tau_r, k, \rho^*$) that would recall each match in $\widetilde{M}$ per regularization configuration $d \in D$.}
    
    \tcc{
        \TTRKJoin with the following changes:
    }
    \tcp{1.~Use provided index}
    \setcounter{AlgoLine}{0}
    $I^+, I^- \gets I^\pm$\;
    \tcp{2.~Initialize conditions}
    $\Theta \gets \{ \emptyset \}^d$\;
    
    \nonl $\cdots$
    
    \setcounter{AlgoLine}{2}
    \tcp{3.~Query $\mathcal{A}_{\widetilde{M}}$ instead of $\mathcal{A}$}
    \nonl$\mathcal{A}_M \gets \Big\{\bm{a} \in \mathcal{A} \mid {\exists b\!\in\!B} \big[ (a,b) \in \widetilde{M} \big] \Big\}$\;
    \ForEach{$a \in \mathcal{A}_{\widetilde{M}}$}{\label{alg:A_M}
        \nonl $\cdots$\\
        \setcounter{AlgoLine}{5}
        \tcp{4.~Use matches to tighten threshold}
        $\dynthresh \gets \max \Big( \tau, (1 - \max D)\min_{b \mid (a,b) \in \widetilde{M}} sim(a,b) \Big)$\label{alg:tighten-with-matches}\;
        \nonl $\cdots$\\
        \SetAlgoNoLine\nonl\DummyBlock{\SetAlgoVlined%
            \SetAlgoNoLine\nonl\DummyBlock{\SetAlgoVlined%
                \SetAlgoNoLine\nonl\DummyBlock{\SetAlgoVlined%
                    \SetAlgoNoLine\nonl\DummyBlock{\SetAlgoVlined%
                        \tcp{5.~Store $\rho$ for each candidate}
                        \setcounter{AlgoLine}{24}
                        $Q.\text{push}((b, \rho), S)$\label{alg:push_rho}\;
                    }
                }
            }
        }
        \nonl $\cdots$\\
        \setcounter{AlgoLine}{33}
        \tcp{5.~Recall conditions instead of candidates}
        $\Theta^d \gets$ extract recall conditions from $Q$ for each $d \in D$\;
    }
    \KwRet{$\Theta$}\;
\end{algorithm}

We can run a modified version of \ttrkjoin once to determine the recall conditions of $\widetilde{M}$.
Algorithm~\ref{alg:find-recall-cond} outlines this modified version, \texttt{FindRecallCond}.
The idea is to search for each $a$ that occur in a match $(a,b) \in \widetilde{M}$,
and then extract the recall conditions from $Q$.
We make sure to store the traversal rank $\rho$ of when $b$ was discovered in $Q$ (Line~\ref{alg:push_rho}).
The recall conditions are then trivially found by sorting in descending similarity order and traversing $Q$ looking for each $b$ that match $a$.
Given $(b, \rho)$ at position $i$ in sorted $Q$, the maximum $\tau$ is $Q[i].sim$, the maximum $\tau_r$ is $Q[i].sim / Q[1].sim$, the minimum $k$ is $i$, and the minimum $\rho^*$ is $\rho$.

We make two optimizations beyond those in \ttrkjoin.
First, we only query $a \in \mathcal{A}$ which are part of at least one match (Line~\ref{alg:A_M}).
Second, we can tighten $\dynthresh$ up front (Line~\ref{alg:tighten-with-matches}) since we know which $b \in \mathcal{B}$ we are interested in finding and token sets with lower similarity will not affect their recall conditions.

\subsubsection{Regularization}\label{sec:regularization}
If we use recall conditions for a limited sized $\widetilde{M}$ to decide hyperparameter configurations we risk overfitting.
Therefore, we propose a simple, yet effective, regularization parameter applied on recall conditions to control overfitting: the similarity margin $d$.
Given a match $m = (a,b)$, the recall conditions $\theta_m^d$ for $m$ with similarity margin $d$ is the recall conditions $m$ would have if the similarity of $a$ and $b$ was $(1 - d) sim(a,b)$ and all other similarities remained unchanged.
Higher values of $d$ give more conservative recall conditions and stronger regularization.

\texttt{FindRecallCond} (Algorithm~\ref{alg:find-recall-cond}) accepts a set $D$ of similarity margins and returns the recall conditions for all $m \in \widetilde{M}$ for each $d \in D$.
The recall conditions for different $d$ are extracted from the sorted $Q$ by simulating the effect it would have if the similarity was $(1 - d) \cdot sim(a,b)$.
This is straightforward for maximum $\tau$, maximum $\tau_r$, and minimum $k$,
but the effect on minimum $\rho^*$ is not easily defined.
As a heuristic, we set the minimum $\rho^*$ to the maximum of the traversal rank of $b$ and the candidate at the position in $Q$ we would replace if we lowered the similarity of $b$ to $(1 - d) \cdot sim(a,b)$.
Note that in order to be sure $Q$ contains enough candidates to see the effect of each $d \in D$ we factor them in when tightening $\dynthresh$ with the matches on line~\ref{alg:tighten-with-matches}.

\subsection{Search Trajectories}

One uncomplicated way to estimate the number of retrieved pairs and runtime for a hyperparameter configuration without running the full algorithm is to randomly sample a subset $\widetilde{\mathcal{A}} \subseteq \mathcal{A}$, run \ttrkjoin on it, and then multiply the resulting number of pairs and runtime with $|\mathcal{A}| / |\widetilde{\mathcal{A}}|$.
Assuming $|\widetilde{\mathcal{A}}|$ is large enough, it will provide accurate estimations with a $|\widetilde{\mathcal{A}}| / |\mathcal{A}|$ reduction in runtime.
Unfortunately, that might still be too computationally expensive if we want to check many hyperparameter configurations.

Therefore, we propose an approach were we selectively record the state of the search and runtime as we query each $a \in \widetilde{\mathcal{A}}$ once,
and then use this information to quickly simulate approximate runs of \ttrkjoin on $\widetilde{\mathcal{A}}$.
The result of the simulated runs will provide us with pessimistic estimations of the number of returned pairs and runtime.
Additionally, we can estimate the sum of similarities for the returned pairs, which will be important when doing approximate joins.

Let a search trajectory checkpoint reflect the state of the search at some traversal rank $\rho$ and consist of six parts:
\begin{enumerate}
    \item $CH_P$: Cumulative histogram quantifying the number of pairs with similarity below $|CH_P|$ linearly spaced similarity levels between $0$ and the highest possible similarity\footnote{The highest possible similarity is $1$ for Jaccard, Cosine, and Dice --- while for Overlap it is $\min(\omega(a), \max_{b \in \mathcal{B}} \omega(b))$.}.
    \item $CH_S$: Cumulative histogram quantifying the summed similarity for pairs with similarity below $|CH_S|$ linearly spaced similarity levels between $0$ and the highest possible similarity.
    \item $S^-_{\log k}$: Lower bound on similarities for the top $k$ pairs for exponentially spaced values of $k$ such that $k = \argmax_{\hat{k}} \Big[ \lceil \log \hat{k} \rceil = \lceil \log k \rceil \Big]$.
    \item $rt$: Empirically measured runtime since the start of query for $a$.
    \item $S^*$: Highest similarity among the candidates.
    \item $s_a$: The current prefix of the query $a$ we have traversed so far.
\end{enumerate}
Let a search trajectory $\psi$ for a token set $a$ on index $I^\pm$ be a list of checkpoints for exponentially spaced values of $\rho$.
\begin{algorithm}[tbhp]
    \caption{RecordTrajectories$(\widetilde{\mathcal{A}}, I^\pm, [\tau, \tau_r, k, \rho^*]; sim)$}
    \label{alg:record-trajectories}
    \footnotesize
    \SetKwInput{KwConstants}{Constants}
    \SetKw{Continue}{\textbf{continue}}
    \SetKw{Break}{\textbf{break}}
    \SetKwFunction{TTRKJoin}{TTRKJoin}
    \SetKwBlock{DummyBlock}{\vspace{-1em}}{}
    \KwIn{
        $\widetilde{\mathcal{A}}$ is a collection of weighted token sets already sorted by $\mathcal{O}$.
        $I^\pm$ is a PPS index of $\mathcal{B}$.
        $\tau, \tau_r, k, \rho^*$ are the most permissive values of the \TTRKJoin parameters that the trajectories will cover (defaults to $[0,0,\infty,\infty]$).
    }
    \KwOut{$\Psi$, search trajectory for each $a \in \widetilde{\mathcal{A}}$.}
    \KwConstants{
        $|H_P| = |H_S| = |CH_P| = |CH_S| = 100$;
        $r_\psi = 1.1$;
        Base number of $\log k$: 1.1;
    }
    
    \tcc{
        \TTRKJoin with the following changes:
    }
    \tcp{1.~Use provided index}
    \setcounter{AlgoLine}{0}
    $I^+, I^- \gets I^\pm$\;
    \tcp{2.~Initialize trajectories}
    $\Psi \gets \emptyset$\;
    
    \nonl $\cdots$
    
    \ForEach{$a \in \widetilde{\mathcal{A}}$}{
        \setcounter{AlgoLine}{5}
        \tcp{3.~Initialize trajectory bookkeeping}
        $\psi \gets $empty array, $\rho_\psi \gets 1$, $S^* \gets 0$\;
        \nonl $H_P, H_S \gets$ empty histograms, $start \gets now()$\;
        \nonl $\cdots$\\
        \SetAlgoNoLine\nonl\DummyBlock{\SetAlgoVlined%
            \SetAlgoNoLine\nonl\DummyBlock{\SetAlgoVlined%
                \setcounter{AlgoLine}{15}
                \ForEach{$(b, j, s_b) \in I_t[start..end]$}{%
                    \tcp{4.~Store trajectory checkpoints}
                    \nonl\If{$\rho \geq \rho_\psi$}{
                        \nonl$\rho_\psi \gets \lceil r_\psi \cdot \rho_\psi \rceil$\;
                        \nonl$rt \gets now() - start$\;
                        \nonl$CH_P, CH_S, S^-_{\log k} \gets$ from $H_P$ and $H_S$\;
                        \nonl$\psi.push((CH_P, CH_S, S^-_{\log k}, rt, S^*, s_a))$\;
                    }
                    
                    \nonl $\cdots$
                    
                    \SetAlgoNoLine\nonl\DummyBlock{\SetAlgoVlined%
                        \setcounter{AlgoLine}{24}
                        \tcp{5.~Maintain $H_P$, $H_S$, and $S^*$}
                        $Q.\text{push}(b, S)$, $H.add(S)$\;
                        $S^* \gets \max(S^*, S)$\;
                        \If{$|Q| > k$}{%
                            \nonl $S_{pop} \gets Q.pop().sim$\;
                            \nonl Remove $S_{pop}$ from $H_P$ and $H_S$\;
                        }%
                        
                        \nonl $\cdots$
                        
                        \SetAlgoNoLine\nonl\DummyBlock{\SetAlgoVlined%
                            \While{$Q.top.sim < \dynthresh$}{%
                                \nonl $S_{pop} \gets Q.pop().sim$\;
                                \nonl Remove $S_{pop}$ from $H_P$ and $H_S$\;
                            }%
                        }
                    }
                }
            }
        }
        \nonl $\cdots$\\
        \tcp{6.~Last checkpoint}
        \nonl$ts \gets now() - start$\;
        \nonl$CH_P, CH_S, S^-_{\log k} \gets$ from $H_P$ and $H_S$\;
        \nonl$\psi.push((CH_P, CH_S, S^-_{\log k}, rt, S^*, s_a))$\;
        \nonl$\mathcal{S} \gets$ sorted descending similarities from $Q$\;
        \nonl$C\mathcal{S} \gets$ cumulative $\mathcal{S}$\;
        \tcp{7.~Store trajectory}
        \setcounter{AlgoLine}{33}
        $\Psi \gets \Psi \cup \{ (\psi, \mathcal{S}, C\mathcal{S}) \}$\;
    }
    \KwRet{$\Psi$}\;
\end{algorithm}
We can record search trajectories for $\widetilde{\mathcal{A}}$ by running a modified version of \ttrkjoin,
as outlined in Algorithm~\ref{alg:record-trajectories}.
We will now briefly explain how we use the recorded trajectories to estimate the number of returned pairs, runtime, and similarity sum.

\subsubsection{Estimating the Number of Pairs}\label{sec:estimated-number-of-pairs}
In order to be more robust to overfitting we estimate an upper bound of the number of pairs instead of the number of pairs directly.
We do this by first upper bounding the returned pairs from querying $a \in \widetilde{\mathcal{A}}$ and then multiplying the upper bound with $|\mathcal{A}| / |\widetilde{\mathcal{A}}|$ to get an estimated upper bound on the total number of pairs returned.

To get an upper bound on the number of pairs for a single query of $a$ with hyperparameters $(\tau, \tau_r, k, \rho^*)$ using the corresponding search trajectory $\psi$,
first lookup the checkpoint with lowest $\rho$ such that $\rho \geq \rho^*$.
Then lookup the lowest similarity bin of $CH_P$ which interval lies below (or touches) $\max(\tau, \tau_r S^*)$ --- denoted $CH_P[\max(\tau, \tau_r S^*)]$.
Taking $k$ into consideration, the lower bound is $\min(k, CH_P[\max(\tau, \tau_r S^*)])$.
Sum the upper bounds for all $a \in \widetilde{\mathcal{A}}$ and multiple by $|\mathcal{A}| / |\widetilde{\mathcal{A}}|$ to get the estimate for (an upper bound of) the total number of returned pairs.
This is done in $\Theta(|\widetilde{\mathcal{A}}|)$ time.

\subsubsection{Estimating Runtime}\label{sec:estimated-runtime}
To estimate an upper bound for the runtime of a single query $a$ with hyperparameters $(\tau, \tau_r, k, \rho^*)$ using the corresponding search trajectory $\psi$,
we binary search for the checkpoint with lowest $\rho$ that satisfy $\rho \geq \rho^*$ and $s_a \geq \sigma \big(a, \max[\tau, \tau_r S^*, S^-_{\lceil \log k \rceil}] \big)$.
The estimated upper bound for the runtime is then the empirical runtime $rt$ for that checkpoint.
Sum the estimates for all $a \in \widetilde{\mathcal{A}}$ and multiple by $|\mathcal{A}| / |\widetilde{\mathcal{A}}|$ to get the estimate for total runtime of the join\footnote{In addition, divide by number of threads when running in parallel.}.
Since the traversal rank is always bounded by the total number of tokens $\sum_\mathcal{B}$ in $\mathcal{B}$,
and the checkpoints are exponentially spaced,
estimating the runtime is $O(|\widetilde{\mathcal{A}}| \log \log \sum_\mathcal{B})$.

There are many sources of uncertainty when estimating the runtime like this.
Measuring empirical runtime is in itself unreliable,
the \texttt{RecordTrajectories} routine have higher overhead than \ttrkjoin,
and it does not effectively capture the effect of the PPS and positional filter (only prefix) when using different hyperparameters.
However,
we argue that it is sufficient as an heuristic to distinguish different orders of magnitude in runtime when automatically picking hyperparameters.
The reported results of our method will back up this claim.

\section{Approximate Joins}\label{sec:approximate-joins}
One can achieve great runtime performance for even large datasets with similarity joins by picking the similarity threshold aggressively enough~\cite{mannEmpiricalEvaluationSet2016}.
Unfortunately, for datasets that require conservative thresholds (e.g., $\tau < 0.4$ for cosine, or equivalently picking $k$ moderately high for a $k$-join approach) to get high recall it is challenging to avoid approaching quadratic runtime.
The main reason is because the effectiveness of prefix filtering degrades quickly once the similarity threshold is so low we do not prune away high frequency tokens anymore --- we can end up checking most token sets in $\mathcal{B}$.

A popular approach to scalability problems when further impactful algorithmic improvements are difficult to pull off is to relax the requirements and allow approximations that are good enough for practical purposes.
Note that for string similarity-based blocking in particular this is not unreasonable since the similarity measures and thresholds we use are already in reality approximations.
In this section,
we will propose a simple and effective way of performing approximate $(\tau, \tau_r, k)$-joins with \ttrkjoin.

\subsection{Definitions}

Let us start by defining what constitute an approximate join and the quality of it.

\begin{definition}[Approximate $\bm{(\tau, \tau_r, k)}$-Join]
    Let $\mathcal{A}$ and $\mathcal{B}$ be two collections of sets, and $\text{sim}(a,b)$ be some set similarity measure.
    An approximate $(\tau, \tau_r, k)$-join is any routine that will return a set of pairs $P$ such that for all pairs $(a,b) \in P$
    \begin{equation*}
            sim(a,b) \geq \tau \land sim(a,b) \geq \tau_r \max_{(a', b') \in P \mid a' = a} \big[ sim(a',b') \big]
    \end{equation*}
    and that for all $a \in \mathcal{A}$
    \begin{equation*}
        \big| \{(a',b') \in P \mid a' = a\} \big| \leq k
    \end{equation*}
\end{definition}

In other words,
while a $(\tau, \tau_r, k)$-join must return all pairs that are among the $k$ most similar and within $\tau_r$ of the most similar for each $a \in \mathcal{A}$ in addition to having a similarity of at least $\tau$,
an approximate $(\tau, \tau_r, k)$-join must only return some set of pairs that is internally consistent.
That is,
there can not be more than $k$ pairs for each $a \in \mathcal{A}$,
two pairs $(a,b)$ and $(a, b')$ such that $sim(a,b) < \tau_r \cdot sim(a,b')$,
or any pairs with similarity below $\tau$.
In order to quantify how well an approximate $(\tau, \tau_r, k)$-join approximates an exact $(\tau, \tau_r, k)$-join we define the quality of an approximate join.

\begin{definition}[Quality of Approximate Join]
Let $\mathcal{B}_a$ be the token sets from $\mathcal{B}$ matched to some token set $a \in \mathcal{A}$ by a $(\tau, \tau_r, k)$-join and let $S_a^* = \max_{b \in \mathcal{B}_a} sim(a, b)$.
Furthermore, let $\widehat{\mathcal{B}}_a$ be the token sets matched to $a$ from an approximate $(\tau, \tau_r, k)$-join.
The quality of this approximate $(k, \tau, \tau_r)$-join on $\mathcal{A}$ and $\mathcal{B}$ is defined as
\begin{equation*}
    \frac{1}{|\mathcal{A}|} \sum_{a \in \mathcal{A}} \frac{ \sum_{b \in \widehat{\mathcal{B}}_a \mid sim(a,b) \geq \tau_r S^*_a} sim(a, b) }{ \sum_{b \in \mathcal{B}_a} sim(a, b) }
\end{equation*}
\end{definition}

The quality of an exact $(k, \tau, \tau_r)$-join is trivially always $1$.
The intuition is that approximations that miss high-similarity pairs that stick out and can not be replaced are of low quality,
while approximations that miss some pairs but have found others with comparable similarity are of high quality.
Note that we only include $b \in \widehat{\mathcal{B}}_a$ when the similarity is at least $\tau_r$ of the highest similarity that exist to $a$.
This is so approximate joins can not be rewarded for not including the highest similarity pair for any $a \in \mathcal{A}$ (and achieving a quality above 1).

\begin{example}
Assume $\tau = 0.2, \tau_r = 0.5, k = 4$ and let $\mathcal{A} = \{ a_1, a_2 \}$ and $\mathcal{B} = \{ b_1, \dots, b_6 \}$.
The similarity according to some similarity measure is
\begin{center}
    \begin{tabular}{c|cccccc}
              & $b_1$ & $b_2$ & $b_3$ & $b_4$ & $b_5$ & $b_6$ \\ \midrule
        $a_1$ &   0.5 &   0.8 &   0.1 &   0.4 &   0.6 &   0.6 \\
        $a_2$ &   0.4 &   0.3 &   0.5 &   0.9 &   0.2 &   0.4 \\
    \end{tabular}
\end{center}
The result of a $(\tau, \tau_r, k)$-join would be
\begin{equation*}
    P = \{ (a_1, b_1), (a_1, b_2), (a_1, b_5), (a_1, b_6), (a_2, b_3), (a_2, b_4) \}
\end{equation*}
Let $P_1$ and $P_2$ be the result of two different approximate joins:
\begin{align*}
    P_1 = \{ & (a_1, b_2), (a_1, b_4), (a_1, b_5), (a_1, b_6), (a_2, b_4) \} \\
    P_2 = \{ & (a_1, b_1), (a_1, b_4), (a_1, b_5), (a_1, b_6), (a_2, b_1), (a_2, b_2), \\ & (a_2, b_3), (a_2, b_6) \}
\end{align*}
The corresponding quality is
\begin{align*}
    q_1 = & \frac{1}{2} \Bigg( \frac{0.8+0.4+0.6+0.6}{0.5+0.8+0.6+0.6} + \frac{0.9}{0.5+0.9} \Bigg) = 0.80 \\
    q_2 = & \frac{1}{2} \Bigg( \frac{0.4+0.5+0.6+0.6}{0.5+0.8+0.6+0.6} + \frac{0.5}{0.5+0.9} \Bigg) = 0.60
\end{align*}
\end{example}

We are specifically interested in approximate joins that achieve a certain level of quality with some probability.

\begin{definition}[$\bm{(q,q_p)}$-Approximate $\bm{(\tau, \tau_r, k)}$-Join]
A $(q,q_p)$-approximate $(\tau, \tau_r, k)$-join is an approximate $(\tau, \tau_r, k)$-join with quality of at least $q$ with probability $q_p$.
\end{definition}

We will now look at how to perform a $(q, q_p)$-approximated join with \ttrkjoin.

\subsection{Approximation with \ttrkjoin}

A reasonable hypothesis for most real world data is that matching token sets will have at least one rare token in common with high probability~\cite{ohareHighValueTokenBlockingEfficient2021}.
Prefix filtering already exploits this for runtime performance but only within the constraints of still guaranteeing finding all pairs above some threshold.
On the assumption that most matches have at least one rare token in common,
prefix filtering with low thresholds must drudge through a disproportionately long tail of token sets that only have ubiquitous tokens in common compared to how likely it is to find any matches.
Taking inspiration from information retrieval and \citet{ohareHighValueTokenBlockingEfficient2021},
a natural solution is to simply cut off the search early for each query $a \in \mathcal{A}$.
Trading in some small degradation of the result for reduced runtime.
Unfortunately, setting the cutoff condition is hard because its impact so heavily rely on the dataset and other hyperparameters.
We do not have any control over how much quality we give up.

The key insight is that \ttrkjoin always performs an approximate $(\tau, \tau_r, k)$-join,
no matter the value of the traversal rank cutoff $\rho^*$,
and that any approximation quality for a dataset $(\mathcal{A}$, $\mathcal{B})$ can be achieved by setting $\rho^*$ high enough.
In other words, for any dataset $(\mathcal{A}, \mathcal{B})$, join conditions $(\tau, \tau_r, k)$, and approximation quality $q$ there exist a finite minimum $\rho^*$ that will make \ttrkjoin do an approximate join with quality of at least $q$.
Importantly, the converse is also true.
That is, any \ttrkjoin will yield an approximate join of some quality $q$.
This opens up two main ways to utilize \ttrkjoin for approximate joins.
We can either target a specific quality level and try to determine the necessary $\rho^*$,
or we can determine a suitable $\rho^*$ by other means (e.g., validation/supervised learning) and then try to determine the approximation quality.
We cover the former since it most relevant,
but we note the latter is easily achieved in similar spirit.

\subsection{Choosing $\rho^*$ to achieve quality $q$}

\begin{algorithm}[tbhp]
    \caption{QualityToRank$(q, q_p, \tau, \tau_r, k, \Psi)$}
    \label{alg:quality-to-rank}
    \footnotesize
    \SetKw{Continue}{\textbf{continue}}
    \SetKw{Break}{\textbf{break}}
    \SetKwFunction{TTRKJoin}{TTRKJoin}
    \SetKwFunction{SimSumLB}{SimSumLB}
    \SetKwProg{Fn}{Function}{}{}
    \KwIn{
        $q$ and $q_p$ is the desired approximation quality and probability of achieving at least that.
        $\tau, \tau_r, k$ is the join conditions of a $(\tau, \tau_r, k)$-join.
        $\Psi$ are a set of randomly sampled \ttrkjoin search trajectories from the dataset of interest.
    }
    \KwOut{$\rho^*$, a max traversal rank that will make \ttrkjoin perform a $(q,q_p)$-approximated $(\tau, \tau_r, k)$-join.}
    
    \vspace{0.5em}
    
    \Fn{\SimSumLB($\tau, \tau_r, k, \rho^*, \psi$)}{%
        $CH_P, CH_S \gets $ from the earliest checkpoint in $\psi$ with $\rho \geq \rho^*$ (or last)\;
        $S^* \gets $ from last checkpoint in $\psi$\;
        $\dynthresh \gets \max(\tau, \tau_r \cdot S^*)$\;
        $i \gets$ lowest bin of $CH_P$ whose entire interval is at least $\dynthresh$\;
        \uIf(\tcp*[f]{Tight similarity constraint}){$CH_P[i] \leq k$}{%
            \KwRet{$CH_S[i]$}\;
        }
        \Else(\tcp*[f]{Tight cardinality constraint}){%
            $j \gets $ lowest bin $j \geq i$ such that $CH_P[j] \geq k$\;
            $\tau^+ \gets$ upper similarity bound of bin $j$\;
            \KwRet{$CH_S[j] - \tau^+ (CH_P[j] - k)$}\;
        }
    }
    
    \vspace{0.5em}
    
    \ForEach{$(\psi, \mathcal{S}, C\mathcal{S}) \in \Psi$}{%
        $i \gets $ binary search lowest $\mathcal{S}[i] \geq \max(\tau, \tau_r \mathcal{S}[1])$\;
        $\Sigma_\psi \gets C\mathcal{S}[\min(i, k)]$\;
    }
    
    $R \gets$ empty array\;
    \ForEach(\tcp*[f]{$N_B$ resamples}){bootstrap resample $\widetilde{\Psi}$ of $\Psi$}{%
        $\hat{\rho}^* \gets $ binary search on log scale lowest such that \\
        \phantom{$\rho^* \gets $} $\frac{1}{|\widetilde{\Psi}|} \sum_{\psi \in \widetilde{\Psi}} \frac{\SimSumLB(\tau, \tau_r, k, \rho^*, \psi)}{\Sigma_\psi} \geq q$\;
        $R.push(\hat{\rho}^*)$\;
    }
    Sort $R$ ascending\;
    \KwRet{$R\Big[ \big\lceil q_p \cdot |R| \big\rceil \Big]$}\;
\end{algorithm}

If we want to achieve a certain approximation quality by choosing a finite $\rho^*$,
and without actually running a full exact join,
we need to accept some level of uncertainty.
Therefore, we propose to use a sample of search trajectories to find a $\rho^*$ that will yield a $(q, q_p)$-approximated join.

Assume we want to do a $(q,q_p)$-approximated $(\tau, \tau_r, k)$-join between $\mathcal{A}$ and $\mathcal{B}$.
Let $\Psi$ be a set of recorded search trajectories for a random subset $\widetilde{\mathcal{A}} \subseteq \mathcal{A}$.
We can quickly determine an upper bound of the lowest $\rho^*$ that is guaranteed to achieve quality $q$ with \ttrkjoin between $\widetilde{\mathcal{A}}$ and $\mathcal{B}$ using $\Psi$.
It is mostly a matter of binary searching the exponentially spaced traversal ranks of the trajectories in $\Psi$ because $CH_S$ can be used to lower bound the quality.
Furthermore,
we can use this as an estimator for a $\rho^*$ that achieve quality $q$ on the full join with $\mathcal{A}$.
Since we do not know the underlying distribution,
we use bootstrapping to find the $q_p$ quantile.
Resample the already existing search trajectories $\Psi$ $N_B$ times to produce $N_B$ estimates and pick the empirical $q_p$ quantile.
The resulting $\rho^*$ makes \ttrkjoin do a $(q,q_p)$-approximated join.
Algorithm~\ref{alg:quality-to-rank} outlines the proposed approach.

\section{ShallowBlocker}\label{sec:shallowblocker}

\begin{figure}[tbhp]
    \centering
    \includegraphics{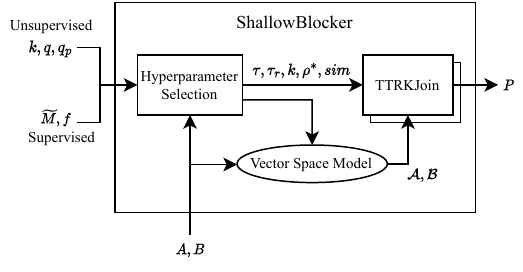}
    \caption{Illustration of the high-level conceptual flow of ShallowBlocker.}
    \label{fig:shallow-blocker}
\end{figure}

We are now ready to assemble everything and present our proposed method for blocking, ShallowBlocker.
Fundamentally,
our method performs one or more $(\tau, \tau_r, k)$-joins using \ttrkjoin.
Therefore, the focus in this section is mainly how ShallowBlocker chooses hyperparameters and applies \ttrkjoin.
Figure~\ref{fig:shallow-blocker} illustrates the high-level flow consisting of two main steps.
First, we perform hyperparameter selection --- which not only includes hyperparameters to \ttrkjoin and choice of set similarity measure, but also the token set model we use to go from strings to weighted token sets.
Then, secondly, we run one or more configurations of \ttrkjoin.

Note that we intentionally present the method in this fashion to highlight a key characteristic, namely its interpretability.
At its core,
the proposed method is a string similarity join with clearly defined conditions dictating what pairs to produce,
and these join conditions can be inspected and interpreted to understand how the end result is produced.
Specifically,
since ShallowBlocker runs \ttrkjoin it will always perform exact or approximated $(\tau, \tau_r, k)$-joins.
In cases where $\rho^*$ is finite we can infer some $q$ and $q_p$ for which it is true that the performed join is a $(q, q_p)$-approximate $(\tau,\tau_r,k)$-join (see Section~\ref{sec:approximate-joins}).

We perform hyperparameter selection differently depending on whether examples of true matches are available or not.
So we describe the unsupervised and supervised approach separately for ease of presentation.

\subsection{Unsupervised Blocking}

\begin{algorithm}[tbhp]
    \caption{ShallowBlocker$(A, B, k, q)$, Unsupervised}
    \label{alg:unsupervised-shallowblocker}
    \footnotesize
    \SetKw{Continue}{\textbf{continue}}
    \SetKw{Break}{\textbf{break}}
    \KwIn{
        $A$ and $B$ are collections of strings.
        $k$ bounds to maximum number of pairs to $k \cdot \min(|\mathcal{A}|, |\mathcal{B}|)$,
        and $q$ is the quality of underlying joins.
    }
    \KwOut{Pairs $C \subseteq A \times B$ such that $|P| \leq k \cdot \min(|A|,|B|)$.}
    
    \SetKwFunction{BalancedTTRKJoin}{BalancedTTRKJoin}
    \SetKwFunction{flipPairs}{flipPairs}
    
    \If{$|A| > |B|$}{\label{alg:unsupervised-budget-start}%
        $\text{swap}(A,B)$\tcp*[r]{Also flip pairs in resulting $C$}
    }
    $k_{ba} \gets \Big\lfloor \frac{k|A|}{2 |B|} \Big\rfloor$\;
    $k_{ab} \gets \Big\lfloor \frac{k|A| - k_{ba} |B|}{|A|} \Big\rfloor$ \label{alg:unsupervised-budget-end}\;
    
    Precompute TF-IDF weights and $\mathcal{O}$ for $A$ and $B$ using all tokenizers\;
    
    $P_{ab} \gets \BalancedTTRKJoin(A, B, k_{ab}, q)$\;
    $P_{ba} \gets \BalancedTTRKJoin(B, A, k_{ba}, q)$\;
    
    $P \gets P_{ab} \cup \flipPairs(P_{ba})$\;
    \KwRet{$P$}\;
\end{algorithm}

Algorithm~\ref{alg:unsupervised-shallowblocker} outlines the unsupervised approach.
It accepts two hyperparameters:
\begin{itemize}
    \item $k$:
    Bounds the number of pairs to $|P| \leq k \cdot \min(|A|,|B|)$.
    This lets the user adjust trade-off between recall and number of returned pairs.
    A reasonable default is 10.
    \item $q$:
    The approximation quality of the joins performed (with $q_p$ probability).
    This lets the user adjust trade-off between precision and runtime.
    This is a valuable parameter for quick iterations early in an entity matching development workflow.
    A reasonable default is 1 (no approximation).
\end{itemize}
While it is possible to construct unsupervised methods without any user-provided parameters~\cite{ohareHighValueTokenBlockingEfficient2021},
we find that unpractical in real-world use cases.
No method is perfect on all datasets for all use cases and it is in practice necessary to be able to adjust the precision-recall and/or precision-runtime tradeoff.
Existing blocking methods vary in the parameters they expose to let the user adjust this.
We argue, based on our own experience, that a user-friendly way to adjust these trade-offs is to control the pair budget and some level of approximation.
Our reasoning is twofold:
1) The parameters are not very sensitive to small changes\footnote{Contrast this to similarity threshold parameters, for which it is easy to fall into the failure modes of producing almost no pairs or way too many.}.
2) The parameters are not very sensitive to dataset-specific characteristics, and therefore it is possible to provide reasonable defaults.

Since $(\tau, \tau_r, k)$-joins are directional it matters which way to perform \ttrkjoin.
For robust results we perform one join in each direction --- denoted $ab$ when indexing $B$ and querying $A$ and $ba$ for the other way around.
We divide the pair budget between them by fixing each joins $k$ parameter, which guarantees we will not exceed the total budget.
The budget is divided as even as possible (see line~\ref{alg:unsupervised-budget-start}-\ref{alg:unsupervised-budget-end}).

\begin{algorithm}[tbhp]
    \caption{BalancedTTRKJoin$(A, B, k, q)$}
    \label{alg:balanced-ttrkjoin}
    \footnotesize
    \SetKwInput{KwIn}{Constants}
    \SetKw{Continue}{\textbf{continue}}
    \SetKw{Break}{\textbf{break}}
    
    \SetKwFunction{TTRKJoin}{TTRKJoin}
    \SetKwFunction{BuildTTRKIndex}{BuildTTRKIndex}
    \SetKwFunction{tsm}{ToTokenSets}
    \SetKwFunction{RecordTrajectories}{RecordTrajectories}
    \SetKwFunction{QualityToRank}{QualityToRank}
    \SetKwFunction{DiscriminatePower}{DiscriminatePower}
    \SetKwFunction{EstimateNumPairs}{EstimateNumPairs}
    
    \KwIn{$N_A=1000$; $k_{dp} = 10$; $q_p=0.95$;}
    
    $\widetilde{A} \gets$ sample max $N_A$ strings from $A$ \label{alg:pick-discriminate-start}\;
    
    $dp^* \gets -1$\;
    $\Psi^* \gets \emptyset$\;
    \ForEach{tokenizer \texttt{tok}}{%
        $\widetilde{\mathcal{A}} \gets \tsm(\widetilde{A}, \texttt{tok}, \text{TF-IDF})$\;
        $\mathcal{B} \gets \tsm(B, \texttt{tok}, \text{TF-IDF})$\;
        \ForEach{similarity measure $sim$}{%
            $I^\pm \gets \BuildTTRKIndex(\mathcal{B}, 0; sim)$\;
            $\Psi \gets \RecordTrajectories(I^\pm, \widetilde{\mathcal{A}}, \infty)$\;
            \uIf{$q = 1$}{%
                $\rho^* \gets \infty$\;
            }
            \Else{%
                $\rho^* \gets \QualityToRank(q, q_p, 0, 0, k_{dp}, \Psi)$\;
            }
            $\widetilde{P} \gets \TTRKJoin(\widetilde{\mathcal{A}}, \mathcal{B}, 0, 0, k_{dp}, \rho^*; sim)$ using $I^\pm$\;
            $dp \gets \DiscriminatePower(\widetilde{P})$\;
            \If{$dp > dp^*$}{%
                $dp^* \gets dp$\;
                $\Psi^* \gets \Psi$\;
            }
        }
        \label{alg:pick-discriminate-end}
    }
    
    $\tau \gets \max \{ \tau \mid \EstimateNumPairs(\tau, 0, \infty, \Psi^*) \geq k |A| \}$ \label{alg:unsupervised-tau}\;
    $\tau_r \gets \max \{ \tau_r \mid \EstimateNumPairs(0, \tau_r, \infty, \Psi^*) \geq k |A| \}$ \label{alg:unsupervised-tau-r}\;
    
    $\rho^* \gets \QualityToRank(q, \tau, \tau_r, k, \Psi^*)$\;
    $P \gets \TTRKJoin(\mathcal{A}, \mathcal{B}, \tau, \tau_r, k, \rho^*)$\;
    
    \KwRet{$P$}\;
\end{algorithm}

For both invocations of \ttrkjoin we need to determine the tokenizer, token weighting scheme, the set similarity measure, and join parameters $\tau$ and $\tau_r$.
Algorithm~\ref{alg:balanced-ttrkjoin} outlines the proposed method for doing so.

\subsubsection{Choosing Token Set Model and Similarity Measure}

To determine the token set model and set similarity measure we pick the combination we heuristically find to be the most discriminate (see line~\ref{alg:pick-discriminate-start}-\ref{alg:pick-discriminate-end}).
We define the measure for discriminatory power of a token set model and similarity measure $sim$ on dataset $(A, B)$ as
\begin{equation*}
    dp = 1 - \frac{1}{|\mathcal{A}|} \sum_{a \in \mathcal{A}} \Bigg[ \frac{1}{k_{dp}-1} \sum_{b \in \mathcal{B}_a} \bigg[ \frac{sim(a, b)}{\max_{b' \in \mathcal{B}} sim(a,b')} \bigg] - 1 \Bigg]
\end{equation*}
where $\mathcal{A}$ and $\mathcal{B}$ is $A$ and $B$ converted to token sets and $\mathcal{B}_a$ is the $k_{\mathrm{dp}}$ most similar token sets to $a$.
We can run \ttrkjoin with $\tau = \tau_r = 0$ and $k = k_{dp}$ to get the necessary pairs.
To avoid a costly evaluation we estimate the discriminatory power by only using a subset $\widetilde{A}$ of $A$ and with the desired approximation quality.
Table~\ref{tab:unsupervised-configurations} shows the six configurations that are evaluated.
We consider these to be robust options for most datasets.
Note that Dice similarity measure is left out because it will, per definition, have the same discriminatory power as Jaccard.

\begin{table}[htbp]
    \scriptsize
    \centering
    \begin{tabular}{ccccc}
        \toprule
        \multicolumn{3}{c}{Token Set Model} & & \\
        Tokenization & & Token Weight & & Set Similarity Measure \\
        \midrule
        \begin{tabular}{@{}c@{}}3-gram$^*$ \\ word\end{tabular} &
        $\times$ &
        \begin{tabular}{@{}c@{}}TF-IDF\end{tabular}&
        $\times$ &
        \begin{tabular}{@{}c@{}}Jaccard \\ Cosine \\ Overlap\end{tabular}
        \\
        \bottomrule
    \end{tabular}
    \caption{Configuration space for unsupervised ShallowBlocker. $^*$Only if the total number of characters in the dataset are below $100\,000\,000$.}
    \label{tab:unsupervised-configurations}
\end{table}

\subsubsection{Choosing Hybrid Join Conditions}
We argued in Section~\ref{sec:hybrid-join} that there are strong potential benefits of mixing different types of join conditions.
In an unsupervised setting we have limited signals to guide how we determine such conditions.
We argue that a reasonable approach is to balance the pruning power of the absolute similarity threshold, relative similarity threshold, and local cardinality threshold in a $(\tau, \tau_r, k)$-join so that each condition would in isolation produce the same number of pairs.
That way we dynamically adapt to data and we get to exploit the complementary strengths of the three join types.
The combined conditions will consistently filter out pairs of low absolute similarity, relative similarity, and similarity order.

Specifically,
we determine $\tau$, $\tau_r$, and $k$ that would hit the pair budget ($k |A|$) if used in isolation.
Obviously, this is per definition the case for the provided $k$.
For $\tau$ and $\tau_r$, we binary search for the highest value that produce at least $k |A|$ estimated number of pairs according to a sample of recorded trajectories as described in Section~\ref{sec:estimated-number-of-pairs} (see line~\ref{alg:unsupervised-tau} and \ref{alg:unsupervised-tau-r}).
Note that the $k$ parameter of the \ttrkjoin will still guarantee that we stay within the pair budget no matter what $\tau$ and $\tau_r$ is set to.

\subsection{Supervised}

In a supervised setting we assume a subset of known matches $\widetilde{M} \subseteq M$.
This enables us to estimate the recall of a solution.
Furthermore, using the proposed techniques in Section~\ref{sec:estimation-framework} we can quickly estimate the recall, an upper bound of $|P|$, and an upper bound of the runtime for any run of \ttrkjoin.
Our proposed method exploits this to tune the parameters of \ttrkjoin to optimize a user-defined objective function $f$ based on recall, $|P|$, and runtime.
The only requirement of $f$ is that its codomain is totally ordered.
This is powerful because it lets users define their desired trade-off of blocking behaviour expressed explicitly in terms of these three main ways in which we judge blocking methods.

Algorithm~\ref{alg:supervised-shallowblocker} describes our proposed supervised version of ShallowBlocker --- which we will denote AutoShallowBlocker when we want to explicitly distinguish it from the unsupervised version.
The goal is to do two invocations of \ttrkjoin, one in each direction.
We first acquire the recall conditions and some randomly sampled search trajectories for each configuration of token set model and set similarity measures (see Table~\ref{tab:supervised-configurations}) in both directions (see line~\ref{alg:supervised-analyze-start}-\ref{alg:supervised-analyze-end}).
Then we jointly optimize the parameters for both joins (line~\ref{alg:supervised-optimize}) before we execute them (line~\ref{alg:supervised-ttrkjoin-start}-\ref{alg:supervised-ttrkjoin-end}).
Note that since $\rho^*$ are among the parameters we optimize we also let ShallowBlocker choose the degree of approximation.
All details except the optimization routine have already been covered.

\begin{table}[htbp]
    \scriptsize
    \centering
    \begin{tabular}{ccc}
        \toprule
        Token Set Model & & Set Similarity Measure \\
        \midrule
        \begin{tabular}{@{}l@{}}Binary weighted 3-grams$^*$ \\ TF-IDF weighted 3-grams$^*$ \\ Binary weighted words \\ TF-IDF weighted words \end{tabular} &
        $\times$ &
        \begin{tabular}{@{}l@{}}Jaccard \\ Dice \\ Cosine \\ Overlap\end{tabular}
        \\
        \bottomrule
    \end{tabular}
    \caption{Configuration space for supervised ShallowBlocker. $^*$Only if the total number of characters in the dataset are below $100\,000\,000$.}
    \label{tab:supervised-configurations}
\end{table}

\begin{algorithm}[tbh]
    \caption{ShallowBlocker$(A, B, \widetilde{M}, f)$, Supervised}
    \label{alg:supervised-shallowblocker}
    \footnotesize
    \SetKwInput{KwConstants}{Constants}
    \SetKw{Continue}{\textbf{continue}}
    \SetKw{Break}{\textbf{break}}
    \SetKwFunction{TTRKJoin}{TTRKJoin}
    \SetKwFunction{BuildTTRKIndex}{BuildTTRKIndex}
    \SetKwFunction{tsm}{ToTokenSets}
    \SetKwFunction{RecordTrajectories}{RecordTrajectories}
    \SetKwFunction{FindRecallCond}{FindRecallCond}
    \SetKwFunction{OptimizeTTRKJoins}{OptimizeTTRKJoins}
    \KwIn{%
        $A$ and $B$ are collections of strings.
        $\widetilde{M} \subseteq M$ are known matches and $f$ is an objective function that accepts recall, $|P|$, and runtime and returns an a value from a totally ordered set.
    }
    \KwOut{Pairs $P \subseteq A \times B$ produced such that $f(\text{recall}, |P|, \text{runtime})$ is high by best effort.}
    
    \SetKwFunction{OneWayUnsupSB}{OneWayUnsupSB}
    \SetKwFunction{flipPairs}{flipPairs}
    
    \KwConstants{$N=500; q_p = 0.95;$ $D = \{0, 0.01, 0.05, 0.1, 0.2, 0.3, 0.4, 0.5\};$}

    Precompute TF-IDF weights and $\mathcal{O}$ for $A$ and $B$ using all tokenizers\;
    $\Psi_{ab}, \Psi_{ba} \gets$ empty maps from $(\texttt{tms}, sim)$ to sampled TTRK trajectories \label{alg:supervised-analyze-start}\;
    $\Theta_{ab}, \Theta_{ba} \gets$ empty maps from $(\texttt{tms}, sim)$ to recall conditions\;
    \ForEach{token set model \texttt{tsm}}{%
        $\mathcal{A} \gets \tsm(A, \texttt{tms})$\;
        $\mathcal{B} \gets \tsm(B, \texttt{tms})$\;
        $\widetilde{\mathcal{A}} \gets$ sample $N$ sets from $\mathcal{A}$\;
        $\widetilde{\mathcal{B}} \gets$ sample $N$ sets from $\mathcal{B}$\;
        \ForEach{similarity measure $sim$}{%
            $I^\pm \gets \BuildTTRKIndex(\mathcal{B}, 0; sim)$\;
            $\Psi_{ab}[\texttt{tsm}, sim] \gets \RecordTrajectories(I^\pm, \widetilde{\mathcal{A}}; sim)$\;
            $\Theta_{ab}[\texttt{tsm}, sim] \gets \FindRecallCond(\mathcal{A}, I^\pm, \widetilde{M}, D; sim)$\;
            $I^\pm \gets \BuildTTRKIndex(\mathcal{A}, 0; \texttt{sim})$\;
            $\Psi_{ba}[\texttt{tsm}, sim] \gets \RecordTrajectories(I^\pm, \widetilde{\mathcal{B}}; sim)$\;
            $\Theta_{ba}[\texttt{tsm}, sim] \gets \FindRecallCond(\mathcal{B}, I^\pm, \widetilde{M}, D; sim)$\;
        }
    }\label{alg:supervised-analyze-end}

    $\texttt{tsm}_{ab}, sim_{ab}, \tau_{ab}, \tau_{r.ab}, k_{ab}, \rho^*_{ab}$, \\
    $\texttt{tsm}_{ba}, sim_{ba}, \tau_{ba}, \tau_{r.ba}, k_{ba}, \rho^*_{ba} \gets \OptimizeTTRKJoins(f, \Psi_{ab}, \Psi_{ba}, \Theta_{ab}, \Theta_{ba})$ \label{alg:supervised-optimize}\;
    $\mathcal{A}_{ab} \gets \tsm(A, \texttt{tsm}_{ab})$ \label{alg:supervised-ttrkjoin-start}\;
    $\mathcal{B}_{ab} \gets \tsm(B, \texttt{tsm}_{ab})$\;
    $P_{ab} \gets \TTRKJoin(\mathcal{A}_{ab}, \mathcal{B}_{ab}, \tau_{ab}, \tau_{r.ab}, k_{ab}, \rho^*_{ab}; sim_{ab})$\;
    $\mathcal{A}_{ba} \gets \tsm(A, \texttt{tsm}_{ba})$\;
    $\mathcal{B}_{ba} \gets \tsm(B, \texttt{tsm}_{ba})$\;
    $P_{ba} \gets \TTRKJoin(\mathcal{B}_{ba}, \mathcal{A}_{ba}, \tau_{ba}, \tau_{r.ba}, k_{ba}, \rho^*_{ba}; sim_{ba})$\;
    $P \gets P_{ab} \cup \flipPairs(P_{ba})$ \label{alg:supervised-ttrkjoin-end}\;
    \KwRet{$P$}\;
\end{algorithm}

\subsubsection{Optimization}

The optimization problem has four categorical and eight continuous variables in total for the two joins.
We evaluate a solution by estimating recall, $|P|$, and runtime in both directions with the corresponding recall conditions and search trajectories, combining them, and feeding them to the objective function $f$.
Since the objective function $f$ is user-defined and only required to produce values from a totally ordered set it is a black-box optimization problem.
The number of variables is manageable and evaluation is quick.
Furthermore, it is reasonable to assume, given the nature of the problem, that the objective function is not overly sensitive.
Our experience is therefore that most random and local search heuristics are effective enough in finding good solutions.

We opt for a relatively simple approach.
Draw ten random solutions, pick the best one, and then hill climb.
Repeat 32 times (can be done in parallel) and pick the best solution.
In hill climbing, try all exponentially spaced increments/decrements of the numeric parameters.
It is important to note that $k_{ab}$ or $k_{ba}$ can be zero,
so there might not be a \ttrkjoin in both directions.

There is a considerable chance of overfitting on recall if the number of known matches $|\widetilde{M}|$ is not large enough.
Therefore,
we acquire the recall conditions with varying regularization parameter values $D$ as described in Section~\ref{sec:regularization}.
The regularization parameter $d \in D$ we end up using is picked through 3-fold cross validation of all values in $D$.
Note that $|P|$ and runtime is not as prone to overfitting because we are in control of the number of samples those estimates are based on and can simply choose a large enough number of samples.
While for recall we can not control the number of known matches.

\subsection{Deduplication}

So far we have only discussed blocking across two sets of records.
A special case of this is deduplication,
where we only have one set of records $A$ and want to find all duplicates within $A$.
To avoid unnecessary clutter we have not incorporated this within the presentation of our proposed techniques and method.
However, they are all trivially adapted to deduplication.
The main differences is that \ttrkjoin must take care to not return self-pairs and that ShallowBlocker should only do \ttrkjoin in one direction (from $\mathcal{A}$ to $\mathcal{A}$).

\subsection{Constants}
There are several constant parameters used throughout ShallowBlocker.
This is not contradictory to the method being hands-off.
For all practical purposes these are part of the method.
They do not need tuning, are not user-provided, and are in essence no different from the network architecture and training details of a deep learning method or similar details from other methods.
None of the constants are, by their nature, sensitive as long as they are large enough (e.g., sample sizes).
So we have simply picked round numbers of reasonable sizes that we observe provide stable results.

\section{Experimental Setup}\label{sec:experimental-setup}

\subsection{Datasets}\label{sec:datasets}
\begin{table*}[htbp]
    \scriptsize
    \centering
    \begin{tabular}{lrrrrrrrrrrrr}
        \toprule
        & \multicolumn{2}{c}{\# Records} & \multicolumn{2}{c}{\# Matches} & \multicolumn{4}{c}{\# Characters} & \multicolumn{4}{c}{\# Words} \\
        \cmidrule(lr){2-3}
        \cmidrule(lr){4-5}
        \cmidrule(lr){6-9}
        \cmidrule(lr){10-13}
        Dataset & \multicolumn{1}{c}{$|A|$} & \multicolumn{1}{c}{$|B|$} & Train & Test & Min & Median & Mean & Max & Min & Median & Mean & Max \\
        \midrule
            Amazon-Google         & \num{1363}    & \num{3226}    & \num{699}    & \num{234}    & \num{10} & \num{58} & \num{60} & \num{222} & \num{2} & \num{9} & \num{9.6} & \num{33} \\
            Beer                  & \num{4345}    & \num{3000}    & \num{40}     & \num{100}$^+$    & \num{25} & \num{68} & \num{69} & \num{153} & \num{5} & \num{11} & \num{11.3} & \num{29} \\
            DBLP-ACM$^*$          & \num{2616}    & \num{2294}    & \num{1332}   & \num{444}    & \num{17} & \num{129} & \num{132} & \num{421} & \num{4} & \num{19} & \num{19.3} & \num{66} \\
            DBLP-GoogleScholar$^*$& \num{2616}    & \num{64263}   & \num{3207}   & \num{1070}   & \num{7} & \num{110} & \num{111} & \num{344} & \num{2} & \num{17} & \num{16.8} & \num{58} \\
            iTunes-Amazon$^*$     & \num{6907}    & \num{55923}   & \num{78}     & \num{100}$^+$    & \num{54} & \num{141} & \num{152} & \num{606} & \num{10} & \num{24} & \num{25.8} & \num{88} \\
            Walmart-Amazon$^*$    & \num{2554}    & \num{22074}   & \num{576}    & \num{193}    & \num{19} & \num{97} & \num{102} & \num{913} & \num{3} & \num{16} & \num{17.1} & \num{172} \\
            Abt-Buy               & \num{1081}    & \num{1092}    & \num{616}    & \num{206}    & \num{13} & \num{173} & \num{193} & \num{555} & \num{2} & \num{28} & \num{30.7} & \num{88} \\
            Company               & \num{28200}   & \num{28200}   & \num{10000}  & \num{5640}   & \num{0} & \num{4311} & \num{5177} & \num{29035} & \num{0} & \num{640} & \num{792.4} & \num{2007} \\
            Songs                 & \num{1000000} &               & \num{10000}  & \num{136011} & \num{31} & \num{86} & \num{90} & \num{440} & \num{6} & \num{12} & \num{13.0} & \num{72} \\
            Citeseer-DBLP         & \num{1823978} & \num{2512927} & \num{10000}  & \num{548787} & \num{1} & \num{122} & \num{126} & \num{2541} & \num{0} & \num{17} & \num{18.2} & \num{312} \\
         \bottomrule
    \end{tabular}
    \caption{
        The different datasets used in our experiments.
        Note that some datasets have been slightly adapted to fit our experimental setup.
        See Section~\ref{sec:datasets}.
        $^*$These datasets has a "dirty" version~\cite{mudgalDeepLearningEntity2018}.
        $^+$We have labelled additional matches to get 100 test matches.
    }
    \label{tab:datasets}
\end{table*}

We use the DeepMatcher~\cite{mudgalDeepLearningEntity2018} and Falcon~\cite{dasFalconScalingHandsOff2017} datasets from the Magellan Data Repository~\cite{dasMagellanDataRepository2022}.
Table~\ref{tab:datasets} shows key characteristics for the different datasets.
We ignore the known non-matches for all datasets since they are not used by any method in our experiments or relevant for measuring recall.

We use the provided train/test split for the DeepMatcher datasets.
The validation sets are not used in the experiments.
All supervised methods only get the training matches and must do validation within those.
We omit the Fodors-Zagats dataset because it is tiny, trivial for all tested methods, and do not have enough matches (even if we label more) to reliably measure recall.
For Beer and iTunes-Amazon we manually label additional matches to get 100 test matches.

The Songs and Citeseer-DBLP datasets from Falcon~\cite{dasFalconScalingHandsOff2017} do not have any train/test split.
We therefore split of \num{10000} matches to training and the rest to testing.
Songs is a deduplication dataset --- i.e., one set of records that is matched to itself.
Importantly, we remove self-matches from the dataset since they are trivial.

\subsection{Baseline Methods}

\begin{table*}[htbp]
    \scriptsize
    \centering
    \begin{tabular}{ll>{\centering\arraybackslash}p{2cm}>{\centering\arraybackslash}p{2cm}l}
        \toprule
        & & \multicolumn{2}{c}{Default hyperparameter values} & \vspace{0.5em} \\
        & & \multicolumn{2}{c}{Recall target} & \\
        \cmidrule{3-4}
        Method & Hyperparameters & 0.90 & 0.98 & Hyperparameter space  \\
        \midrule
        Token Blocking & - & - & - & - \\
        \midrule
        HVTB & - & - & - & - \\
        \midrule
        \multirow{4}{*}{Minhash LSH}
        & $\tau$ & 0.08 & 0.04 & $[0, 1]$ \\
        & Tokenization & \multicolumn{2}{c}{Word} & 3-gram, word \\
        & $b$ & \multicolumn{2}{c}{Optimize eq. false} & $[1, \dots, \text{numPerm}]$ \\
        & $r$ & \multicolumn{2}{c}{positive and negative} & $[1, \dots, \lfloor \text{numPerm} / b \rfloor]$ \\
        \midrule
        \multirow{3}{*}{PPJoin}
        & $\tau$ & 0.20 & 0.14 & $[0,\infty)$ \\
        & Similarity measure & \multicolumn{2}{c}{Cosine} & Jaccard, Cosine, Dice, Overlap \\
        & Tokenization & \multicolumn{2}{c}{Word} & 3-gram, word \\
        \midrule
        \multirow{4}{*}{$\tau$-join}
        & $\tau$ & 0.22 & 0.16 & $[0,\infty)$ \\
        & Similarity measure & \multicolumn{2}{c}{Cosine} & Jaccard, Cosine, Dice, Overlap \\
        & Tokenization & \multicolumn{2}{c}{Word} & 3-gram, word \\
        & Token weighting & \multicolumn{2}{c}{TF-IDF} & Binary, TF-IDF \\
        \midrule
        DeepBlocker & $k$ & 240 & Not possible & $[1, \dots, \max(|A|, |B|)]$ \\
        \midrule
        AutoBlock & $k$ & - & - & $[1, \dots, \max(|A|, |B|)]$ \\
        \midrule
        \multirow{4}{*}{$k$-join}
        & $k$ & 10 & 32 & $[1, \dots, \max(|A|, |B|)]$ \\
        & Similarity measure & \multicolumn{2}{c}{Cosine} & Jaccard, Cosine, Dice, Overlap \\
        & Tokenization & \multicolumn{2}{c}{Word} & 3-gram, word \\
        & Token weighting & \multicolumn{2}{c}{TF-IDF} & Binary, TF-IDF \\
         \bottomrule
    \end{tabular}
    \caption{
        Baseline methods and their hyperparameters.
        Also included is the default hyperparameter values when used in an unsupervised setting and the hyperparameter space.
    }
    \label{tab:baselines}
\end{table*}

We perform our experiments on a variety of baseline methods covering different method categories.
See them listed in Table~\ref{tab:baselines}.
First, we briefly introduce the different methods.
Then we discuss how we determine hyperparameters in an unsupervised and supervised context.
For all methods were it is relevant the largest string collection is indexed and the smallest queried (i.e., we ensure $|A| \leq |B|$).

\subsubsection{Token Blocking}
It is one of the most ubiquitous blocking methods and is considered a naive approach.
Simply produce all pairs with at least one overlapping token.
We use word tokenization and run it using \ttrkjoin with $\tau=1, \tau_r=0, k=\infty,\rho^*=\infty$ and Overlap similarity.

\subsubsection{HVTB~\cite{ohareHighValueTokenBlockingEfficient2021}}
A significant improvement over Token Blocking for most cases,
and is reported to achieve, or be comparable to, state-of-the-art results.
Instead of producing all pairs with at least one overlapping token,
HVTB only considers non-unique tokens with above average TF-IDF score and thereby pruning away pairs that only have less impactful tokens in common.
Furthermore, pairs with below average number of overlapping tokens are pruned.
The original source code is not available so we implemented according the the authors specification.
For a fair comparison, we also include our own optimized version: HVTB$^+$.
The three main improvements are using integer representation of tokens, parallelization, and eagerly pruning pairs with one common token upon construction if overlaps greater than one have already been observed.
It produces the same pairs but with greatly reduced runtime and memory consumption.

\subsubsection{$\tau$-join}
Classical absolute threshold-based similarity join achieved by running \ttrkjoin with $\tau_r = 0, k = \infty, \rho^* = \infty$ fixed.

\subsubsection{$k$-join}
Classical local cardinality threshold-based similarity join achieved by running \ttrkjoin with $\tau = 0, \tau_r = 0, \rho^* = \infty$ fixed.

\subsubsection{PPJoin~\cite{xiaoEfficientSimilarityJoins2008,xiaoEfficientSimilarityJoins2011}}
Popular state-of-the-art $\tau$-join blocking method~\cite{papadakisBlockingFilteringTechniques2021}.
We use the highly optimized implementation from \citet{mannEmpiricalEvaluationSet2016}.
However, note that this implementation uses, and is optimized for, unweighted sets --- as is common for PPJoin.
We include this in our experiments because it is a widely used method and point of comparison.

\subsubsection{Minhash LSH}
Well-known technique using locality-sensitive hashing~\cite{gionisSimilaritySearchHigh1999} for approximate similarity joins~\cite{broderResemblanceContainmentDocuments1997,broderMinWiseIndependentPermutations2000}.
We use the implementation from the datasketch\footnote{\url{https://github.com/ekzhu/datasketch}} package.
This is what the authors of AutoBlock~\cite{zhangAutoBlockHandsoffBlocking2020} used in their experiments.
For fair comparison we implement a parallelization layer on top to exploit all available CPU cores.
We use the recommended default setting of 128 permutations.

\subsubsection{DeepBlocker\cite{thirumuruganathanDeepLearningBlocking2021}}
State-of-the-art unsupervised deep learning-based blocker.
It avoids the need for labeled training data by using either a encoder-decoder setup which is trained for reconstruction or by generating synthetic labeled data.
We use the authors provided implementation\footnote{\url{https://github.com/qcri/DeepBlocker}}.
However, note that we implemented a GPU accelerated top-k Cosine similarity search using FAISS~\cite{johnsonBillionScaleSimilaritySearch2021} as the described by the authors since the provided source code only includes a brute force CPU version.
Based on the reported results and recommendation of the authors we use the Hybrid model for the "textual" datasets (Walmart-Amazon, Abt-Buy, and Company) and AutoEncoder for the rest.

\subsubsection{AutoBlock\cite{zhangAutoBlockHandsoffBlocking2020}}
Supervised deep learning-based blocker that embeds records and uses locality-sensitive hashing to find similar embeddings.
Known labeled matches are used to train the deep network,
but one must still specify the local cardinality threshold $k$.
The original source code is not available so we implement it according to the authors description using PyTorch and FALCONN.
We also parallalize the LSH nearest neighbor search for a fair comparison.

\subsection{Unsupervised Baselines}\label{sec:unsupervised-baselines}
All baseline methods except AutoBlock are unsupervised.
However, most of them still have one or more hyperparameters that must be set manually.
In an unsupervised setting we are not able to tune them per dataset and must rely on defaults.
To give all methods an equal footing we chose the hyperparameters for each method that yields the best overall results across all datasets on the test sets.
We find Cosine similarity measure, word tokenization, and TF-IDF token weighting to be optimal for methods that have any of those hyperparameters.
For Minhash LSH we set $b$ and $r$ automatically to the configuration that balances false positives and negatives equally.

Most baseline methods have a similarity threshold $\tau$ or local cardinality threshold $k$ as the main hyperparameter balancing recall against number of returned pairs.
For the experiments that target a specific recall level,
we set this parameter to the most aggressive (highest $\tau$, lowest $k$) that achieves the recall target on all datasets\footnote{Except the Company dataset since it contains so many wrongly labelled matches that it is unrealistic to reach high recall.} on the test set.
For $\tau$ we only consider increments of 0.01.
Table~\ref{tab:baselines} shows the hyperparameters for recall targets of 0.90 and 0.98.
It is important to note that this way of determining hyperparameters yields best case results for the baseline methods by exploiting knowledge about the test set and not incurring any runtime cost for picking hyperparameter values.
Therefore, it is not strictly comparable to results from supervised methods.

\subsection{Supervised Baselines}\label{sec:supervised-baselines}
The Supervised version of ShallowBlocker (denoted AutoShallowBlocker) uses known matches $\widetilde{M}$ and an objective function $f$ to set all parameters.
The only supervised baseline method, AutoBlock, has a hyperparameter that must be set: $k$.
Furthermore, most of the other unsupervised methods can be tuned.
So in order to compare against the baseline methods fairly we extend them with a supervised approach similar to ShallowBlocker for determining their hyperparameters from the training set.
We will denote these modified supervised baselines with a Sup- prefix (for supervised) --- e.g., Sup-DeepBlocker.

Assume we are given training matches $\widetilde{M}$ and a recall target $R$.
We can determine the hyperparameters by enumerating all combinations of categorical parameters, picking the threshold ($\tau$ or $k$) for each combination that achieves recall $R$ on $\widetilde{M}$, and then picking the combination that will return the fewest pairs.
In order to avoid overfitting, and thereby missing the recall target, we use regularization.
$\widetilde{M}$ is split into 3-fold cross-validation splits.
We determine the hyperparameters using the train split of each fold with some threshold regularization margin $d$.
The regularization margin $d$ is increased until the average recall of validation splits is at least $R$ with 95\% confidence (not adjusting for multiple testing).
The final hyperparameters are picked using the entire $\widetilde{M}$ and the chosen $d$.

For the methods with a similarity threshold $\tau$ we can simply calculate the similarity for all train matches and pick the $R$ quantile of the similarities to determine the highest threshold that would recall $R$ of the matches $\widetilde{M}$.
We estimate the number of returned pairs by running the method on a sample of \num{1000} strings from $A$ and the entire $B$ and multiplying the number of pairs with $|A| / 1000$.
The margin $d$ adjusts the threshold to $\tau \cdot (1 - d)$ and uses the same values as ShallowBlocker,
For Minhash LSH we pick the $b$ and $r$ that minimizes probability of false positives while also reaching the recall target.

For the methods with a local cardinality threshold $k$ we can determine the smallest threshold that would recall $R$ of the training matches by doubling $k$ and running the method on the subset of $A$ that are in the training matches and the entire $B$ until at least $R$ of the matches are found.
Then using the result to find the minimum between $k/2$ and $k$ that still recall at least $R$ of the matches.
The configuration that return the fewest pairs are simply the one with the lowest $k$.
The margin $d$ adjusts the threshold to $k + d$ and is increased in powers of two.

Note that we make sure to avoid redundant work where possible for the different methods in this supervised approach.
For example, we only train the deep learning model and compute the embeddings for each record once for DeepBlocker --- not for every fold and value of $k$.
Additionally,
note that we do not do this for PPJoin since Sup-$\tau$-join would be a superset of Sup-PPJoin.

\subsection{Hardware}
All CPU-only experiments are run on an AMD Ryzen 5950X system with 64GB RAM.
Because of practical resource constraints all experiments with deep learning models using GPU are run on a different machine with a Nvidia V100 GPU and Intel Xeon Gold 6148 CPU with 128GB.
While the systems are not equivalent,
and therefore not directly comparable,
there will always be a divide between the CPU-only and GPU accelerated methods since the latter exploits a fundamentally different compute resource.
The GPU machine is the more capable and expensive,
so at least the deep learning methods can not be said to be hampered by the lack of compute and memory.

\section{Experiments}\label{sec:experiments}

\subsection{Recall Target}\label{sec:experiment-recall-target}

\begin{table*}[htbp]
    \setlength{\tabcolsep}{2pt}
    \tiny
    \centering
    \vspace{-2em}

\begin{tabular}{llcccccccccccccc}
    \toprule
    & & \multicolumn{14}{c}{Dataset} \\
    \cmidrule{3-16} \\
    Resource Setting & Method & AG & B & DA & DG & IA & WA & D-DA & D-DG & D-IA & D-WA & AB & C & S & CD \\
    \midrule
    \multicolumn{16}{c}{Recall} \\
    \midrule
    \multirow{11}{1.9cm}{Unsupervised}
    & Token Blocking & 1.000 & 1.000 & 1.000 & 0.999 & 1.000 & 1.000 & 1.000 & 0.999 & 1.000 & 1.000 & 0.995 & 0.964 & {\color{gray}OOM} & {\color{gray}OOM} \\
    & HVTB & {\color{gray}0.868} & {\color{gray}0.210} & 1.000 & 0.993 & {\color{gray}0.860} & 0.990 & 1.000 & 0.993 & {\color{gray}0.200} & 0.990 & {\color{gray}0.893} & 0.923 & {\color{gray}0.576} & {\color{gray}OOM} \\
    & HVTB+ & {\color{gray}0.868} & {\color{gray}0.210} & 1.000 & 0.993 & {\color{gray}0.860} & 0.990 & 1.000 & 0.993 & {\color{gray}0.200} & 0.990 & {\color{gray}0.893} & 0.923 & {\color{gray}0.576} & 0.998 \\
    & MinHash LSH & 1.000 & 1.000 & 1.000 & 0.999 & 1.000 & 0.995 & 1.000 & 0.999 & 1.000 & 0.995 & 0.937 & {\color{gray}0.699} & {\color{gray}OOM} & {\color{gray}OOM} \\
    & PPJoin & 0.991 & 1.000 & 1.000 & 0.996 & 1.000 & 1.000 & 1.000 & 0.996 & 1.000 & 1.000 & 0.917 & {\color{gray}OOM} & {\color{gray}OOM} & {\color{gray}OOM} \\
    & $\tau$-join & 0.983 & 1.000 & 1.000 & 0.993 & 1.000 & 1.000 & 1.000 & 0.993 & 1.000 & 1.000 & 0.903 & {\color{gray}0.455} & 0.998 & 1.000 \\
    & DeepBlocker & 0.996 & 1.000 & 0.998 & 0.979 & 0.988 & 0.991 & 0.998 & 0.983 & 0.994 & 0.992 & 0.995 & {\color{gray}0.714} & 0.901 & {\color{gray}TIME} \\
    & $k$-join & 0.974 & 1.000 & 1.000 & 0.968 & 1.000 & 0.984 & 1.000 & 0.968 & 1.000 & 0.984 & 0.966 & {\color{gray}0.863} & 0.906 & 0.996 \\
    & ShallowBlocker ($q\!=\!0.95$) & 0.980 & 1.000 & 1.000 & 0.975 & 0.956 & 0.968 & 1.000 & 0.976 & 0.956 & 0.968 & 0.995 & {\color{gray}0.767} & 0.904 & 0.990 \\
    & ShallowBlocker ($q\!=\!0.99$) & 0.981 & 1.000 & 1.000 & 0.974 & 0.970 & 0.964 & 1.000 & 0.976 & 0.970 & 0.964 & 0.995 & {\color{gray}0.666} & 0.906 & 0.989 \\
    & ShallowBlocker ($q\!=\!1$) & 0.977 & 1.000 & 1.000 & 0.970 & 0.964 & 0.962 & 1.000 & 0.972 & 0.964 & 0.962 & 0.995 & {\color{gray}0.809} & 0.900 & 0.988 \\
    \cmidrule{1-16}
    \multirow{6}{1.9cm}{Supervised \\ $\leq$ 100 matches}
    & Sup-MinHash LSH & 0.978 & 0.996 & 0.942 & 0.956 & 0.960 & 0.973 & 0.924 & 0.948 & 0.960 & 0.973 & 0.923 & 0.949 & {\color{gray}OOM} & {\color{gray}OOM} \\
    & Sup-$\tau$-join & 0.974 & 1.000 & 0.935 & 0.953 & 0.984 & 0.981 & 0.935 & 0.941 & 0.984 & 0.981 & 0.939 & 0.940 & 0.980 & 0.979 \\
    & Sup-AutoBlock & {\color{gray}0.754} & {\color{gray}0.706} & {\color{gray}0.712} & {\color{gray}0.853} & 1.000 & 0.940 & {\color{gray}0.655} & {\color{gray}0.807} & 0.926 & {\color{gray}0.742} & 0.900 & {\color{gray}0.243} & {\color{gray}OOM} & {\color{gray}OOM} \\
    & Sup-DeepBlocker & 0.956 & 1.000 & 0.984 & 0.954 & 0.990 & 0.981 & 0.984 & 0.963 & 0.986 & 0.983 & 0.976 & {\color{gray}ERR} & {\color{gray}ERR} & 0.936$^*$ \\
    & Sup-$k$-join & 0.989 & 1.000 & 0.993 & 0.976 & 0.996 & 0.980 & 0.991 & 0.980 & 0.994 & 0.967 & 0.933 & 0.962 & 0.985 & 0.977 \\
    & ShallowBlocker & 0.992 & 1.000 & 0.940 & 0.964 & 0.986 & 0.985 & 0.920 & 0.950 & 0.988 & 0.988 & 0.938 & 0.924 & 0.978 & 0.967 \\
    \cmidrule{1-16}
    \multirow{6}{1.9cm}{Supervised \\ $\leq$ 1000 matches}
    & Sup-MinHash LSH & 0.938 & - & 0.928 & 0.940 & - & 0.921 & 0.928 & 0.947 & - & 0.921 & 0.932 & 0.942 & {\color{gray}OOM} & {\color{gray}OOM} \\
    & Sup-$\tau$-join & 0.965 & - & 0.936 & 0.933 & - & 0.928 & 0.936 & 0.932 & - & 0.928 & 0.935 & 0.926 & 0.930 & 0.944 \\
    & Sup-AutoBlock & {\color{gray}0.891} & - & {\color{gray}0.825} & {\color{gray}0.899} & - & 0.906 & {\color{gray}0.882} & {\color{gray}0.862} & - & {\color{gray}0.878} & 0.907 & {\color{gray}0.506} & 0.975$^*$ & {\color{gray}OOM} \\
    & Sup-DeepBlocker & 0.932 & - & 0.985 & 0.930 & - & 0.931 & 0.983 & 0.930 & - & 0.933 & 0.938 & {\color{gray}ERR} & {\color{gray}ERR} & 0.932$^*$ \\
    & Sup-$k$-join & 0.964 & - & 0.993 & 0.936 & - & 0.914 & 0.991 & 0.930 & - & 0.921 & 0.965 & 0.930 & 0.964 & 0.979 \\
    & ShallowBlocker & 0.982 & - & 0.955 & 0.933 & - & 0.940 & 0.937 & 0.943 & - & 0.947 & 0.913 & 0.926 & 0.930 & 0.948 \\
    \cmidrule{1-16}
    \multirow{6}{1.9cm}{Supervised \\ $\leq$ \num{10000} matches}
    & Sup-MinHash LSH & - & - & - & - & - & - & - & - & - & - & - & 0.935 & {\color{gray}OOM} & {\color{gray}OOM} \\
    & Sup-$\tau$-join & - & - & - & - & - & - & - & - & - & - & - & 0.918 & 0.918 & 0.928 \\
    & Sup-AutoBlock & - & - & - & - & - & - & - & - & - & - & - & {\color{gray}0.577} & 0.959$^*$ & {\color{gray}OOM} \\
    & Sup-DeepBlocker & - & - & - & - & - & - & - & - & - & - & - & {\color{gray}ERR} & 0.931$^*$ & 0.925 \\
    & Sup-$k$-join & - & - & - & - & - & - & - & - & - & - & - & 0.915 & 0.956 & 0.979 \\
    & ShallowBlocker & - & - & - & - & - & - & - & - & - & - & - & 0.920 & 0.926 & 0.924 \\
    \midrule
    \multicolumn{16}{c}{$\widetilde{k}$} \\
    \midrule
    \multirow{11}{1.9cm}{Unsupervised}
    & Token Blocking & 513 & 3685 & 1848 & 34879 & 51042 & 3987 & 1848 & 34526 & 51042 & 3987 & 426 & 27078 & {\color{gray}OOM} & {\color{gray}OOM} \\
    & HVTB & {\color{gray}5.39} & {\color{gray}27.7} & 13.2 & 41.0 & {\color{gray}1384} & 41.2 & 13.2 & 41.8 & {\color{gray}278} & 41.2 & {\color{gray}9.00} & \textbf{5887} & {\color{gray}25.6} & {\color{gray}OOM} \\
    & HVTB+ & {\color{gray}5.39} & {\color{gray}27.7} & 13.2 & 41.0 & {\color{gray}1384} & 41.2 & 13.2 & 41.8 & {\color{gray}278} & 41.2 & {\color{gray}9.00} & \textbf{5887} & {\color{gray}25.6} & 150 \\
    & MinHash LSH & 418 & 3078 & 997 & 18528 & 31742 & 2332 & 997 & 18292 & 31742 & 2332 & 178 & {\color{gray}13438} & {\color{gray}OOM} & {\color{gray}OOM} \\
    & PPJoin & 45.8 & 2168 & 80.8 & 958 & 2431 & 107 & 80.8 & 912 & 2431 & 107 & 12.1 & {\color{gray}OOM} & {\color{gray}OOM} & {\color{gray}OOM} \\
    & $\tau$-join & 11.0 & 5.85 & 4.99 & 16.2 & 165 & 26.6 & 4.99 & 15.9 & 165 & 26.6 & 5.49 & {\color{gray}0.626} & 25.9 & 50.3 \\
    & DeepBlocker & 240 & 240 & 240 & 240 & 240 & 240 & 240 & 240 & 240 & 240 & 240 & {\color{gray}240} & 184 & {\color{gray}TIME} \\
    & $k$-join & 9.99 & 9.99 & 10.00 & 10.00 & 10.00 & 10.00 & 10.00 & 10.00 & 10.00 & 10.00 & 10.00 & {\color{gray}10.00} & 6.61 & 10.00 \\
    & ShallowBlocker ($q\!=\!0.95$) & 4.92 & 3.80 & 4.87 & 6.23 & \underline{\textbf{2.49}} & \textbf{4.21} & 4.87 & 6.48 & \underline{\textbf{2.49}} & \textbf{4.21} & 6.12 & {\color{gray}2.80} & 6.92 & 2.32 \\
    & ShallowBlocker ($q\!=\!0.99$) & 4.07 & 3.75 & \textbf{4.22} & 5.30 & 2.95 & 4.98 & \textbf{4.22} & 5.43 & 2.95 & 4.98 & 5.09 & {\color{gray}1.90} & 5.71 & \textbf{2.02} \\
    & ShallowBlocker ($q\!=\!1$) & \textbf{3.54} & \underline{\textbf{3.46}} & \textbf{4.12} & \textbf{4.90} & 2.73 & 4.72 & \textbf{4.12} & \textbf{5.12} & 2.73 & 4.72 & \textbf{4.59} & {\color{gray}2.57} & \underline{\textbf{5.35}} & \textbf{1.92} \\
    \cmidrule{1-16}
    \multirow{6}{1.9cm}{Supervised \\ $\leq$ 100 matches}
    & Sup-MinHash LSH & 72.8 & 1535 & 1.48 & 286 & 3053 & 570 & 1.85 & 440 & 3053 & 570 & 165 & 25952 & {\color{gray}OOM} & {\color{gray}OOM} \\
    & Sup-$\tau$-join & 6.85 & 21.6 & 1.01 & 3.81 & 15.4 & 10.3 & 1.01 & 3.59 & 15.4 & 10.3 & 7.29 & 10337 & 17.8 & 0.927 \\
    & Sup-AutoBlock & {\color{gray}595} & {\color{gray}2.90} & {\color{gray}6.61} & {\color{gray}4763} & 39168 & 1660 & {\color{gray}10.7} & {\color{gray}27626} & 24459 & {\color{gray}4555} & 527 & {\color{gray}444} & {\color{gray}OOM} & {\color{gray}OOM} \\
    & Sup-DeepBlocker & 61.6 & 106 & 1.000 & 252 & 202 & 120 & 1.000 & 134 & 240 & 133 & 126 & {\color{gray}ERR} & {\color{gray}ERR} & 7.00$^*$ \\
    & Sup-$k$-join & 10.6 & 56.9 & 1.000 & 11.4 & 57.0 & \textbf{7.40} & 1.000 & 12.2 & 58.6 & \textbf{4.40} & \textbf{1.80} & 25580 & 27.1 & 2.00 \\
    & ShallowBlocker & \textbf{4.73} & \textbf{4.41} & \underline{\textbf{0.951}} & \textbf{2.68} & \textbf{5.69} & 12.8 & \underline{\textbf{0.923}} & \textbf{2.56} & \textbf{6.35} & 11.8 & 1.91 & \textbf{424} & \textbf{10.4} & \textbf{0.721} \\
    \cmidrule{1-16}
    \multirow{6}{1.9cm}{Supervised \\ $\leq$ 1000 matches}
    & Sup-MinHash LSH & 43.7 & - & 1.47 & 173 & - & 194 & 1.97 & 250 & - & 194 & 175 & 25325 & {\color{gray}OOM} & {\color{gray}OOM} \\
    & Sup-$\tau$-join & 5.11 & - & \textbf{1.00} & 2.86 & - & 2.88 & 1.00 & 2.84 & - & \underline{\textbf{2.88}} & 6.50 & 1198 & 8.68 & 0.532 \\
    & Sup-AutoBlock & {\color{gray}532} & - & {\color{gray}31.0} & {\color{gray}914} & - & 353 & {\color{gray}23.5} & {\color{gray}17682} & - & {\color{gray}8252} & 222 & {\color{gray}593} & 99.1$^*$ & {\color{gray}OOM} \\
    & Sup-DeepBlocker & 22.0 & - & \textbf{1.000} & 15.0 & - & 15.6 & 1.000 & 15.4 & - & 16.0 & 49.2 & {\color{gray}ERR} & {\color{gray}ERR} & 2.75$^*$ \\
    & Sup-$k$-join & 5.20 & - & \textbf{1.000} & 6.40 & - & 2.60 & 1.000 & 6.20 & - & \underline{\textbf{2.80}} & 3.40 & 1725 & 11.2 & 2.00 \\
    & ShallowBlocker & \underline{\textbf{2.85}} & - & \underline{\textbf{0.961}} & \underline{\textbf{2.28}} & - & \underline{\textbf{2.29}} & \underline{\textbf{0.937}} & \underline{\textbf{2.32}} & - & \underline{\textbf{2.82}} & \underline{\textbf{1.16}} & \textbf{237} & \textbf{5.93} & \textbf{0.459} \\
    \cmidrule{1-16}
    \multirow{6}{1.9cm}{Supervised \\ $\leq$ \num{10000} matches}
    & Sup-MinHash LSH & - & - & - & - & - & - & - & - & - & - & - & 24796 & {\color{gray}OOM} & {\color{gray}OOM} \\
    & Sup-$\tau$-join & - & - & - & - & - & - & - & - & - & - & - & 620 & 7.69 & 0.470 \\
    & Sup-AutoBlock & - & - & - & - & - & - & - & - & - & - & - & {\color{gray}505} & 77.7$^*$ & {\color{gray}OOM} \\
    & Sup-DeepBlocker & - & - & - & - & - & - & - & - & - & - & - & {\color{gray}ERR} & 471$^*$ & 2.00 \\
    & Sup-$k$-join & - & - & - & - & - & - & - & - & - & - & - & 97.9 & 9.46 & 2.00 \\
    & ShallowBlocker & - & - & - & - & - & - & - & - & - & - & - & \underline{\textbf{70.1}} & \textbf{6.10} & \underline{\textbf{0.389}} \\
    \midrule
    \multicolumn{16}{c}{Runtime} \\
    \midrule
    \multirow{11}{1.9cm}{Unsupervised}
    & Token Blocking & 26ms & 191ms & 96ms & 1.65s & 6.31s & 214ms & 95ms & 1.62s & 6.29s & 219ms & 24ms & 1m21s & {\color{gray}OOM} & {\color{gray}OOM} \\
    & HVTB & {\color{gray}44ms} & {\color{gray}49ms} & 66ms & 748ms & {\color{gray}5.11s} & 243ms & 71ms & 757ms & {\color{gray}1.23s} & 244ms & {\color{gray}31ms} & 18m17s & {\color{gray}5m41s} & {\color{gray}OOM} \\
    & HVTB+ & {\color{gray}33ms} & {\color{gray}24ms} & 25ms & 168ms & {\color{gray}356ms} & 70ms & 26ms & \underline{\textbf{150ms}} & {\color{gray}134ms} & 70ms & {\color{gray}14ms} & \underline{\textbf{28.19s}} & {\color{gray}10.83s} & 1m10s \\
    & MinHash LSH & 380ms & 982ms & 447ms & 8.53s & 19.95s & 2.05s & 439ms & 8.44s & 19.73s & 2.04s & 240ms & {\color{gray}32.47s} & {\color{gray}OOM} & {\color{gray}OOM} \\
    & PPJoin & 60ms & 5.00s & 210ms & 2.85s & 16.81s & 360ms & 200ms & 2.72s & 16.80s & 350ms & 30ms & {\color{gray}OOM} & {\color{gray}OOM} & {\color{gray}OOM} \\
    & $\tau$-join & 17ms & \underline{\textbf{20ms}} & \underline{\textbf{20ms}} & \underline{\textbf{158ms}} & 213ms & \underline{\textbf{65ms}} & \underline{\textbf{23ms}} & 159ms & 220ms & 73ms & 19ms & {\color{gray}7.36s} & 13.41s & 4m49s \\
    & DeepBlocker & 47.66s & 1m5s & 47.94s & 9m44s & 8m58s & 3m49s & 51.67s & 9m4s & 8m53s & 53m59s & 5m30s & {\color{gray}4h52m} & 44m6s & {\color{gray}TIME} \\
    & $k$-join & \underline{\textbf{15ms}} & \underline{\textbf{21ms}} & 25ms & \underline{\textbf{155ms}} & \underline{\textbf{140ms}} & 69ms & 25ms & \underline{\textbf{154ms}} & \underline{\textbf{145ms}} & \underline{\textbf{64ms}} & \underline{\textbf{17ms}} & {\color{gray}18.21s} & 4.30s & 3m37s \\
    & ShallowBlocker ($q\!=\!0.95$) & 760ms & 866ms & 1.01s & 2.00s & 2.11s & 1.06s & 1.02s & 1.99s & 2.17s & 1.07s & 921ms & {\color{gray}11.07s} & \underline{\textbf{3.18s}} & \textbf{45.69s} \\
    & ShallowBlocker ($q\!=\!0.99$) & 760ms & 874ms & 1.02s & 2.11s & 1.92s & 803ms & 1.02s & 2.11s & 1.99s & 813ms & 919ms & {\color{gray}15.25s} & \underline{\textbf{3.33s}} & 2m26s \\
    & ShallowBlocker ($q\!=\!1$) & 392ms & 447ms & 682ms & 2.85s & 2.34s & 688ms & 676ms & 2.83s & 2.46s & 686ms & 493ms & {\color{gray}29.01s} & 3.43s & 4m14s \\
    \cmidrule{1-16}
    \multirow{6}{1.9cm}{Supervised \\ $\leq$ 100 matches}
    & Sup-MinHash LSH & 11.60s & 22.58s & 13.06s & 40.20s & 36.74s & 25.54s & 13.12s & 34.38s & 38.22s & 25.86s & 6.83s & 6m54s & {\color{gray}OOM} & {\color{gray}OOM} \\
    & Sup-$\tau$-join & 1.46s & 3.70s & 1.41s & \textbf{7.07s} & \textbf{8.63s} & 4.87s & 1.46s & \textbf{6.59s} & \textbf{9.32s} & 4.91s & 2.16s & 6m8s & 1m19s & 2m17s \\
    & Sup-AutoBlock & {\color{gray}2m49s} & {\color{gray}3m52s} & {\color{gray}6m2s} & {\color{gray}5m36s} & 26m25s & 4m37s & {\color{gray}3m43s} & {\color{gray}6m42s} & 14m45s & {\color{gray}3m54s} & 5m37s & {\color{gray}48m40s} & {\color{gray}OOM} & {\color{gray}OOM} \\
    & Sup-DeepBlocker & 1m0s & 1m21s & 52.51s & 8m57s & 8m46s & 3m26s & 49.03s & 9m2s & 8m54s & 55m15s & 5m15s & {\color{gray}ERR} & {\color{gray}ERR} & 1h40m$^*$ \\
    & Sup-$k$-join & \textbf{755ms} & \textbf{854ms} & \textbf{419ms} & 8.14s & 9.76s & \textbf{2.00s} & \textbf{430ms} & 8.46s & 10.57s & \textbf{2.00s} & \textbf{652ms} & 2m29s & 42.51s & 13m31s \\
    & ShallowBlocker & 2.93s & 3.24s & 2.00s & 7.97s & 10.53s & 4.14s & 2.00s & 8.05s & 10.90s & 4.16s & 3.18s & \textbf{45.76s} & \textbf{13.46s} & \underline{\textbf{23.96s}} \\
    \cmidrule{1-16}
    \multirow{6}{1.9cm}{Supervised \\ $\leq$ 1000 matches}
    & Sup-MinHash LSH & 7.56s & - & 11.71s & 35.53s & - & 16.37s & 11.47s & 37.97s & - & 16.73s & 7.74s & 5m1s & {\color{gray}OOM} & {\color{gray}OOM} \\
    & Sup-$\tau$-join & 1.33s & - & 1.28s & \textbf{5.56s} & - & 3.86s & 1.33s & \textbf{5.64s} & - & 3.90s & 2.11s & 5m29s & 52.66s & 1m41s \\
    & Sup-AutoBlock & {\color{gray}2m19s} & - & {\color{gray}5m53s} & {\color{gray}6m2s} & - & 4m35s & {\color{gray}3m56s} & {\color{gray}9m33s} & - & {\color{gray}5m16s} & 3m49s & {\color{gray}48m21s} & 2h51m$^*$ & {\color{gray}OOM} \\
    & Sup-DeepBlocker & 52.01s & - & 53.83s & 9m5s & - & 3m22s & 50.37s & 9m0s & - & 56m5s & 5m16s & {\color{gray}ERR} & {\color{gray}ERR} & 1h38m$^*$ \\
    & Sup-$k$-join & \textbf{821ms} & - & \textbf{480ms} & 22.17s & - & \textbf{2.58s} & \textbf{488ms} & 22.63s & - & \textbf{2.68s} & \textbf{1.04s} & 6m46s & 56.57s & 11m43s \\
    & ShallowBlocker & 4.08s & - & 3.58s & 12.22s & - & 5.36s & 3.62s & 12.24s & - & 5.36s & 4.38s & \textbf{1m13s} & \textbf{9.60s} & \textbf{27.56s} \\
    \cmidrule{1-16}
    \multirow{6}{1.9cm}{Supervised \\ $\leq$ \num{10000} matches}
    & Sup-MinHash LSH & - & - & - & - & - & - & - & - & - & - & - & \textbf{4m19s} & {\color{gray}OOM} & {\color{gray}OOM} \\
    & Sup-$\tau$-join & - & - & - & - & - & - & - & - & - & - & - & 4m36s & 39.86s & 1m25s \\
    & Sup-AutoBlock & - & - & - & - & - & - & - & - & - & - & - & {\color{gray}59m39s} & 2h22m$^*$ & {\color{gray}OOM} \\
    & Sup-DeepBlocker & - & - & - & - & - & - & - & - & - & - & - & {\color{gray}ERR} & 49m31s$^*$ & 1h41m \\
    & Sup-$k$-join & - & - & - & - & - & - & - & - & - & - & - & 55m57s & 41.48s & 12m27s \\
    & ShallowBlocker & - & - & - & - & - & - & - & - & - & - & - & 4m53s & \textbf{16.99s} & \textbf{1m7s} \\
    \bottomrule
\end{tabular}

    \caption{
        Performance of all methods across all datasets in both unsupervised and different supervised settings when the recall target is 90\%.
        See Table~\ref{tab:low-recall-target-memory} for memory usage.
        The highest performing method for each dataset and resource setting according to each measure is underlined, while methods that are within 5\% of this is bold.
        Methods that failed to reach to recall target is grayed out.
        OOM: Out of memory.
        ERR: Hitting limits of FAISS.
        TIME: Exceeding 24h time limit.
        $^*$One or more runs crashed.
    }
    \label{tab:low-recall-target}
\end{table*}

\begin{table*}[htbp]
    \setlength{\tabcolsep}{2pt}
    \tiny
    \centering
    \vspace{-2em}

\begin{tabular}{llcccccccccccccc}
    \toprule
    & & \multicolumn{14}{c}{Dataset} \\
    \cmidrule{3-16} \\
    Resource Setting & Method & AG & B & DA & DG & IA & WA & D-DA & D-DG & D-IA & D-WA & AB & C & S & CD \\
    \midrule
    \multicolumn{16}{c}{Recall} \\
    \midrule
    \multirow{11}{1.9cm}{Unsupervised}
    & Token Blocking & 1.000 & 1.000 & 1.000 & 0.999 & 1.000 & 1.000 & 1.000 & 0.999 & 1.000 & 1.000 & 0.995 & {\color{gray}0.964} & {\color{gray}OOM} & {\color{gray}OOM} \\
    & HVTB & {\color{gray}0.868} & {\color{gray}0.210} & 1.000 & 0.993 & {\color{gray}0.860} & 0.990 & 1.000 & 0.993 & {\color{gray}0.200} & 0.990 & {\color{gray}0.893} & {\color{gray}0.923} & {\color{gray}0.576} & {\color{gray}OOM} \\
    & HVTB+ & {\color{gray}0.868} & {\color{gray}0.210} & 1.000 & 0.993 & {\color{gray}0.860} & 0.990 & 1.000 & 0.993 & {\color{gray}0.200} & 0.990 & {\color{gray}0.893} & {\color{gray}0.923} & {\color{gray}0.576} & 0.998 \\
    & MinHash LSH & 1.000 & 1.000 & 1.000 & 0.999 & 1.000 & 1.000 & 1.000 & 0.999 & 1.000 & 1.000 & 0.985 & {\color{gray}0.862} & {\color{gray}OOM} & {\color{gray}OOM} \\
    & PPJoin & 1.000 & 1.000 & 1.000 & 0.999 & {\color{gray}OOM} & 1.000 & 1.000 & 0.999 & {\color{gray}OOM} & 1.000 & 0.981 & {\color{gray}OOM} & {\color{gray}OOM} & {\color{gray}OOM} \\
    & $\tau$-join & 0.987 & 1.000 & 1.000 & 0.997 & 1.000 & 1.000 & 1.000 & 0.997 & 1.000 & 1.000 & 0.981 & {\color{gray}0.660} & 0.999 & 1.000 \\
    & DeepBlocker & - & - & - & - & - & - & - & - & - & - & - & - & - & - \\
    & $k$-join & 0.991 & 1.000 & 1.000 & 0.993 & 1.000 & 0.990 & 1.000 & 0.993 & 1.000 & 0.990 & 0.990 & {\color{gray}0.892} & 0.980 & 0.998 \\
    & ShallowBlocker ($q\!=\!0.95$) & 0.996 & 1.000 & 1.000 & 0.990 & 0.998 & 1.000 & 1.000 & 0.989 & 0.998 & 1.000 & 1.000 & {\color{gray}0.865} & 0.980 & 0.997 \\
    & ShallowBlocker ($q\!=\!0.99$) & 0.987 & 1.000 & 1.000 & 0.988 & 0.980 & 0.992 & 1.000 & 0.988 & 0.980 & 0.992 & 1.000 & {\color{gray}0.796} & 0.981 & 0.995 \\
    & ShallowBlocker ($q\!=\!1$) & 0.987 & 1.000 & 1.000 & 0.987 & 0.980 & 0.994 & 1.000 & 0.988 & 0.980 & 0.994 & 0.995 & {\color{gray}0.853} & 0.980 & 0.995 \\
    \cmidrule{1-16}
    \multirow{6}{1.9cm}{Supervised \\ $\leq$ 100 matches}
    & Sup-MinHash LSH & 0.997 & 1.000 & 0.994 & 0.991 & 0.992 & 0.995 & 0.994 & 0.999 & 0.992 & 0.995 & {\color{gray}0.969} & {\color{gray}0.949} & {\color{gray}OOM} & {\color{gray}OOM} \\
    & Sup-$\tau$-join & 0.987 & 1.000 & 0.998 & 0.989 & 0.980 & 1.000 & 0.998 & 0.991 & 0.980 & 1.000 & {\color{gray}0.971} & {\color{gray}0.963} & 0.999 & 0.993 \\
    & Sup-AutoBlock & {\color{gray}0.830} & {\color{gray}0.792} & {\color{gray}0.742} & {\color{gray}0.925} & 1.000 & {\color{gray}0.952} & {\color{gray}0.701} & {\color{gray}0.809} & {\color{gray}0.950} & {\color{gray}0.802} & {\color{gray}0.932} & {\color{gray}0.243} & {\color{gray}OOM} & {\color{gray}OOM} \\
    & Sup-DeepBlocker & 0.998 & {\color{gray}ERR} & 0.995 & 0.990$^*$ & 1.000 & 1.000$^*$ & 0.996 & 0.989$^*$ & 1.000 & 1.000$^*$ & 1.000 & {\color{gray}ERR} & {\color{gray}ERR} & {\color{gray}0.975} \\
    & Sup-$k$-join & 1.000 & 1.000 & 1.000 & 0.995 & 1.000 & 1.000 & 1.000 & 0.995 & 1.000 & 1.000 & {\color{gray}0.961} & {\color{gray}0.964} & 0.998 & 0.992 \\
    & ShallowBlocker & 0.995 & 1.000 & 0.980 & 0.989 & 0.994 & 0.993 & 0.987 & 0.989 & 0.992 & 0.992 & {\color{gray}0.972} & {\color{gray}0.937} & 0.995 & {\color{gray}0.974} \\
    \cmidrule{1-16}
    \multirow{6}{1.9cm}{Supervised \\ $\leq$ 1000 matches}
    & Sup-MinHash LSH & 0.993 & - & 0.989 & 0.991 & - & 0.991 & 0.992 & 0.990 & - & 0.991 & 0.983 & {\color{gray}0.949} & {\color{gray}OOM} & {\color{gray}OOM} \\
    & Sup-$\tau$-join & 0.987 & - & 0.993 & 0.989 & - & 0.992 & 0.993 & 0.989 & - & 0.992 & 0.989 & {\color{gray}0.963} & 0.995 & 0.995 \\
    & Sup-AutoBlock & {\color{gray}0.924} & - & {\color{gray}0.898} & {\color{gray}0.953} & - & {\color{gray}0.951} & {\color{gray}0.964} & {\color{gray}0.869} & - & {\color{gray}0.885} & {\color{gray}0.928} & {\color{gray}0.506} & {\color{gray}OOM} & {\color{gray}OOM} \\
    & Sup-DeepBlocker & 0.997 & - & 0.994 & 0.992 & - & 0.982 & 0.992 & 0.990$^*$ & - & 0.988 & 0.994 & {\color{gray}ERR} & {\color{gray}ERR} & {\color{gray}ERR} \\
    & Sup-$k$-join & 1.000 & - & 0.993 & 0.995 & - & {\color{gray}0.979} & 0.991 & 0.996 & - & 0.991 & {\color{gray}0.979} & {\color{gray}0.964} & 0.997 & 0.992 \\
    & ShallowBlocker & 0.997 & - & 0.989 & 0.992 & - & 0.992 & 0.991 & 0.992 & - & 0.993 & 0.987 & {\color{gray}0.962} & 0.992 & 0.990 \\
    \cmidrule{1-16}
    \multirow{6}{1.9cm}{Supervised \\ $\leq$ \num{10000} matches}
    & Sup-MinHash LSH & - & - & - & - & - & - & - & - & - & - & - & {\color{gray}0.949} & {\color{gray}OOM} & {\color{gray}OOM} \\
    & Sup-$\tau$-join & - & - & - & - & - & - & - & - & - & - & - & {\color{gray}0.963} & 0.986 & 0.986 \\
    & Sup-AutoBlock & - & - & - & - & - & - & - & - & - & - & - & {\color{gray}0.577} & {\color{gray}OOM} & {\color{gray}OOM} \\
    & Sup-DeepBlocker & - & - & - & - & - & - & - & - & - & - & - & {\color{gray}ERR} & {\color{gray}ERR} & 0.982$^*$ \\
    & Sup-$k$-join & - & - & - & - & - & - & - & - & - & - & - & {\color{gray}0.964} & 0.992 & 0.994 \\
    & ShallowBlocker & - & - & - & - & - & - & - & - & - & - & - & {\color{gray}0.964} & 0.990 & 0.988 \\
    \midrule
    \multicolumn{16}{c}{$\widetilde{k}$} \\
    \midrule
    \multirow{11}{1.9cm}{Unsupervised}
    & Token Blocking & 513 & 3685 & 1848 & 34879 & 51042 & 3987 & 1848 & 34526 & 51042 & 3987 & 426 & {\color{gray}27078} & {\color{gray}OOM} & {\color{gray}OOM} \\
    & HVTB & {\color{gray}5.39} & {\color{gray}27.7} & 13.2 & 41.0 & {\color{gray}1384} & 41.2 & 13.2 & 41.8 & {\color{gray}278} & 41.2 & {\color{gray}9.00} & {\color{gray}5887} & {\color{gray}25.6} & {\color{gray}OOM} \\
    & HVTB+ & {\color{gray}5.39} & {\color{gray}27.7} & 13.2 & 41.0 & {\color{gray}1384} & 41.2 & 13.2 & 41.8 & {\color{gray}278} & 41.2 & {\color{gray}9.00} & {\color{gray}5887} & {\color{gray}25.6} & 150 \\
    & MinHash LSH & 487 & 3587 & 1381 & 27536 & 41566 & 3322 & 1381 & 27160 & 41566 & 3322 & 276 & {\color{gray}19166} & {\color{gray}OOM} & {\color{gray}OOM} \\
    & PPJoin & 124 & 2963 & 310 & 4580 & {\color{gray}OOM} & 287 & 310 & 4422 & {\color{gray}OOM} & 287 & 26.7 & {\color{gray}OOM} & {\color{gray}OOM} & {\color{gray}OOM} \\
    & $\tau$-join & 22.0 & 13.3 & \textbf{10.3} & 54.0 & 337 & 60.4 & \textbf{10.3} & 52.6 & 337 & 60.4 & \textbf{10.9} & {\color{gray}1.49} & 40.4 & 210 \\
    & DeepBlocker & - & - & - & - & - & - & - & - & - & - & - & - & - & - \\
    & $k$-join & 31.8 & 31.9 & 32.0 & 32.0 & 32.0 & 32.0 & 32.0 & 32.0 & 32.0 & 32.0 & 32.0 & {\color{gray}32.0} & 21.2 & 32.0 \\
    & ShallowBlocker ($q\!=\!0.95$) & 16.2 & 16.3 & 16.1 & 11.4 & 9.84 & 16.3 & 16.1 & 11.8 & 9.84 & 16.3 & 21.2 & {\color{gray}8.81} & 21.4 & 7.38 \\
    & ShallowBlocker ($q\!=\!0.99$) & \textbf{10.6} & 9.65 & 10.8 & 11.3 & \underline{\textbf{5.71}} & 10.5 & 10.8 & 11.7 & \underline{\textbf{5.71}} & 10.5 & 14.6 & {\color{gray}5.60} & 16.9 & 5.24 \\
    & ShallowBlocker ($q\!=\!1$) & \textbf{10.2} & \underline{\textbf{8.96}} & \textbf{9.92} & \textbf{10.7} & \underline{\textbf{5.58}} & \underline{\textbf{9.87}} & \textbf{9.92} & \textbf{11.0} & \underline{\textbf{5.58}} & \underline{\textbf{9.87}} & 13.6 & {\color{gray}6.11} & \textbf{15.4} & \textbf{4.90} \\
    \cmidrule{1-16}
    \multirow{6}{1.9cm}{Supervised \\ $\leq$ 100 matches}
    & Sup-MinHash LSH & 351 & 2789 & 23.1 & 2427 & 10017 & 2272 & 23.1 & 10091 & 10017 & 2332 & {\color{gray}239} & {\color{gray}25952} & {\color{gray}OOM} & {\color{gray}OOM} \\
    & Sup-$\tau$-join & 21.2 & 1524 & 1.82 & 8.42 & 17.0 & 54.8 & 1.82 & 8.82 & 17.0 & 54.8 & {\color{gray}8.82} & {\color{gray}27056} & 27.4 & \textbf{1.69} \\
    & Sup-AutoBlock & {\color{gray}1074} & {\color{gray}24.6} & {\color{gray}10.9} & {\color{gray}23984} & 39168 & {\color{gray}2017} & {\color{gray}17.3} & {\color{gray}27912} & {\color{gray}28541} & {\color{gray}7081} & {\color{gray}648} & {\color{gray}444} & {\color{gray}OOM} & {\color{gray}OOM} \\
    & Sup-DeepBlocker & 1041 & {\color{gray}ERR} & 5.40 & 1246$^*$ & 1314 & 940$^*$ & 5.80 & 649$^*$ & 903 & 951$^*$ & \textbf{552} & {\color{gray}ERR} & {\color{gray}ERR} & {\color{gray}111} \\
    & Sup-$k$-join & 63.8 & 185 & 17.0 & 33.6 & 112 & 45.6 & 17.0 & 41.0 & 112 & 58.4 & {\color{gray}3.00} & {\color{gray}27078} & 56.4 & 3.00 \\
    & ShallowBlocker & \textbf{11.4} & \textbf{25.4} & \underline{\textbf{1.02}} & \textbf{6.76} & \textbf{15.4} & \textbf{23.8} & \textbf{1.09} & \textbf{6.67} & \textbf{11.5} & \textbf{24.0} & {\color{gray}2.37} & {\color{gray}11870} & \textbf{19.6} & {\color{gray}0.999} \\
    \cmidrule{1-16}
    \multirow{6}{1.9cm}{Supervised \\ $\leq$ 1000 matches}
    & Sup-MinHash LSH & 390 & - & 6.95 & 2213 & - & 1294 & 7.34 & 2111 & - & 1588 & 274 & {\color{gray}25952} & {\color{gray}OOM} & {\color{gray}OOM} \\
    & Sup-$\tau$-join & 23.9 & - & 1.29 & 12.9 & - & 22.4 & 1.29 & 12.7 & - & 22.4 & 32.3 & {\color{gray}27056} & 22.8 & 2.01 \\
    & Sup-AutoBlock & {\color{gray}1186} & - & {\color{gray}62.3} & {\color{gray}21349} & - & {\color{gray}2275} & {\color{gray}330} & {\color{gray}19356} & - & {\color{gray}8527} & {\color{gray}415} & {\color{gray}593} & {\color{gray}OOM} & {\color{gray}OOM} \\
    & Sup-DeepBlocker & 700 & - & 2.40 & 1194 & - & 149 & 2.40 & 947$^*$ & - & 199 & 275 & {\color{gray}ERR} & {\color{gray}ERR} & {\color{gray}ERR} \\
    & Sup-$k$-join & 22.8 & - & \underline{\textbf{1.000}} & 19.4 & - & {\color{gray}7.40} & \underline{\textbf{1.000}} & 17.8 & - & \textbf{10.4} & {\color{gray}5.60} & {\color{gray}27078} & 53.3 & 3.00 \\
    & ShallowBlocker & \underline{\textbf{9.01}} & - & \underline{\textbf{1.01}} & \underline{\textbf{4.66}} & - & \textbf{16.5} & \underline{\textbf{1.01}} & \underline{\textbf{5.02}} & - & 17.8 & \underline{\textbf{3.25}} & {\color{gray}25303} & \textbf{16.8} & \underline{\textbf{0.805}} \\
    \cmidrule{1-16}
    \multirow{6}{1.9cm}{Supervised \\ $\leq$ \num{10000} matches}
    & Sup-MinHash LSH & - & - & - & - & - & - & - & - & - & - & - & {\color{gray}25952} & {\color{gray}OOM} & {\color{gray}OOM} \\
    & Sup-$\tau$-join & - & - & - & - & - & - & - & - & - & - & - & {\color{gray}27056} & 18.6 & 1.10 \\
    & Sup-AutoBlock & - & - & - & - & - & - & - & - & - & - & - & {\color{gray}505} & {\color{gray}OOM} & {\color{gray}OOM} \\
    & Sup-DeepBlocker & - & - & - & - & - & - & - & - & - & - & - & {\color{gray}ERR} & {\color{gray}ERR} & 17.8$^*$ \\
    & Sup-$k$-join & - & - & - & - & - & - & - & - & - & - & - & {\color{gray}27078} & 35.1 & 4.00 \\
    & ShallowBlocker & - & - & - & - & - & - & - & - & - & - & - & {\color{gray}26921} & \underline{\textbf{14.4}} & \underline{\textbf{0.782}} \\
    \midrule
    \multicolumn{16}{c}{Runtime} \\
    \midrule
    \multirow{11}{1.9cm}{Unsupervised}
    & Token Blocking & 26ms & 191ms & 96ms & 1.65s & 6.31s & 214ms & 95ms & 1.62s & 6.29s & 219ms & 24ms & {\color{gray}1m21s} & {\color{gray}OOM} & {\color{gray}OOM} \\
    & HVTB & {\color{gray}44ms} & {\color{gray}49ms} & 66ms & 748ms & {\color{gray}5.11s} & 243ms & 71ms & 757ms & {\color{gray}1.23s} & 244ms & {\color{gray}31ms} & {\color{gray}18m17s} & {\color{gray}5m41s} & {\color{gray}OOM} \\
    & HVTB+ & {\color{gray}33ms} & {\color{gray}24ms} & 25ms & 168ms & {\color{gray}356ms} & \underline{\textbf{70ms}} & 26ms & \underline{\textbf{150ms}} & {\color{gray}134ms} & \underline{\textbf{70ms}} & {\color{gray}14ms} & {\color{gray}28.19s} & {\color{gray}10.83s} & \textbf{1m10s} \\
    & MinHash LSH & 731ms & 1.58s & 835ms & 16.09s & 29.71s & 3.79s & 820ms & 15.94s & 29.83s & 3.77s & 294ms & {\color{gray}48.91s} & {\color{gray}OOM} & {\color{gray}OOM} \\
    & PPJoin & 150ms & 6.78s & 600ms & 10.04s & {\color{gray}OOM} & 710ms & 600ms & 9.69s & {\color{gray}OOM} & 710ms & 40ms & {\color{gray}OOM} & {\color{gray}OOM} & {\color{gray}OOM} \\
    & $\tau$-join & 18ms & \underline{\textbf{18ms}} & 24ms & \underline{\textbf{150ms}} & 299ms & 74ms & \underline{\textbf{23ms}} & \underline{\textbf{158ms}} & 311ms & \underline{\textbf{73ms}} & 19ms & {\color{gray}10.71s} & 24.59s & 8m48s \\
    & DeepBlocker & - & - & - & - & - & - & - & - & - & - & - & - & - & - \\
    & $k$-join & \underline{\textbf{16ms}} & 22ms & \underline{\textbf{23ms}} & 161ms & \underline{\textbf{168ms}} & \underline{\textbf{71ms}} & 25ms & 159ms & \underline{\textbf{173ms}} & \underline{\textbf{73ms}} & \underline{\textbf{17ms}} & {\color{gray}21.45s} & 10.29s & 5m34s \\
    & ShallowBlocker ($q\!=\!0.95$) & 789ms & 909ms & 1.08s & 2.54s & 2.66s & 1.26s & 1.07s & 2.55s & 2.73s & 1.25s & 944ms & {\color{gray}13.26s} & \underline{\textbf{4.28s}} & 2m18s \\
    & ShallowBlocker ($q\!=\!0.99$) & 786ms & 912ms & 1.07s & 2.41s & 2.64s & 1.23s & 1.06s & 2.41s & 2.71s & 1.24s & 952ms & {\color{gray}19.76s} & \underline{\textbf{4.43s}} & 4m20s \\
    & ShallowBlocker ($q\!=\!1$) & 413ms & 487ms & 716ms & 3.04s & 3.01s & 937ms & 703ms & 3.03s & 3.15s & 949ms & 521ms & {\color{gray}33.77s} & 6.20s & 6m43s \\
    \cmidrule{1-16}
    \multirow{6}{1.9cm}{Supervised \\ $\leq$ 100 matches}
    & Sup-MinHash LSH & 16.02s & 23.96s & 19.11s & 42.57s & 51.04s & 23.71s & 19.21s & 45.05s & 53.67s & 24.28s & {\color{gray}7.44s} & {\color{gray}7m30s} & {\color{gray}OOM} & {\color{gray}OOM} \\
    & Sup-$\tau$-join & 1.75s & 5.55s & 2.64s & 9.91s & \textbf{9.16s} & 5.79s & 2.82s & 9.86s & \textbf{9.54s} & 5.80s & {\color{gray}2.35s} & {\color{gray}2m18s} & 1m39s & \textbf{2m40s} \\
    & Sup-AutoBlock & {\color{gray}2m52s} & {\color{gray}4m35s} & {\color{gray}6m16s} & {\color{gray}6m55s} & 26m17s & {\color{gray}4m51s} & {\color{gray}3m57s} & {\color{gray}6m47s} & {\color{gray}15m22s} & {\color{gray}4m0s} & {\color{gray}5m37s} & {\color{gray}48m26s} & {\color{gray}OOM} & {\color{gray}OOM} \\
    & Sup-DeepBlocker & 1m5s & {\color{gray}ERR} & 51.13s & 9m13s$^*$ & 9m1s & 3m35s$^*$ & 50.57s & 9m1s$^*$ & 8m51s & 55m42s$^*$ & \textbf{5m20s} & {\color{gray}ERR} & {\color{gray}ERR} & {\color{gray}1h40m} \\
    & Sup-$k$-join & \textbf{1.04s} & \textbf{1.64s} & \textbf{574ms} & 10.12s & 13.31s & \textbf{2.81s} & \textbf{592ms} & 10.40s & 14.31s & \textbf{2.87s} & {\color{gray}849ms} & {\color{gray}2m38s} & 1m11s & 14m6s \\
    & ShallowBlocker & 2.99s & 3.17s & 2.00s & \textbf{8.03s} & 10.75s & 4.27s & 2.00s & \textbf{8.07s} & 11.14s & 4.28s & {\color{gray}3.38s} & {\color{gray}1m17s} & \textbf{12.41s} & {\color{gray}25.02s} \\
    \cmidrule{1-16}
    \multirow{6}{1.9cm}{Supervised \\ $\leq$ 1000 matches}
    & Sup-MinHash LSH & 16.36s & - & 13.71s & 40.27s & - & 34.43s & 13.46s & 38.18s & - & 30.76s & 9.76s & {\color{gray}7m31s} & {\color{gray}OOM} & {\color{gray}OOM} \\
    & Sup-$\tau$-join & 1.74s & - & 2.20s & \textbf{8.09s} & - & \textbf{5.32s} & 2.31s & \textbf{8.34s} & - & 5.44s & \textbf{2.97s} & {\color{gray}2m18s} & 1m25s & 2m39s \\
    & Sup-AutoBlock & {\color{gray}3m4s} & - & {\color{gray}6m25s} & {\color{gray}11m19s} & - & {\color{gray}5m56s} & {\color{gray}5m4s} & {\color{gray}9m54s} & - & {\color{gray}5m24s} & {\color{gray}4m3s} & {\color{gray}47m41s} & {\color{gray}OOM} & {\color{gray}OOM} \\
    & Sup-DeepBlocker & 1m3s & - & 50.17s & 9m8s & - & 3m44s & 50.16s & 9m7s$^*$ & - & 54m6s & 5m36s & {\color{gray}ERR} & {\color{gray}ERR} & {\color{gray}ERR} \\
    & Sup-$k$-join & \textbf{1.37s} & - & \textbf{472ms} & 35.64s & - & {\color{gray}3.99s} & \textbf{482ms} & 36.81s & - & \textbf{4.47s} & {\color{gray}1.59s} & {\color{gray}10m23s} & 1m23s & 13m53s \\
    & ShallowBlocker & 4.52s & - & 3.99s & 13.82s & - & 5.90s & 4.00s & 13.92s & - & 5.88s & 4.89s & {\color{gray}2m0s} & \textbf{17.93s} & \underline{\textbf{38.05s}} \\
    \cmidrule{1-16}
    \multirow{6}{1.9cm}{Supervised \\ $\leq$ \num{10000} matches}
    & Sup-MinHash LSH & - & - & - & - & - & - & - & - & - & - & - & {\color{gray}7m48s} & {\color{gray}OOM} & {\color{gray}OOM} \\
    & Sup-$\tau$-join & - & - & - & - & - & - & - & - & - & - & - & {\color{gray}2m20s} & 50.69s & 1m35s \\
    & Sup-AutoBlock & - & - & - & - & - & - & - & - & - & - & - & {\color{gray}59m30s} & {\color{gray}OOM} & {\color{gray}OOM} \\
    & Sup-DeepBlocker & - & - & - & - & - & - & - & - & - & - & - & {\color{gray}ERR} & {\color{gray}ERR} & 1h36m$^*$ \\
    & Sup-$k$-join & - & - & - & - & - & - & - & - & - & - & - & {\color{gray}1h31m} & 2m14s & 16m30s \\
    & ShallowBlocker & - & - & - & - & - & - & - & - & - & - & - & {\color{gray}5m51s} & \textbf{32.85s} & \textbf{1m16s} \\
    \bottomrule
\end{tabular}

    \caption{
        Performance of all methods across all datasets in both unsupervised and different supervised settings when the recall target is 98\%.
        See Table~\ref{tab:high-recall-target-memory} for memory usage.
        The highest performing method for each dataset and resource setting according to each measure is underlined, while methods that are within 5\% of this is bold.
        Methods that failed to reach to recall target is grayed out.
        OOM: Out of memory.
        ERR: Hitting limits of FAISS.
        $^*$One or more runs crashed.
    }
    \label{tab:high-recall-target}
\end{table*}

\begin{table*}[htb]
    \setlength{\tabcolsep}{3pt}
    \tiny
    \centering

\begin{tabular}{llcccccccccccccc}
    \toprule
    & & \multicolumn{14}{c}{Dataset} \\
    \cmidrule{3-16} \\
    Resource Setting & Method & AG & B & DA & DG & IA & WA & D-DA & D-DG & D-IA & D-WA & AB & C & S & CD \\
    \midrule
    \multicolumn{16}{c}{Memory Usage (GB)} \\
    \midrule
    \multirow{11}{1.9cm}{Unsupervised}
    & Token Blocking & 0.14 & 0.32 & 0.21 & 1.85 & 6.04 & 0.40 & 0.21 & 1.84 & 6.04 & 0.38 & 0.13 & 18.7 & {\color{gray}OOM} & {\color{gray}OOM} \\
    & HVTB & {\color{gray}0.12} & {\color{gray}0.14} & 0.12 & 0.28 & {\color{gray}0.98} & 0.17 & 0.13 & 0.28 & {\color{gray}0.39} & 0.17 & {\color{gray}0.12} & 8.00 & {\color{gray}38.3} & {\color{gray}OOM} \\
    & HVTB+ & {\color{gray}0.11} & {\color{gray}0.13} & 0.12 & 0.24 & {\color{gray}0.44} & 0.15 & 0.12 & 0.23 & {\color{gray}0.22} & 0.15 & {\color{gray}0.11} & \textbf{5.10} & {\color{gray}2.23} & 9.30 \\
    & MinHash LSH & 0.17 & 0.45 & 0.21 & 1.62 & 4.69 & 0.43 & 0.21 & 1.60 & 4.72 & 0.43 & 0.15 & {\color{gray}10.1} & {\color{gray}OOM} & {\color{gray}OOM} \\
    & PPJoin & \underline{\textbf{0.01}} & \underline{\textbf{0.07}} & \underline{\textbf{0.01}} & \underline{\textbf{0.05}} & 0.29 & \underline{\textbf{0.01}} & \underline{\textbf{0.01}} & \underline{\textbf{0.05}} & 0.29 & \underline{\textbf{0.01}} & \underline{\textbf{0.01}} & {\color{gray}OOM} & {\color{gray}OOM} & {\color{gray}OOM} \\
    & $\tau$-join & 0.12 & 0.12 & 0.12 & 0.29 & 0.24 & 0.19 & 0.12 & 0.29 & 0.24 & 0.19 & 0.12 & {\color{gray}1.86} & 2.16 & 5.77 \\
    & DeepBlocker & 13.1 & 13.2 & 13.2 & 13.4 & 13.6 & 13.3 & 13.2 & 13.4 & 13.6 & 14.8 & 13.2 & {\color{gray}18.7} & 39.9 & {\color{gray}TIME} \\
    & $k$-join & 0.12 & 0.12 & 0.12 & 0.29 & \underline{\textbf{0.21}} & 0.18 & 0.12 & 0.29 & \underline{\textbf{0.21}} & 0.18 & 0.12 & {\color{gray}1.87} & \underline{\textbf{1.34}} & \underline{\textbf{4.05}} \\
    & ShallowBlocker ($q\!=\!0.95$) & 0.41 & 0.43 & 0.47 & 0.76 & 0.76 & 0.55 & 0.47 & 0.76 & 0.76 & 0.55 & 0.42 & {\color{gray}2.14} & \underline{\textbf{1.38}} & \underline{\textbf{3.88}} \\
    & ShallowBlocker ($q\!=\!0.99$) & 0.41 & 0.43 & 0.47 & 0.76 & 0.77 & 0.56 & 0.47 & 0.76 & 0.77 & 0.56 & 0.42 & {\color{gray}2.14} & \underline{\textbf{1.37}} & \underline{\textbf{3.89}} \\
    & ShallowBlocker ($q\!=\!1$) & 0.41 & 0.43 & 0.47 & 0.77 & 0.78 & 0.56 & 0.47 & 0.77 & 0.78 & 0.56 & 0.42 & {\color{gray}2.12} & \underline{\textbf{1.35}} & \underline{\textbf{3.89}} \\
    \cmidrule{1-16}
    \multirow{6}{1.9cm}{Supervised \\ $\leq$ 100 matches}
    & Sup-MinHash LSH & 0.21 & 0.37 & 0.22 & 1.21 & 1.08 & 0.56 & 0.22 & 1.01 & 1.06 & 0.56 & 0.21 & 18.7 & {\color{gray}OOM} & {\color{gray}OOM} \\
    & Sup-$\tau$-join & \textbf{0.17} & 0.30 & \textbf{0.17} & \textbf{0.36} & \textbf{0.24} & \textbf{0.26} & \textbf{0.18} & \textbf{0.36} & \textbf{0.25} & \textbf{0.26} & \textbf{0.18} & 9.16 & 2.81 & \underline{\textbf{4.03}} \\
    & Sup-AutoBlock & {\color{gray}11.1} & {\color{gray}10.9} & {\color{gray}10.9} & {\color{gray}13.5} & 66.2 & 12.0 & {\color{gray}10.9} & {\color{gray}23.9} & 38.1 & {\color{gray}13.3} & 11.1 & {\color{gray}15.6} & {\color{gray}OOM} & {\color{gray}OOM} \\
    & Sup-DeepBlocker & 13.0 & 13.1 & 13.1 & 13.4 & 13.5 & 13.2 & 13.1 & 13.4 & 13.6 & 14.8 & 13.1 & {\color{gray}ERR} & {\color{gray}ERR} & 45.1$^*$ \\
    & Sup-$k$-join & \textbf{0.18} & \textbf{0.18} & 0.19 & 1.05 & 0.92 & 0.45 & 0.19 & 1.12 & 0.91 & 0.45 & \textbf{0.17} & 18.1 & 3.75 & 9.93 \\
    & ShallowBlocker & 1.74 & 2.60 & 2.03 & 2.92 & 3.32 & 2.37 & 2.03 & 2.96 & 3.31 & 2.37 & 1.95 & \textbf{4.11} & \textbf{2.18} & 5.30 \\
    \cmidrule{1-16}
    \multirow{6}{1.9cm}{Supervised \\ $\leq$ 1000 matches}
    & Sup-MinHash LSH & 0.20 & - & 0.22 & 1.17 & - & 0.40 & 0.22 & 1.18 & - & 0.40 & 0.21 & 18.4 & {\color{gray}OOM} & {\color{gray}OOM} \\
    & Sup-$\tau$-join & \textbf{0.17} & - & \textbf{0.18} & \textbf{0.36} & - & \textbf{0.26} & \textbf{0.18} & \textbf{0.37} & - & \textbf{0.26} & \textbf{0.18} & \underline{\textbf{2.78}} & \textbf{2.19} & \underline{\textbf{4.02}} \\
    & Sup-AutoBlock & {\color{gray}11.1} & - & {\color{gray}11.1} & {\color{gray}12.0} & - & 11.4 & {\color{gray}11.0} & {\color{gray}19.6} & - & {\color{gray}14.8} & 10.9 & {\color{gray}15.6} & 54.2$^*$ & {\color{gray}OOM} \\
    & Sup-DeepBlocker & 13.0 & - & 13.1 & 13.4 & - & 13.2 & 13.1 & 13.4 & - & 14.8 & 13.1 & {\color{gray}ERR} & {\color{gray}ERR} & 45.1$^*$ \\
    & Sup-$k$-join & 0.19 & - & 0.20 & 1.10 & - & 0.48 & 0.20 & 1.12 & - & 0.49 & \textbf{0.18} & 4.93 & 3.22 & 10.0 \\
    & ShallowBlocker & 1.82 & - & 2.33 & 3.11 & - & 2.40 & 2.33 & 3.10 & - & 2.40 & 2.06 & 4.13 & \textbf{2.30} & 5.50 \\
    \cmidrule{1-16}
    \multirow{6}{1.9cm}{Supervised \\ $\leq$ \num{10000} matches}
    & Sup-MinHash LSH & - & - & - & - & - & - & - & - & - & - & - & 17.7 & {\color{gray}OOM} & {\color{gray}OOM} \\
    & Sup-$\tau$-join & - & - & - & - & - & - & - & - & - & - & - & \underline{\textbf{2.92}} & \textbf{2.14} & \underline{\textbf{4.02}} \\
    & Sup-AutoBlock & - & - & - & - & - & - & - & - & - & - & - & {\color{gray}15.6} & 52.7$^*$ & {\color{gray}OOM} \\
    & Sup-DeepBlocker & - & - & - & - & - & - & - & - & - & - & - & {\color{gray}ERR} & 73.9$^*$ & 45.0 \\
    & Sup-$k$-join & - & - & - & - & - & - & - & - & - & - & - & 8.16 & 3.14 & 10.1 \\
    & ShallowBlocker & - & - & - & - & - & - & - & - & - & - & - & 4.18 & 2.67 & 6.17 \\
    \bottomrule
\end{tabular}

    \caption{
        Memory usage of all methods across all datasets in both unsupervised and different supervised settings when the recall target is 90\%.
        This complements Table~\ref{tab:low-recall-target}.
        The lowest memory usage each dataset and resource setting is underlined, while methods that are within 5\% of this is bold.
        Methods that failed to reach to recall target is grayed out.
        OOM: Out of memory.
        ERR: Hitting limits of FAISS.
        TIME: Exceeding 24h time limit.
        $^*$ One or more runs crashed.
    }
    \label{tab:low-recall-target-memory}
\end{table*}

\begin{table*}[htb]
    \setlength{\tabcolsep}{3pt}
    \tiny
    \centering

\begin{tabular}{llcccccccccccccc}
    \toprule
    & & \multicolumn{14}{c}{Dataset} \\
    \cmidrule{3-16} \\
    Resource Setting & Method & AG & B & DA & DG & IA & WA & D-DA & D-DG & D-IA & D-WA & AB & C & S & CD \\
    \midrule
    \multicolumn{16}{c}{Memory Usage (GB)} \\
    \midrule
    \multirow{11}{1.9cm}{Unsupervised}
    & Token Blocking & 0.14 & 0.32 & 0.21 & 1.85 & 6.04 & 0.40 & 0.21 & 1.84 & 6.04 & 0.38 & 0.13 & {\color{gray}18.7} & {\color{gray}OOM} & {\color{gray}OOM} \\
    & HVTB & {\color{gray}0.12} & {\color{gray}0.14} & 0.12 & 0.28 & {\color{gray}0.98} & 0.17 & 0.13 & 0.28 & {\color{gray}0.39} & 0.17 & {\color{gray}0.12} & {\color{gray}8.00} & {\color{gray}38.3} & {\color{gray}OOM} \\
    & HVTB+ & {\color{gray}0.11} & {\color{gray}0.13} & 0.12 & 0.24 & {\color{gray}0.44} & 0.15 & 0.12 & 0.23 & {\color{gray}0.22} & 0.15 & {\color{gray}0.11} & {\color{gray}5.10} & {\color{gray}2.23} & 9.30 \\
    & MinHash LSH & 0.19 & 0.47 & 0.24 & 2.06 & 5.53 & 0.58 & 0.24 & 2.11 & 5.46 & 0.58 & 0.16 & {\color{gray}13.8} & {\color{gray}OOM} & {\color{gray}OOM} \\
    & PPJoin & \underline{\textbf{0.01}} & 0.14 & \underline{\textbf{0.02}} & \underline{\textbf{0.15}} & {\color{gray}OOM} & \underline{\textbf{0.02}} & \underline{\textbf{0.02}} & \underline{\textbf{0.15}} & {\color{gray}OOM} & \underline{\textbf{0.02}} & \underline{\textbf{0.01}} & {\color{gray}OOM} & {\color{gray}OOM} & {\color{gray}OOM} \\
    & $\tau$-join & 0.12 & \underline{\textbf{0.12}} & 0.12 & 0.29 & 0.28 & 0.19 & 0.12 & 0.29 & 0.29 & 0.19 & 0.12 & {\color{gray}1.88} & 3.40 & 12.6 \\
    & DeepBlocker & - & - & - & - & - & - & - & - & - & - & - & - & - & - \\
    & $k$-join & 0.12 & \underline{\textbf{0.12}} & 0.13 & 0.29 & \underline{\textbf{0.21}} & 0.18 & 0.13 & 0.29 & \underline{\textbf{0.22}} & 0.18 & 0.12 & {\color{gray}1.87} & \underline{\textbf{1.69}} & 4.89 \\
    & ShallowBlocker ($q\!=\!0.95$) & 0.42 & 0.46 & 0.48 & 0.77 & 0.78 & 0.57 & 0.48 & 0.77 & 0.78 & 0.57 & 0.44 & {\color{gray}2.12} & 2.34 & \underline{\textbf{4.02}} \\
    & ShallowBlocker ($q\!=\!0.99$) & 0.42 & 0.45 & 0.48 & 0.77 & 0.78 & 0.57 & 0.48 & 0.78 & 0.78 & 0.57 & 0.43 & {\color{gray}2.14} & 1.89 & \underline{\textbf{3.97}} \\
    & ShallowBlocker ($q\!=\!1$) & 0.42 & 0.45 & 0.48 & 0.78 & 0.79 & 0.56 & 0.48 & 0.78 & 0.79 & 0.56 & 0.43 & {\color{gray}2.12} & 1.81 & \underline{\textbf{3.97}} \\
    \cmidrule{1-16}
    \multirow{6}{1.9cm}{Supervised \\ $\leq$ 100 matches}
    & Sup-MinHash LSH & 0.23 & 0.43 & 0.23 & 1.33 & 2.04 & 0.45 & 0.23 & 1.69 & 2.05 & 0.46 & {\color{gray}0.24} & {\color{gray}18.9} & {\color{gray}OOM} & {\color{gray}OOM} \\
    & Sup-$\tau$-join & \textbf{0.17} & 0.34 & \textbf{0.18} & \textbf{0.37} & \textbf{0.25} & \textbf{0.26} & \textbf{0.18} & \textbf{0.37} & \textbf{0.25} & \textbf{0.26} & {\color{gray}0.18} & {\color{gray}18.4} & 3.59 & \underline{\textbf{4.02}} \\
    & Sup-AutoBlock & {\color{gray}11.3} & {\color{gray}10.9} & {\color{gray}10.9} & {\color{gray}22.2} & 66.3 & {\color{gray}12.1} & {\color{gray}10.9} & {\color{gray}24.8} & {\color{gray}45.0} & {\color{gray}14.5} & {\color{gray}11.1} & {\color{gray}15.6} & {\color{gray}OOM} & {\color{gray}OOM} \\
    & Sup-DeepBlocker & 13.3 & {\color{gray}ERR} & 13.1 & 13.7$^*$ & 14.2 & 13.4$^*$ & 13.1 & 13.6$^*$ & 14.0 & 14.8$^*$ & 13.2 & {\color{gray}ERR} & {\color{gray}ERR} & {\color{gray}59.4} \\
    & Sup-$k$-join & \textbf{0.18} & \textbf{0.21} & 0.19 & 1.10 & 0.81 & 0.43 & \textbf{0.19} & 1.06 & 0.77 & 0.43 & {\color{gray}0.17} & {\color{gray}18.9} & 4.52 & 10.1 \\
    & ShallowBlocker & 1.74 & 2.61 & 2.03 & 2.94 & 3.31 & 2.37 & 2.03 & 2.96 & 3.31 & 2.37 & {\color{gray}1.95} & {\color{gray}11.5} & \textbf{2.71} & {\color{gray}5.31} \\
    \cmidrule{1-16}
    \multirow{6}{1.9cm}{Supervised \\ $\leq$ 1000 matches}
    & Sup-MinHash LSH & 0.23 & - & 0.22 & 1.18 & - & 0.62 & 0.22 & 1.17 & - & 0.44 & 0.25 & {\color{gray}18.7} & {\color{gray}OOM} & {\color{gray}OOM} \\
    & Sup-$\tau$-join & \textbf{0.18} & - & \textbf{0.18} & \textbf{0.36} & - & \textbf{0.26} & \textbf{0.18} & \textbf{0.37} & - & \textbf{0.26} & \textbf{0.19} & {\color{gray}18.5} & 3.26 & \underline{\textbf{4.01}} \\
    & Sup-AutoBlock & {\color{gray}11.4} & - & {\color{gray}11.2} & {\color{gray}21.9} & - & {\color{gray}12.2} & {\color{gray}11.5} & {\color{gray}20.0} & - & {\color{gray}14.8} & {\color{gray}11.0} & {\color{gray}15.6} & {\color{gray}OOM} & {\color{gray}OOM} \\
    & Sup-DeepBlocker & 13.2 & - & 13.1 & 13.7 & - & 13.2 & 13.1 & 13.7$^*$ & - & 14.8 & 13.2 & {\color{gray}ERR} & {\color{gray}ERR} & {\color{gray}ERR} \\
    & Sup-$k$-join & 0.19 & - & 0.20 & 1.08 & - & {\color{gray}0.48} & 0.20 & 1.12 & - & 0.48 & {\color{gray}0.19} & {\color{gray}19.1} & 4.18 & 10.0 \\
    & ShallowBlocker & 1.81 & - & 2.33 & 3.10 & - & 2.39 & 2.33 & 3.09 & - & 2.40 & 2.06 & {\color{gray}19.3} & \textbf{2.74} & 5.51 \\
    \cmidrule{1-16}
    \multirow{6}{1.9cm}{Supervised \\ $\leq$ \num{10000} matches}
    & Sup-MinHash LSH & - & - & - & - & - & - & - & - & - & - & - & {\color{gray}18.9} & {\color{gray}OOM} & {\color{gray}OOM} \\
    & Sup-$\tau$-join & - & - & - & - & - & - & - & - & - & - & - & {\color{gray}18.7} & \textbf{2.79} & \underline{\textbf{4.03}} \\
    & Sup-AutoBlock & - & - & - & - & - & - & - & - & - & - & - & {\color{gray}15.6} & {\color{gray}OOM} & {\color{gray}OOM} \\
    & Sup-DeepBlocker & - & - & - & - & - & - & - & - & - & - & - & {\color{gray}ERR} & {\color{gray}ERR} & 45.0$^*$ \\
    & Sup-$k$-join & - & - & - & - & - & - & - & - & - & - & - & {\color{gray}19.7} & 4.04 & 10.1 \\
    & ShallowBlocker & - & - & - & - & - & - & - & - & - & - & - & {\color{gray}20.3} & 2.97 & 6.17 \\
    \bottomrule
\end{tabular}

    \caption{
        Memory usage of all methods across all datasets in both unsupervised and different supervised settings when the recall target is 98\%.
        This complements Table~\ref{tab:high-recall-target}.
        The lowest memory usage each dataset and resource setting is underlined, while methods that are within 5\% of this is bold.
        Methods that failed to reach to recall target is grayed out.
        OOM: Out of memory.
        ERR: Hitting limits of FAISS.
        $^*$ One or more runs crashed.
    }
    \label{tab:high-recall-target-memory}
\end{table*}

The experiment compare the performance characteristics between methods when targeting the same recall level.
Table~\ref{tab:low-recall-target} and \ref{tab:low-recall-target-memory} report results when targeting 90\%,
while Table~\ref{tab:high-recall-target} and \ref{tab:high-recall-target-memory} report results when targeting 98\%.
We consider 90\% and 98\% as representative of low and high recall.
For each method we report recall, the number of returned pairs, runtime, and memory usage.
We normalize the number of returned pairs $|P|$ and report the effective cardinality $\widetilde{k} = |P| / \min(|A|, |B|)$ for easier comparison across dataset sizes.
All reported numbers are the average over five runs,
and all are given a maximum runtime of 24 hours.

When used supervised, ShallowBlocker uses the objective function
\begin{equation*}
    f(\text{recall}, |P|, \text{runtime}) = (\text{recall}, \text{cost} )
\end{equation*}
where cost is
\begin{equation*}
    \text{cost}(|P|, \text{runtime}) = \widetilde{k} + c_{\text{rt}} \text{runtime}
\end{equation*}
and ordering of $x = (\text{recall}_x, \text{cost}_x)$ and $y = (\text{recall}_y, \text{cost}_y)$ is given by
\begin{equation*}
    x \geq y = \begin{cases}
        \text{cost}_x \leq \text{cost}_y & \text{if}~\text{recall}_x \geq R \land \text{recall}_y \geq R \\
        \text{recall}_x \geq \text{recall}_y & \text{otherwise} \\
    \end{cases}
\end{equation*}
We set $c_\text{rt}$ to a conservative value of $0.01$,
meaning we consider a unit increment in effective cardinality and a runtime increase of 100 seconds equal.

The unsupervised methods use the best hyperparameters that achieve the desired recall on all the test sets as described in Section~\ref{sec:unsupervised-baselines}.
We remind the reader that this means the same hyperparameters are used across all datasets, and not tuned per dataset.
For unsupervised ShallowBlocker we report the results when setting the quality $q$ to 0.95, 0.99, and 1.

The supervised methods get a training set of known matches different from the test set.
In order to study how the number of training matches affects the results,
we run each method with maximum \num{100}, \num{1000}, and \num{10000} sampled training matches.

\subsubsection{Unsupervised Pair Effectiveness}
We observe that ShallowBlocker is, with good margin, the method that is able to achieve the desired recall with the least number of returned pairs, both for low and high recall levels.
The contrast is most pronounced with high recall.
DeepBlocker is not able to achieve 98\% recall across all datasets because FAISS is unable to return enough pairs.
Furthermore,
baseline methods using unweighted tokens (Token Blocking, PPJoin, Minhash LSH) struggle significantly more than those relying on weighted tokens (HVTB, $\tau$-join, $k$-join).
We assume this can be attributed to the excellent discriminatory power of TF-IDF.
Unsurprisingly, $k$-join is more stable than $\tau$-join because it explicitly limits the number of pairs.
On the other hand, $\tau$ is able to prune more aggressively on some datasets but end up returning way to many pairs on others.
We see that ShallowBlocker is able avoid exploding $|P|$ while still exploiting the pruning power of similarity thresholds --- showing the strength of using $(\tau, \tau_r, k)$-joins.

\subsubsection{Supervised Pair Effectiveness}
We see that ShallowBlocker achieves the highest effectiveness in most supervised cases,
and often with a substantial margin.
The differences tends to be larger for the high recall target,
which makes sense because effective pruning is harder.
However, note that $\tau$-join and $k$-join are sometimes more effective with few training matches,
and ShallowBlocker slightly misses the high recall target for a few datasets.
This is mainly due to ShallowBlocker having more parameters and expressive power,
and therefore more prone to variance with few training matches.
The attentive reader might have noticed that unsupervised ShallowBlocker actually outperforms all the supervised methods in a few cases.
Here it is important again to remember that the hyperparameters of the unsupervised methods were picked using the test sets to show best case performance.
The supervised methods can only rely on the train set and must pick more conservative parameters to incorporate the resulting uncertainty.

While DeepBlocker significantly outperforms AutoBlock,
both deep learning-based methods are dramatically less effective than ShallowBlocker.
We see that this also make them struggle with the large datasets.

\subsubsection{Runtime}
For small and medium datasets we see that $\tau$-join and $k$-join are overall the fastest.
The reason ShallowBlocker is slower is because of the overhead it incurs for automatically selecting join parameters and performing two joins --- both in the unsupervised and supervised setting.
It is however kept consistently within a few seconds of the fastest baselines,
and importantly, pays of in competitive runtimes for the large datasets Songs and Citeseet-DBLP.
Note that runtime is tightly coupled to the pair effectiveness and the number of returned pairs.

The runtime of PPJoin and Minhash LSH variate a lot more,
and can be high for even moderately sized datasets such as iTunes-Amazon.
Mainly because it needs very permissive thresholds to achieve the target recall.
DeepBlocker is the slowest of all the unsupervised methods by a significant margin,
and the slowest together with AutoBlock in the supervised setting.
The embedding training is the bottleneck for small datasets,
while the pair generation using nearest neighbor search on the embeddings is the bottleneck for larger datasets.
For the largest dataset, Citeseer-DBLP, DeepBlocker is not able to finish within the time limit of 24 hours in the low recall setting.

The overhead of supervised ShallowBlocker is generally more than unsupervised ShallowBlocker,
as shown by the higher runtimes for most datasets.
However, on Citeseer-DBLP we also see that supervised ShallowBlocker can leverage training data to choose join conditions and approximation degree that actually significantly reduce the runtime.

\subsubsection{Approximate Joins}
We see, as expected, using approximate joins instead of exact joins has an overall negative effect on the pair effectiveness of unsupervised ShallowBlocker (except for iTunes-Amazon and Walmart-Amazon in the low recall setting, which we think is mostly noise).
The loss in effectiveness is more noticeable in the high recall setting, especially for $q = 0.95$.

The goal of using approximated joins is to reduce runtime.
For datasets the exact ShallowBlocker already processes fast we actually observe an increase in runtime.
The reason is that extra overhead of determining $\rho^*$ exceeds any reduction in the already low runtime.
However, note that for the largest dataset, Citeseer-DBLP, we are able to significantly reduce the runtime in exchange for moderate increases in $|P|$.
For example, in the high recall setting ShallowBlocker achieve a 36\% (2.5 minutes) reduction in runtime by an increase in $|P|$ of only 7\% when using $q = 0.99$.
So while approximate joins are not a silver bullet for increased speed while achieving the same recall,
and that it makes sense to run with $q = 1$ by default,
we think it is a valuable option to have at disposal for large datasets.

\subsubsection{Memory Usage}
Unsurpricingly, the deep learning-based methods use substantially more memory than classical methods.
PPJoin is the most nimble on a majority of the datasets because of its highly optimized implementation over unweighted tokens.
However, as with runtime, memory usage is heavily coupled to the number of returned pairs --- and PPJoin is therefore not able to finish within the memory limit on the largest datasets.
ShallowBlocker generally has a higher memory overhead on smaller datasets because of its extra bookkeeping for choosing parameters and running two joins.
Nonetheless, for larger datasets this overhead is outweighted by more effective join conditions resulting in low memory usage.

\subsection{Effectiveness Trade-off}

\begin{figure*}[htbp]
    \centering
    \includegraphics{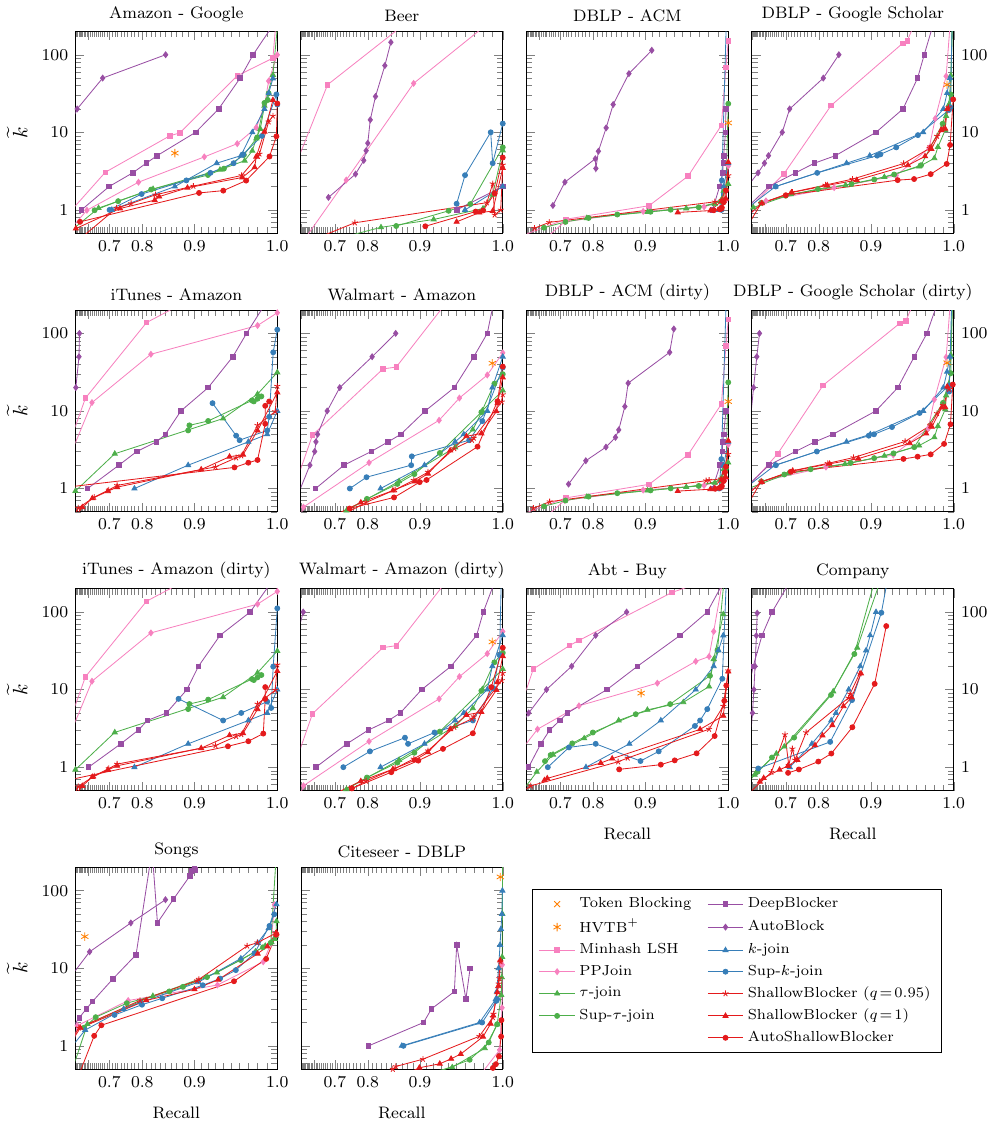}
    \caption{
        Plot for every dataset showing the trade-off between number of returned pairs (expressed as $\widetilde{k} = |P| / \min(|A|, |B|)$) and recall for all methods.
        Note that we have chosen a range of $\widetilde{k}$ and recall we deem interesting,
        which means sometimes methods are not visible in the plot.
        Token Blocking, for example, is not visible in any plot because its $\widetilde{k}$ is so large.
    }
    \label{fig:recall-vs-k}
\end{figure*}

\begin{figure*}[htb]
    \centering
    \includegraphics{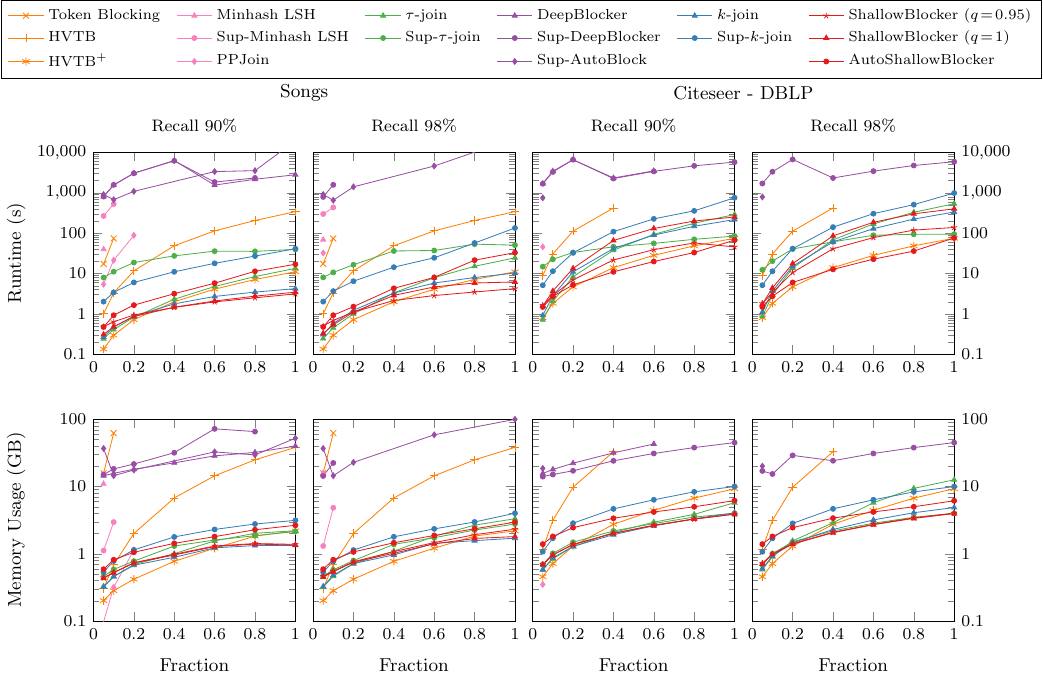}
    \caption{
        Runtime and memory consumption for all methods on Songs and Citeseer-DBLP datasets with 90\% and 98\% recall target when we vary the dataset size.
        Methods with lines that stop early hit the memory limit or similar.
    }
    \label{fig:scalability}
\end{figure*}

We study the trade-off between recall and the number of returned pairs $|P|$ in detail.
Again, for easier comparison across datasets we report the effective cardinality $\widetilde{k}$ instead of $|P|$.
Figure~\ref{fig:recall-vs-k} shows recall plotted against $\widetilde{k}$ across all datasets for each method.
We traverse the trade-off front of each baseline by varying the only threshold (similarity or cardinality) parameter or recall target.
Similarly, for ShallowBlocker we vary $k$ and for AutoShallowBlocker we use the objective function
\begin{equation*}
    f(\text{recall}, |P|, \text{runtime}) = \text{recall} - c_{k} \widetilde{k} - c_{\text{rt}} \text{runtime}
\end{equation*}
and vary $c_{k}$ (using $c_{\text{rt}} = 0.01$ as before).

We observe many of the same trends as in the previous experiment.
ShallowBlocker and AutoShallowBlocker generally demonstrate the best pair effectiveness,
while $\tau$-join and $k$-join are the strongest baselines.
Baselines using unweighted tokens (Token Blocking, Minhash LSH, and PPJoin) are overall considerably less effective,
but PPJoin is effective on datasets like DBLP-ACM, Songs, and Citeseer-DBLP.
The deep learning methods are by far the least effective.
Note that most methods not significantly effected by dirty data,
except for AutoBlock --- which do rely more on attribute boundaries.

\subsection{Scalability}

\begin{figure*}[htbp]
    \centering
    \includegraphics{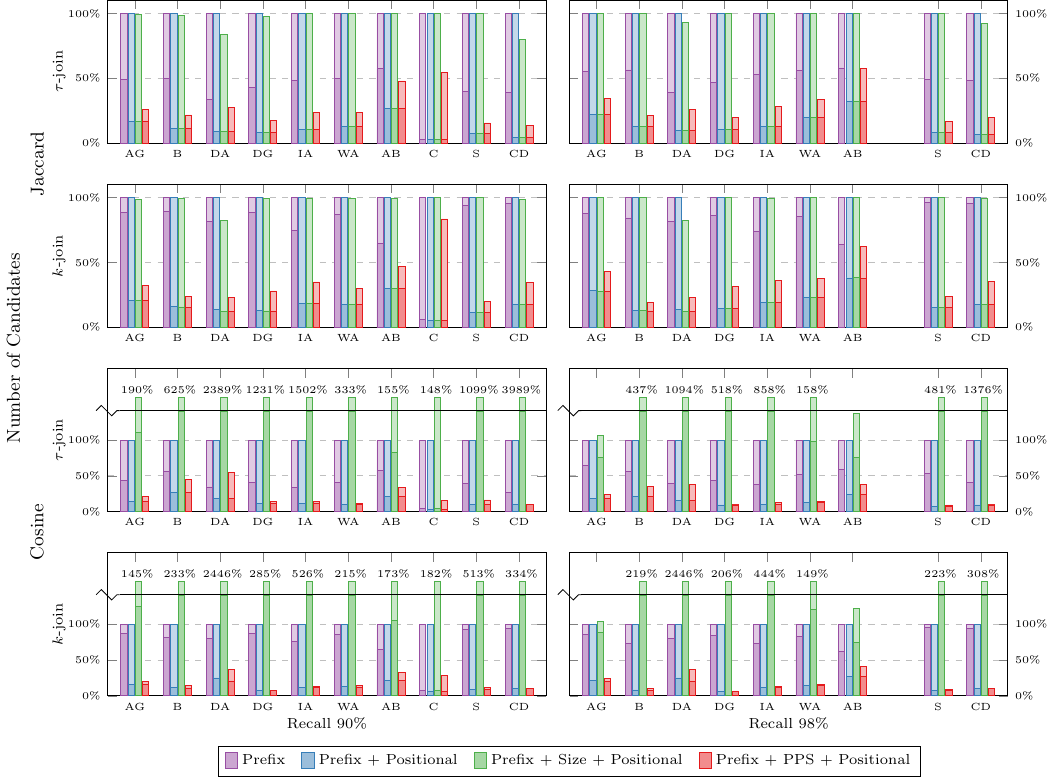}
    \caption{
        The number of pre-candidates and candidates when using different filter configurations across all datasets for $\tau$-join and $k$-join using Jaccard and Cosine with hyperparameters that achieve 90\% and 98\% recall.
        The Company dataset is omitted for 98\% since it is not possible to achieve that recall level.
        The numbers are relative to the number of pre-candidates when using only prefix filtering.
        The full height of each bar is the number of pre-candidates and the bottom is the number of candidates.
    }
    \label{fig:filter-candidates}
\end{figure*}

Figure~\ref{fig:scalability} shows the runtime and memory consumption of different methods when running on different fraction sizes of the two largest datasets, Songs and Citeseer-DBLP.
We report for configurations that achieve both 90\% and 98\% as in Section~\ref{sec:experiment-recall-target}.

From the plots we see clearly that Token Blocking, Minhash LSH, PPJoin, and AutoBlock scales poorly.
They are mostly not able to process the entire datasets.
This is mainly due to low pair effectiveness and the need for conservative hyperparameters to meet the recall target.
On the other hand,
HVTB$^+$, $\tau$-join, $k$-join, ShallowBlocker, and AutoShallowBlocker all scale reasonable well and have similar characteristics.
Note that the improved HVTB$^*$ is considerably more memory efficient than HVTB,
which is not able to process Citeseer-DBLP within the memory constraints.
As we saw in Section~\ref{sec:experiment-recall-target},
ShallowBlocker and AutoShallowBlocker have a larger overhead than many of the leaner baselines.
They have higher runtime and memory consumption for the smallest fraction,
but scale well due to aggressive pruning when the datasets grow.
The deep learning methods have a much more extreme overhead.
They are dominated by long training time for smaller fractions and then increasingly dominated by nearest neighbor search on embeddings for the larger ones.
Unsupervised DeepBlocker and Sup-AutoBlock are not able process the whole Citeseer-DBLP dataset.

\subsection{Prefix-Partitioned Suffix Filtering}

We examine the effect of our new proposed pre-candidate filter: Prefix-Partitioned Suffix (PPS) Filter.

\subsubsection{Number of Pre-Candidates}

\begin{table*}[htbp]
    \setlength{\tabcolsep}{3pt}
    \centering
    \scriptsize

\begin{tabular}{lllcccccccccccc}
\toprule
& & & \multicolumn{6}{c}{$\tau$-join} & \multicolumn{6}{c}{$k$-join} \\
\cmidrule(r){4-9}
\cmidrule(l){10-15}
Measure & Recall & Filters & DG & IA & WA & C & S & CD & DG & IA & WA & C & S & CD \\
\midrule
\multirow{8}{*}{Jaccard} & 
\multirow{4}{*}{90\%} & 
Prefix
& \textbf{153ms} & 158ms & \textbf{67.3ms} & 1m9s & 6.66s & 27.2s & 187ms & 212ms & \textbf{71.8ms} & 49.9s & 19.6s & 27m17s \\
& & Prefix + Pos.
& \underline{\textbf{150ms}} & 132ms & \underline{\textbf{67.2ms}} & 1m9s & 4.07s & 13.5s & 177ms & \textbf{163ms} & \underline{\textbf{69.2ms}} & \textbf{46.6s} & 6.25s & 6m58s \\
& & Prefix + Size + Pos.
& \textbf{151ms} & 138ms & \textbf{69.2ms} & 1m8s & 4.05s & 12.9s & 184ms & 171ms & \textbf{69.4ms} & \textbf{47.5s} & 6.51s & 7m11s \\
& & Prefix + PPS + Pos.
& \textbf{157ms} & \underline{\textbf{125ms}} & \textbf{69.1ms} & \underline{\textbf{1m5s}} & \underline{\textbf{3.57s}} & \underline{\textbf{10.4s}} & \underline{\textbf{164ms}} & \underline{\textbf{161ms}} & \textbf{69.3ms} & \underline{\textbf{46.5s}} & \underline{\textbf{4.91s}} & \underline{\textbf{4m38s}} \\
\cmidrule{2-15}
& \multirow{4}{*}{98\%} & 
Prefix
& 167ms & 184ms & \textbf{71.3ms} & - & 15.7s & 1m15s & 203ms & 230ms & 74.9ms & - & 1m28s & 31m35s \\
& & Prefix + Pos.
& \textbf{159ms} & 150ms & \textbf{70.6ms} & - & 7.05s & 30.0s & 183ms & \textbf{176ms} & \underline{\textbf{67.6ms}} & - & 19.3s & 8m10s \\
& & Prefix + Size + Pos.
& \textbf{156ms} & 155ms & \underline{\textbf{68.5ms}} & - & 7.15s & 30.3s & 184ms & 183ms & 73.2ms & - & 20.2s & 8m28s \\
& & Prefix + PPS + Pos.
& \underline{\textbf{155ms}} & \underline{\textbf{138ms}} & \textbf{70.4ms} & - & \underline{\textbf{5.83s}} & \underline{\textbf{21.4s}} & \underline{\textbf{167ms}} & \underline{\textbf{170ms}} & \textbf{70.5ms} & - & \underline{\textbf{13.9s}} & \underline{\textbf{5m28s}} \\
\cmidrule{1-15}
\multirow{8}{*}{Cosine} & 
\multirow{4}{*}{90\%} & 
Prefix
& \textbf{155ms} & 127ms & \textbf{70.5ms} & 32.6s & 5.58s & 13.1s & 180ms & 172ms & 73.7ms & 26.6s & 23.0s & 21m52s \\
& & Prefix + Pos.
& \textbf{152ms} & \textbf{119ms} & \underline{\textbf{68.4ms}} & 28.8s & 3.62s & \textbf{9.39s} & \underline{\textbf{157ms}} & \textbf{133ms} & \underline{\textbf{69.5ms}} & \textbf{25.1s} & 4.76s & 2m13s \\
& & Prefix + Size + Pos.
& 184ms & 434ms & 72.1ms & 1m3s & 46.8s & 12m18s & 260ms & 563ms & 78.4ms & 48.1s & 1m58s & 1h12m \\
& & Prefix + PPS + Pos.
& \underline{\textbf{151ms}} & \underline{\textbf{116ms}} & \textbf{68.5ms} & \underline{\textbf{26.2s}} & \underline{\textbf{3.40s}} & \underline{\textbf{9.04s}} & \textbf{157ms} & \underline{\textbf{132ms}} & \textbf{69.9ms} & \underline{\textbf{24.0s}} & \underline{\textbf{4.32s}} & \underline{\textbf{1m51s}} \\
\cmidrule{2-15}
& \multirow{4}{*}{98\%} & 
Prefix
& \textbf{156ms} & 136ms & 72.7ms & - & 18.5s & 46.3s & 207ms & 191ms & 76.6ms & - & 2m13s & 26m8s \\
& & Prefix + Pos.
& \underline{\textbf{152ms}} & \textbf{124ms} & \underline{\textbf{68.8ms}} & - & 6.20s & 18.1s & \underline{\textbf{159ms}} & 141ms & \textbf{71.9ms} & - & 13.0s & 2m38s \\
& & Prefix + Size + Pos.
& 219ms & 490ms & 81.2ms & - & 1m38s & 19m3s & 293ms & 590ms & 81.8ms & - & 5m2s & 1h20m \\
& & Prefix + PPS + Pos.
& \textbf{153ms} & \underline{\textbf{121ms}} & \textbf{70.5ms} & - & \underline{\textbf{5.38s}} & \underline{\textbf{16.3s}} & \textbf{160ms} & \underline{\textbf{134ms}} & \underline{\textbf{71.3ms}} & - & \underline{\textbf{10.4s}} & \underline{\textbf{2m14s}} \\
\bottomrule
\end{tabular}

    \caption{
        The runtime when using different filter configurations across all datasets for $\tau$-join and $k$-join using Jaccard and Cosine with hyperparameters that achieve 90\% and 98\% recall.
    }
    \label{tab:filter-runtimes}
\end{table*}

\begin{table}[htbp]
    \setlength{\tabcolsep}{1.5pt}
    \centering
    \scriptsize

\begin{tabular}{clcccccc}
\toprule
& & \multicolumn{6}{c}{Dataset} \\
\cmidrule{3-8}
Recall & \multicolumn{1}{c}{Filters} & DG & IA & WA & C & S & CD \\
\midrule
\multirow{4}{*}{90\%} & 
Prefix
& 3.26s & 2.64s & 710ms & 52.6s & 12.6s & 9m19s \\
& Prefix + Pos.
& \textbf{2.80s} & \textbf{2.26s} & \textbf{649ms} & 48.7s & 4.49s & 2m18s \\
& Prefix + Size + Pos.
& 6.12s & 13.3s & 1.18s & 2m5s & 4.70s & 9m54s \\
& Prefix + PPS + Pos.
& \underline{\textbf{2.71s}} & \underline{\textbf{2.22s}} & \underline{\textbf{642ms}} & \underline{\textbf{44.7s}} & \underline{\textbf{3.62s}} & \underline{\textbf{1m12s}} \\
\cmidrule{2-8}
\multirow{4}{*}{98\%} & 
Prefix
& 3.58s & 3.35s & 1.05s & - & 31.8s & 23m38s \\
& Prefix + Pos.
& \textbf{3.09s} & \textbf{2.90s} & \textbf{971ms} & - & 8.99s & 5m53s \\
& Prefix + Size + Pos.
& 6.69s & 21.9s & 2.08s & - & 9.35s & 15m33s \\
& Prefix + PPS + Pos.
& \underline{\textbf{2.97s}} & \underline{\textbf{2.84s}} & \underline{\textbf{954ms}} & - & \underline{\textbf{6.35s}} & \underline{\textbf{3m24s}} \\
\bottomrule
\end{tabular}

    \caption{
        The runtime when using different filter configurations across all datasets for unsupervised ShallowBlocker with hyperparameters that achieve 90\% and 98\% recall.
    }
    \label{tab:shallowblocker-filter-runtime}
\end{table}

Figure~\ref{fig:filter-candidates} shows the relative number of pre-candidates and candidates for $\tau$-join and $k$-join using Jaccard and Cosine with thresholds achieving 90\% and 98\% recall per dataset.
Reporting results for $\tau$-join and $k$-join instead of ShallowBlocker let us control for the effect of different constraint types and similarity measures.
We compare the number of pre-candidates and candidates when using prefix filter only, prefix and positional filter, PPJoin-style filters (prefix, size, positional), and \ttrkjoin filters (prefix, PPS, positional).
Note that in order to use size filtering for Cosine we must use $L_1$ based bounds and thereby the weaker prefix bound from PPJoin.
The effect of using the improved L2AP~\cite{anastasiuL2APFastCosine2014} prefix bound without PPS filtering can therefore be observed by looking at the results for using only prefix and positional filtering.

Overall,
we observe a reduction between approximately 20\% and 90\% of the number of pre-candidates when applying PPS filtering.
For the majority of cases the reduction is 70\% or more.
This is significantly more than for PPJoin-style size filtering,
which we see have negligible effect for almost all datasets and at most reduce the number of pre-candidates with around 20\%.
For Cosine the weaker $L_1$ prefix bound necessary to use size filtering makes it hard to compare PPS against size filtering directly.
However, it clearly shows that the $L_2$ based prefix bound from L2AP is a large improvement.

From the results we observe no major difference in behavior between thresholds for high and low recall.
Comparing $\tau$-join and $k$-join we see the main difference is the relative number of pre-candidates and candidates.
$\tau$-join has fewer candidates compared to pre-candidates than $k$-join.
This is mostly because $\tau$-join is able to exploit the similarity threshold to prune when indexing,
and thereby reducing the number of duplicates when looking up the inverted lists query time.
Therefore, we would expect the runtime effect of PPS filtering to be greater for $k$-join.
Comparing Jaccard against Cosine we see that PPS filtering is most effective for Cosine --- most likely because Jaccard relies on the looser $\text{minPrefix}$ of the inverted lists as bound on $p_b$ to get similarly tight suffix bounds.

\subsubsection{Runtime}

Table~\ref{tab:filter-runtimes} lists the runtime of the experiments above from Figure~\ref{fig:filter-candidates}.
For clarity and conciseness we only report runtimes for datasets with runtime above 50ms.
The remaining datasets are so small that the runtime is practically indistinguishable between different filtering techniques.

Previous work has shown that filtering techniques that generate fewer candidates than PPJoin or AllPairs often ends up being slower because of the extra overhead~\cite{mannEmpiricalEvaluationSet2016}.
However, we see from the reported runtimes that PPS filtering either improves or matches size filtering across the board.
The improvements are most noticeable for the largest datasets,
while the overhead of any filtering beyond prefix provides no or minimal gains for smaller datasets.
Furthermore, we observe the enormous negative impact PPJoin style $L_1$ size filtering (with corresponding prefix bounds) has when using Cosine.
As expected,
the improvements are greater for $k$-join than $\tau$-join.

In contrast to $\tau$-join and $k$-join,
\ttrkjoin mixes different join conditions and ShallowBlocker mixes different similarity measures.
Table~\ref{tab:shallowblocker-filter-runtime} lists the runtimes for unsupervised ShallowBlocker ($q=1$) using the different filtering configurations for $k$ yielding 90\% and 98\% recall.
These results show that the runtimes from our controlled experiments on $\tau$-join and $k$-join are indicative of the effect it has on general use of \ttrkjoin.

\section{Conclusion}\label{sec:conclusion}
We have introduced, ShallowBlocker, a novel blocking method based on efficient hybrid set similarity joins and shown through extensive experiments that it is state-of-the-art.
Specifically, it is able to achieve the same recall with fewer returned pairs in most cases compared to previous state-of-the-art methods in reasonable time.
While it has a non-trivial runtime overhead for small datasets, experiments show it is more efficient on large datasets than all the baselines.

The need for dataset-specific laborsome human engineering and tuning is often highlighted as a key challenge for blocking.
It is tempting to think that classical blocking methods are too rigid and therefore will always need manual feature engineering for good results,
so we require the automatic feature learning from deep learning to create the best hands-off solutions.
However, our work shows that classical string similarity-based techniques can be used to create hands-off blocking methods that outperform state-of-the-art deep learning based methods --- even on dirty data.
Importantly,
our proposed method requires considerably fewer computational resources and is significantly more interpretable.
Despite recent advances,
our work suggests that classical techniques for blocking are still highly competitive to deep learning methods.

There is still a lot to explore with the proposed techniques and methods.
We think $(\tau,\tau_r,k)$-join is a powerful primitive and we hope to see other fruitful uses of \ttrkjoin outside the scope of ShallowBlocker.
Even though we see ShallowBlocker outperforming deep learning-based methods on the benchmark datasets used in this paper,
it is an inherent limitation that methods based on classical string similarity rely on some level of syntactic similarity.
It would be interesting to explore the possibility of combining elements of both ShallowBlocker and deep learning techniques to exploit the strengths of both approaches.

\section*{Funding}
This work was supported by Cognite and the Research Council of Norway [grant number 298998].


 \bibliographystyle{elsarticle-num-names} 
 \bibliography{bib}

\begin{thebibliography}{26}
\expandafter\ifx\csname natexlab\endcsname\relax\def\natexlab#1{#1}\fi
\providecommand{\url}[1]{\texttt{#1}}
\providecommand{\href}[2]{#2}
\providecommand{\path}[1]{#1}
\providecommand{\DOIprefix}{doi:}
\providecommand{\ArXivprefix}{arXiv:}
\providecommand{\URLprefix}{URL: }
\providecommand{\Pubmedprefix}{pmid:}
\providecommand{\doi}[1]{\href{http://dx.doi.org/#1}{\path{#1}}}
\providecommand{\Pubmed}[1]{\href{pmid:#1}{\path{#1}}}
\providecommand{\bibinfo}[2]{#2}
\ifx\xfnm\relax \def\xfnm[#1]{\unskip,\space#1}\fi
\bibitem[{Christen(2012)}]{christenDataMatchingConcepts2012}
\bibinfo{author}{P.~Christen}, \bibinfo{title}{Data {{Matching}}: {{Concepts}}
  and {{Techniques}} for {{Record Linkage}}, {{Entity Resolution}}, and
  {{Duplicate Detection}}}, \bibinfo{publisher}{{Springer-Verlag}},
  \bibinfo{year}{2012}. \DOIprefix\doi{10.1007/978-3-642-31164-2}.
\bibitem[{Doan et~al.(2012)Doan, Halevy, and
  Ives}]{doanPrinciplesDataIntegration2012}
\bibinfo{author}{A.~Doan}, \bibinfo{author}{A.~Halevy}, \bibinfo{author}{Z.~G.
  Ives}, \bibinfo{title}{Principles of Data Integration},
  \bibinfo{publisher}{{Morgan Kaufmann}}, \bibinfo{year}{2012}.
\bibitem[{Papadakis et~al.(2021)Papadakis, Skoutas, Thanos, and
  Palpanas}]{papadakisBlockingFilteringTechniques2021}
\bibinfo{author}{G.~Papadakis}, \bibinfo{author}{D.~Skoutas},
  \bibinfo{author}{E.~Thanos}, \bibinfo{author}{T.~Palpanas},
\newblock \bibinfo{title}{Blocking and {{Filtering Techniques}} for {{Entity
  Resolution}}: {{A Survey}}},
\newblock \bibinfo{journal}{ACM Comput. Surv.} \bibinfo{volume}{53}
  (\bibinfo{year}{2021}) \bibinfo{pages}{1--42}.
  \DOIprefix\doi{10.1145/3377455}.
\bibitem[{Ebraheem et~al.(2018)Ebraheem, Thirumuruganathan, Joty, Ouzzani, and
  Tang}]{ebraheemDistributedRepresentationsTuples2018}
\bibinfo{author}{M.~Ebraheem}, \bibinfo{author}{S.~Thirumuruganathan},
  \bibinfo{author}{S.~Joty}, \bibinfo{author}{M.~Ouzzani},
  \bibinfo{author}{N.~Tang},
\newblock \bibinfo{title}{Distributed representations of tuples for entity
  resolution},
\newblock \bibinfo{journal}{Proc. VLDB Endow.} \bibinfo{volume}{11}
  (\bibinfo{year}{2018}) \bibinfo{pages}{1454--1467}.
  \DOIprefix\doi{10.14778/3236187.3236198}.
\bibitem[{Thirumuruganathan et~al.(2021)Thirumuruganathan, Li, Tang, Ouzzani,
  Govind, Paulsen, Fung, and Doan}]{thirumuruganathanDeepLearningBlocking2021}
\bibinfo{author}{S.~Thirumuruganathan}, \bibinfo{author}{H.~Li},
  \bibinfo{author}{N.~Tang}, \bibinfo{author}{M.~Ouzzani},
  \bibinfo{author}{Y.~Govind}, \bibinfo{author}{D.~Paulsen},
  \bibinfo{author}{G.~M. Fung}, \bibinfo{author}{A.~Doan},
\newblock \bibinfo{title}{Deep {{Learning}} for {{Blocking}} in {{Entity
  Matching}}: {{A Design Space Exploration}}},
\newblock \bibinfo{journal}{Proceedings of the VLDB Endowment}
  \bibinfo{volume}{14} (\bibinfo{year}{2021}).
\bibitem[{Zhang et~al.(2020)Zhang, Wei, Sisman, Dong, Faloutsos, and
  Page}]{zhangAutoBlockHandsoffBlocking2020}
\bibinfo{author}{W.~Zhang}, \bibinfo{author}{H.~Wei},
  \bibinfo{author}{B.~Sisman}, \bibinfo{author}{X.~L. Dong},
  \bibinfo{author}{C.~Faloutsos}, \bibinfo{author}{D.~Page},
\newblock \bibinfo{title}{{{AutoBlock}}: {{A Hands-off Blocking Framework}} for
  {{Entity Matching}}},
\newblock in: \bibinfo{booktitle}{Proceedings of the 13th {{International
  Conference}} on {{Web Search}} and {{Data Mining}}},
  \bibinfo{publisher}{{ACM}}, \bibinfo{year}{2020}, pp.
  \bibinfo{pages}{744--752}. \DOIprefix\doi{10.1145/3336191.3371813}.
\bibitem[{Papadakis et~al.(2022)Papadakis, Fisichella, Schoger, Mandilaras,
  Augsten, and Nejdl}]{papadakisHowReduceSearch2022}
\bibinfo{author}{G.~Papadakis}, \bibinfo{author}{M.~Fisichella},
  \bibinfo{author}{F.~Schoger}, \bibinfo{author}{G.~Mandilaras},
  \bibinfo{author}{N.~Augsten}, \bibinfo{author}{W.~Nejdl},
\newblock \bibinfo{title}{How to reduce the search space of {{Entity
  Resolution}}: With {{Blocking}} or {{Nearest Neighbor}} search?},
\newblock \bibinfo{journal}{arXiv:2202.12521 [cs]}  (\bibinfo{year}{2022}).
  \href{http://arxiv.org/abs/2202.12521}{{\tt arXiv:2202.12521}}.
\bibitem[{Chaudhuri et~al.(2006)Chaudhuri, Ganti, and
  Kaushik}]{chaudhuriPrimitiveOperatorSimilarity2006}
\bibinfo{author}{S.~Chaudhuri}, \bibinfo{author}{V.~Ganti},
  \bibinfo{author}{R.~Kaushik},
\newblock \bibinfo{title}{A {{Primitive Operator}} for {{Similarity Joins}} in
  {{Data Cleaning}}},
\newblock in: \bibinfo{booktitle}{22nd {{International Conference}} on {{Data
  Engineering}} ({{ICDE}}'06)}, \bibinfo{publisher}{{IEEE}},
  \bibinfo{year}{2006}, pp. \bibinfo{pages}{5--5}.
  \DOIprefix\doi{10.1109/ICDE.2006.9}.
\bibitem[{Bayardo et~al.(2007)Bayardo, Ma, and
  Srikant}]{bayardoScalingAllPairs2007}
\bibinfo{author}{R.~J. Bayardo}, \bibinfo{author}{Y.~Ma},
  \bibinfo{author}{R.~Srikant},
\newblock \bibinfo{title}{Scaling up all pairs similarity search},
\newblock in: \bibinfo{booktitle}{Proceedings of the 16th International
  Conference on {{World Wide Web}} - {{WWW}} '07}, \bibinfo{publisher}{{ACM
  Press}}, \bibinfo{year}{2007}, p. \bibinfo{pages}{131}.
  \DOIprefix\doi{10.1145/1242572.1242591}.
\bibitem[{Xiao et~al.(2009)Xiao, Wang, Lin, and
  Shang}]{xiaoTopkSetSimilarity2009}
\bibinfo{author}{C.~Xiao}, \bibinfo{author}{W.~Wang}, \bibinfo{author}{X.~Lin},
  \bibinfo{author}{H.~Shang},
\newblock \bibinfo{title}{Top-k {{Set Similarity Joins}}},
\newblock in: \bibinfo{booktitle}{2009 {{IEEE}} 25th {{International
  Conference}} on {{Data Engineering}}}, \bibinfo{publisher}{{IEEE}},
  \bibinfo{year}{2009}, pp. \bibinfo{pages}{916--927}.
  \DOIprefix\doi{10.1109/ICDE.2009.111}.
\bibitem[{Xiao et~al.(2011)Xiao, Wang, Lin, Yu, and
  Wang}]{xiaoEfficientSimilarityJoins2011}
\bibinfo{author}{C.~Xiao}, \bibinfo{author}{W.~Wang}, \bibinfo{author}{X.~Lin},
  \bibinfo{author}{J.~X. Yu}, \bibinfo{author}{G.~Wang},
\newblock \bibinfo{title}{Efficient similarity joins for near-duplicate
  detection},
\newblock \bibinfo{journal}{ACM Trans. Database Syst.} \bibinfo{volume}{36}
  (\bibinfo{year}{2011}) \bibinfo{pages}{1--41}.
  \DOIprefix\doi{10.1145/2000824.2000825}.
\bibitem[{Anastasiu and Karypis(2014)}]{anastasiuL2APFastCosine2014}
\bibinfo{author}{D.~C. Anastasiu}, \bibinfo{author}{G.~Karypis},
\newblock \bibinfo{title}{{{L2AP}}: {{Fast}} cosine similarity search with
  prefix {{L-2}} norm bounds},
\newblock in: \bibinfo{booktitle}{2014 {{IEEE}} 30th {{International
  Conference}} on {{Data Engineering}}}, \bibinfo{publisher}{{IEEE}},
  \bibinfo{year}{2014}, pp. \bibinfo{pages}{784--795}.
  \DOIprefix\doi{10.1109/ICDE.2014.6816700}.
\bibitem[{Mann et~al.(2016)Mann, Augsten, and
  Bouros}]{mannEmpiricalEvaluationSet2016}
\bibinfo{author}{W.~Mann}, \bibinfo{author}{N.~Augsten},
  \bibinfo{author}{P.~Bouros},
\newblock \bibinfo{title}{An empirical evaluation of set similarity join
  techniques},
\newblock \bibinfo{journal}{Proc. VLDB Endow.} \bibinfo{volume}{9}
  (\bibinfo{year}{2016}) \bibinfo{pages}{636--647}.
  \DOIprefix\doi{10.14778/2947618.2947620}.
\bibitem[{Barlaug and Gulla(2021)}]{barlaugNeuralNetworksEntity2021}
\bibinfo{author}{N.~Barlaug}, \bibinfo{author}{J.~A. Gulla},
\newblock \bibinfo{title}{Neural {{Networks}} for {{Entity Matching}}: {{A
  Survey}}},
\newblock \bibinfo{journal}{ACM Trans. Knowl. Discov. Data}
  \bibinfo{volume}{15} (\bibinfo{year}{2021}) \bibinfo{pages}{1--37}.
  \DOIprefix\doi{10.1145/3442200}.
\bibitem[{Xiao et~al.(2008)Xiao, Wang, Lin, and
  Yu}]{xiaoEfficientSimilarityJoins2008}
\bibinfo{author}{C.~Xiao}, \bibinfo{author}{W.~Wang}, \bibinfo{author}{X.~Lin},
  \bibinfo{author}{J.~X. Yu},
\newblock \bibinfo{title}{Efficient similarity joins for near duplicate
  detection},
\newblock in: \bibinfo{booktitle}{Proceeding of the 17th International
  Conference on {{World Wide Web}} - {{WWW}} '08}, \bibinfo{publisher}{{ACM
  Press}}, \bibinfo{year}{2008}, p. \bibinfo{pages}{131}.
  \DOIprefix\doi{10.1145/1367497.1367516}.
\bibitem[{Li et~al.(2008)Li, Lu, and Lu}]{liEfficientMergingFiltering2008}
\bibinfo{author}{C.~Li}, \bibinfo{author}{J.~Lu}, \bibinfo{author}{Y.~Lu},
\newblock \bibinfo{title}{Efficient {{Merging}} and {{Filtering Algorithms}}
  for {{Approximate String Searches}}},
\newblock in: \bibinfo{booktitle}{2008 {{IEEE}} 24th {{International
  Conference}} on {{Data Engineering}}}, \bibinfo{publisher}{{IEEE}},
  \bibinfo{year}{2008}, pp. \bibinfo{pages}{257--266}.
  \DOIprefix\doi{10.1109/ICDE.2008.4497434}.
\bibitem[{Li et~al.(2021)Li, Cheng, Chu, He, and
  Chaudhuri}]{liAutoFuzzyJoinAutoProgramFuzzy2021}
\bibinfo{author}{P.~Li}, \bibinfo{author}{X.~Cheng}, \bibinfo{author}{X.~Chu},
  \bibinfo{author}{Y.~He}, \bibinfo{author}{S.~Chaudhuri},
\newblock \bibinfo{title}{Auto-{{FuzzyJoin}}: {{Auto-Program Fuzzy Similarity
  Joins Without Labeled Examples}}},
\newblock in: \bibinfo{booktitle}{Proceedings of the 2021 {{International
  Conference}} on {{Management}} of {{Data}}}, \bibinfo{publisher}{{ACM}},
  \bibinfo{year}{2021}, pp. \bibinfo{pages}{1064--1076}.
  \DOIprefix\doi{10.1145/3448016.3452824}.
\bibitem[{Yang et~al.(2020)Yang, Zheng, Li, Zhao, Zhou, and
  Jensen}]{yangAdaptiveTopkOverlap2020}
\bibinfo{author}{Z.~Yang}, \bibinfo{author}{B.~Zheng}, \bibinfo{author}{G.~Li},
  \bibinfo{author}{X.~Zhao}, \bibinfo{author}{X.~Zhou}, \bibinfo{author}{C.~S.
  Jensen},
\newblock \bibinfo{title}{Adaptive {{Top-k Overlap Set Similarity Joins}}},
\newblock in: \bibinfo{booktitle}{2020 {{IEEE}} 36th {{International
  Conference}} on {{Data Engineering}} ({{ICDE}})},
  \bibinfo{publisher}{{IEEE}}, \bibinfo{year}{2020}, pp.
  \bibinfo{pages}{1081--1092}. \DOIprefix\doi{10.1109/ICDE48307.2020.00098}.
\bibitem[{O'hare et~al.(2021)O'hare, {Jurek-Loughrey}, and
  De~Campos}]{ohareHighValueTokenBlockingEfficient2021}
\bibinfo{author}{K.~O'hare}, \bibinfo{author}{A.~{Jurek-Loughrey}},
  \bibinfo{author}{C.~De~Campos},
\newblock \bibinfo{title}{High-{{Value Token-Blocking}}: {{Efficient Blocking
  Method}} for {{Record Linkage}}},
\newblock \bibinfo{journal}{ACM Trans. Knowl. Discov. Data}
  \bibinfo{volume}{16} (\bibinfo{year}{2021}) \bibinfo{pages}{1--17}.
  \DOIprefix\doi{10.1145/3450527}.
\bibitem[{Mudgal et~al.(2018)Mudgal, Li, Rekatsinas, Doan, Park, Krishnan,
  Deep, Arcaute, and Raghavendra}]{mudgalDeepLearningEntity2018}
\bibinfo{author}{S.~Mudgal}, \bibinfo{author}{H.~Li},
  \bibinfo{author}{T.~Rekatsinas}, \bibinfo{author}{A.~Doan},
  \bibinfo{author}{Y.~Park}, \bibinfo{author}{G.~Krishnan},
  \bibinfo{author}{R.~Deep}, \bibinfo{author}{E.~Arcaute},
  \bibinfo{author}{V.~Raghavendra},
\newblock \bibinfo{title}{Deep {{Learning}} for {{Entity Matching}}: {{A Design
  Space Exploration}}},
\newblock in: \bibinfo{booktitle}{Proc. {{SIGMOD}} 2018},
  \bibinfo{publisher}{{ACM Press}}, \bibinfo{year}{2018}, pp.
  \bibinfo{pages}{19--34}. \DOIprefix\doi{10.1145/3183713.3196926}.
\bibitem[{Das et~al.(2017)Das, G.C., Doan, Naughton, Krishnan, Deep, Arcaute,
  Raghavendra, and Park}]{dasFalconScalingHandsOff2017}
\bibinfo{author}{S.~Das}, \bibinfo{author}{P.~S. G.C.},
  \bibinfo{author}{A.~Doan}, \bibinfo{author}{J.~F. Naughton},
  \bibinfo{author}{G.~Krishnan}, \bibinfo{author}{R.~Deep},
  \bibinfo{author}{E.~Arcaute}, \bibinfo{author}{V.~Raghavendra},
  \bibinfo{author}{Y.~Park},
\newblock \bibinfo{title}{Falcon: {{Scaling Up Hands-Off Crowdsourced Entity
  Matching}} to {{Build Cloud Services}}},
\newblock in: \bibinfo{booktitle}{Proceedings of the 2017 {{ACM International
  Conference}} on {{Management}} of {{Data}}}, \bibinfo{publisher}{{Association
  for Computing Machinery}}, \bibinfo{year}{2017}, pp.
  \bibinfo{pages}{1431--1446}. \DOIprefix\doi{10.1145/3035918.3035960}.
\bibitem[{Das et~al.(2022)Das, Doan, G.~C., Gokhale, Konda, Govind, and
  Paulsen}]{dasMagellanDataRepository2022}
\bibinfo{author}{S.~Das}, \bibinfo{author}{A.~Doan}, \bibinfo{author}{P.~S.
  G.~C.}, \bibinfo{author}{C.~Gokhale}, \bibinfo{author}{P.~Konda},
  \bibinfo{author}{Y.~Govind}, \bibinfo{author}{D.~Paulsen},
  \bibinfo{title}{The {{Magellan Data Repository}}}, \bibinfo{year}{2022}.
\bibitem[{Gionis et~al.(1999)Gionis, Indyk, and
  Motwani}]{gionisSimilaritySearchHigh1999}
\bibinfo{author}{A.~Gionis}, \bibinfo{author}{P.~Indyk},
  \bibinfo{author}{R.~Motwani},
\newblock \bibinfo{title}{Similarity {{Search}} in {{High Dimensions}} via
  {{Hashing}}},
\newblock in: \bibinfo{booktitle}{Proceedings of the 25th {{International
  Conference}} on {{Very Large Data Bases}}}, \bibinfo{publisher}{{Morgan
  Kaufmann Publishers Inc.}}, \bibinfo{year}{1999}, pp.
  \bibinfo{pages}{518--529}.
\bibitem[{Broder(1997)}]{broderResemblanceContainmentDocuments1997}
\bibinfo{author}{A.~Broder},
\newblock \bibinfo{title}{On the resemblance and containment of documents},
\newblock in: \bibinfo{booktitle}{Proceedings. {{Compression}} and
  {{Complexity}} of {{SEQUENCES}} 1997 ({{Cat}}. {{No}}.{{97TB100171}})},
  \bibinfo{year}{1997}, pp. \bibinfo{pages}{21--29}.
  \DOIprefix\doi{10.1109/SEQUEN.1997.666900}.
\bibitem[{Broder et~al.(2000)Broder, Charikar, Frieze, and
  Mitzenmacher}]{broderMinWiseIndependentPermutations2000}
\bibinfo{author}{A.~Z. Broder}, \bibinfo{author}{M.~Charikar},
  \bibinfo{author}{A.~M. Frieze}, \bibinfo{author}{M.~Mitzenmacher},
\newblock \bibinfo{title}{Min-{{Wise Independent Permutations}}},
\newblock \bibinfo{journal}{Journal of Computer and System Sciences}
  \bibinfo{volume}{60} (\bibinfo{year}{2000}) \bibinfo{pages}{630--659}.
  \DOIprefix\doi{10.1006/jcss.1999.1690}.
\bibitem[{Johnson et~al.(2021)Johnson, Douze, and
  Jegou}]{johnsonBillionScaleSimilaritySearch2021}
\bibinfo{author}{J.~Johnson}, \bibinfo{author}{M.~Douze},
  \bibinfo{author}{H.~Jegou},
\newblock \bibinfo{title}{Billion-{{Scale Similarity Search}} with {{GPUs}}},
\newblock \bibinfo{journal}{IEEE Trans. Big Data} \bibinfo{volume}{7}
  (\bibinfo{year}{2021}) \bibinfo{pages}{535--547}.
  \DOIprefix\doi{10.1109/TBDATA.2019.2921572}.

\end{thebibliography}





\end{document}